\documentclass[fleqn,usenatbib]{mnras}

\usepackage{newtxtext,newtxmath}
\usepackage{ulem}

\usepackage[T1]{fontenc}

\usepackage{graphicx}	
\usepackage{amsmath}	
\usepackage{xcolor}
\usepackage{array}
\usepackage{multirow}

\definecolor{pgreen}{RGB}{46, 121, 10}
\setcitestyle{notesep={}}
\newcommand{\shark}{\textsc{shark}}
\newcommand{\sharkfit}{\textsc{shark$_\mathrm{fit}$}}

\newcommand{\prospect}{\textsc{ProSpect}}

\newcommand{\newtext}[1]{\textcolor{orange}{#1}}
\newcommand{\mstar}[1]{$10^{#1}$ M$_\odot$}

\defcitealias{taylor2015}{T15}
\defcitealias{lagos2018}{L18}
\defcitealias{bellstedt2020b}{B20b}
\defcitealias{thorne2021}{T21}

\title[Forensic evolution of galaxy colours]{Forensic reconstruction of galaxy colour evolution and population characterisation}

\author[Bravo et al.]{
Mat\'ias Bravo$^1$\thanks{E-mail:matias.bravo@icrar.org},
Aaron S.~G. Robotham$^{1,2}$,
Claudia del P. Lagos$^{1,2,3}$,
\newauthor{
Luke J. M. Davies$^1$,
Sabine Bellstedt$^1$ and
Jessica E. Thorne$^1$}
\\
$^{1}$International Centre for Radio Astronomy Research (ICRAR), M468, University of Western Australia, 35 Stirling Hwy, Crawley, \\WA 6009, Australia.\\
$^{2}$ARC Centre of Excellence for All Sky Astrophysics in 3 Dimensions (ASTRO 3D).\\
$^{3}$Cosmic Dawn Center (DAWN). 
}

\date{Accepted XXX. Received YYY; in original form ZZZ}

\pubyear{2021}

\begin{document}
\label{firstpage}
\pagerange{\pageref{firstpage}--\pageref{lastpage}}
\maketitle

\begin{abstract}
Mapping the evolution of galaxy colours, from blue star-forming to red passive systems, is fundamental to understand the processes involved in galaxy evolution.
To this end, we reconstruct the colour evolution of low-redshift galaxies, combining stellar templates with star formation and metallicity histories of galaxies from the Galaxy And Mass Assembly survey and \shark\ semi-analytic model.
We use these colour histories to robustly characterise the evolution of red and blue galaxy populations over cosmic time.
Using a Gaussian Mixture Model to characterise the colour distribution at any given epoch and stellar mass, we find both observations and simulations strongly favour a model with only two populations (blue and red), with no evidence for a third "green" population.
We map the evolution of mean, weight, and scatter of the blue and red populations as a function of both stellar mass and lookback time.
Using our simulated galaxy catalogue as a testbed, we find that we can accurately recover galaxies colour histories up to a lookback time of $\sim6$ Gyr.
We find that both populations show little change in the mean colour for low-mass galaxies, while the colours at the massive end become significantly redder with time.
The stellar mass above which the galaxy population is predominantly red decreases by 0.3 dex in the last 5 Gyrs.
We find a good agreement between observations and simulations, with the largest tension being that massive galaxies from \shark\ are too blue (a known issue with many galaxy evolution models).
\end{abstract}


\section{Introduction}\label{sec:intro}

Galaxies in the Local Universe display a bimodal distribution in observed optical colours \citep[e.g.,][]{strateva2001,hogg2002,blanton2003,baldry2004,baldry2006,driver2006}, with a well-defined low-scatter "red "population (the so-called "red sequence") and a broader "blue" population (known as the "blue cloud").
When colours are corrected for dust reddening, which confuses intrinsically red galaxies with blue but dust-obscured galaxies, this bimodality becomes even more striking \citep[][, hereafter T15]{taylor2015}.
This bimodality also shows a marked dependency on stellar mass, with a higher fraction of massive galaxies being red \citep[e.g.,][; \citetalias{taylor2015}]{baldry2004,peng2010}, and a secondary dependence on the environment, where galaxies in low-density environments are more commonly blue \citep[e.g.,][]{kauffmann2004,baldry2006,peng2010,davies2019b}.
However, these populations do not retain the same characteristics with redshift, as the global distribution of galaxies becomes bluer and the dominance of the blue cloud increases with lookback time \citep[e.g.,][]{wolf2003,bell2004,williams2009}.
Conversely, red galaxies are typically more massive and become a larger fraction of the total galaxy population with time, which implies that galaxies form blue and at some point become red as the universe evolves \citep[e.g.,][]{faber2007}.
A fundamental goal for understanding galaxy evolution is explaining how and why this change from blue to red takes place.

For the intrinsic colour of a galaxy to be red, it requires the absence of blue stars, which have far shorter lifespans than their red counterparts \citep[e.g.,][]{schaller1992}.
This happens when a galaxy has stopped forming stars long enough that the last blue stars have exhausted themselves and only the red stars remain.
This process of star formation slowing down and finally ceasing is commonly referred to as "quenching" \citep[e.g.,][]{blanton2006,borch2006,faber2007,fang2013,moustakas2013,peng2015}, which occurs when galaxies no longer have a ready supply of gas for star formation.
There are a variety of physical processes that can facilitate the quenching process, such as active galactic nuclei (AGN) feedback \citep[e.g.,][]{springel2005b,bower2006,croton2006,hopkins2006,somerville2008}, supernovae feedback \citep[e.g.,][]{springel2005b,dallavecchia2012,lagos2013}, ram-pressure stripping \citep[e.g.,][]{crowl2005,machacek2006,kawata2008,mccarthy2008} or strangulation \citep[e.g.,][]{balogh2000,keres2005,dekel2006,peng2015}.
While we have a good understanding of these mechanisms, in practice we know very little about which mechanism dominates the quenching of galaxies in specific populations as defined by e.g. morphology, environment or stellar mass, and how the prevalence of these processes changes with time.
As such, the astrophysical mechanisms that drive galaxies to transition from the blue cloud to the red sequence, and their prevalence as a function of other galaxy properties is far from clear \citep[e.g.,][]{wetzel2013,schawinski2014,hahn2017,belli2019,belli2021}.

The region between the red and blue populations, which is commonly referred to as the "green valley", is sparsely populated \citep[e.g.,][]{martin2007,wyder2007}.
This has been suggested as evidence that, whichever mechanisms are responsible for transforming galaxies from blue to red, must be rapid \citep[relative to cosmic timescales, e.g.,][]{schawinski2014,bremer2018}.
Results from simulations however suggest that different quenching mechanisms have different timescales, such as stellar feedback being significantly slower than AGN feedback or environmental effects \citep[$\sim4$ Gyr compared to $\sim2$ Gyr, e.g.,][]{trayford2016,nelson2018a,wright2019}.
Hence, the study of the timescales of these colour changes, from an observational perspective can provide unique constraints on the physical mechanisms behind star formation quenching.

Although we can observe and measure the colours of different galaxy populations at different lookback times, the inference of the colour transformation of galaxies is difficult, as it is unclear how to connect the observed properties across cosmic time.
In particular, a major challenge is how to connect red galaxies to their blue progenitors, avoiding the effects of progenitor bias \citep[e.g.,][]{vandokkum1996,kaviraj2009,belli2015}.
Instead of trying to establish a connection between galaxy samples at different cosmic times, another possible approach would be to reconstruct the evolutionary history of a given sample.
This evolution is encoded in the light emitted by its stars, which can be extracted with the use of spectral energy distribution (SED) fitting techniques.
These methods use of single stellar population (SSP) spectral templates, with a combination of these SSPs modelling the stellar light emission of galaxies, plus models for other physical processes like dust attenuation and re-emission (e.g., see the reviews by \citealt{walcher2011} and \citealt{conroy2013}).

To limit the number of free parameters required for SED fitting, some model is also required for the star formation history of the galaxy (SFH), which restricts the possible combination of stellar templates.
SFH models are commonly divided between those that assume a single functional form for the SFH \citep[referred to as "parametric", e.g.,][]{carnall2019} and those that assume a more complex model \citep["non-parametric", e.g.,][]{pacifici2012,iyer2017,leja2019a}.
While highly desirable, the inclusion of an evolving gas-phase metallicity (metallicity history, $Z$H) is uncommon in SED fitting tools\footnote{To the authors' knowledge, only in \citet{pacifici2012}, \textsc{beagle} \citep{chevallard2016}, \prospect\ \citep{robotham2020}, and \textsc{prospector} \citep{johnson2021}.} and more so its use in the literature\footnote{To the authors' knowledge, only in \citet{pacifici2012,pacifici2016a,pacifici2016b,robotham2020,bellstedt2020b,bellstedt2021,thorne2021}.}.
Allowing for this is physically well-motivated, as the existence of the stellar mass-star formation-metallicity plane indicates that the metallicity of a galaxy will change as they grow \citep[e.g.,][]{lara-lopez2010,lara-lopez2013,brown2016,brown2018}.
This will lead to different SSP templates being used, compared to assuming a fixed metallicity, which will affect the predicted colour evolution and therefore on the colour transition timescales.
Under the assumption that the IMF does not evolve with time \citep[common in extragalactic studies, e.g.,][]{taylor2011,schaye2015,lagos2018,bellstedt2020b}, the fitted SFH and $Z$H combined with the chosen SSP templates then offer a straightforward reconstruction of the intrinsic galaxy SED at any point backwards in time.
In the particular case of inferring colour transition timescales, a flexible smooth model for the SFH is highly desirable, as a piece-wise SFH can lead to artefacts in reconstructed colour evolution.

In this work we make use of the Galaxy And Mass Assembly \citep[GAMA;][]{driver2011,liske2015} survey and the \shark\ semi-analytic model (SAM) of galaxy formation \citep[, hereafter L18]{lagos2018} to reconstruct the colour evolution of galaxies using SED fitting.
In particular, we aim to answer the following question:
\begin{enumerate}
    \item how have the colours of the local blue and red galaxy populations evolved with time?
\end{enumerate}
This is critical to understand the timescales on which galaxies transition from the blue cloud to the red sequence (the subject of a future paper; Bravo et al. in preparation).

There are also three other important and related questions that we must first address to answer the above:
\begin{enumerate}
    \setcounter{enumi}{1}
    \item how well we can reconstruct the colour evolution of galaxies from their panchromatic SEDs?
    \item how can we best define the blue and red populations across cosmic time?
    \item is the green valley the superposition of the blue and red populations, or a third population on its own?
\end{enumerate}
The use of simulations to answer question (ii) is crucial, as they offer a test-bed for our reconstruction techniques and a way of identifying and quantifying their limitations.
The best approach is to characterise the galaxy populations in both observations and simulations with the same technique and quantify the differences.
Although the presence of two colour populations is well documented, there is no clear agreement across literature examples on a quantitative definition (e.g., \citealt{schawinski2014}; \citetalias{taylor2015}; \citealt{trayford2016}; \citealt{bremer2018}; \citealt{nelson2018a}).

This is especially critical, as the details of the adopted methodology can have a strong effect when inferring the colour transition timescales.
Strongly connected to this is the nature of the "green valley", as the inclusion of a third "green" population will impact any quantitative description of both blue and red populations.
Evidence suggests that it is not a population of its own, but the overlap of the blue and red populations \citep[e.g.,][; \citetalias{taylor2015}]{schawinski2014}.
However, it is worth exploring again, using our exact methodology, whether we find evidence of a third population across cosmic time.
Finally, as we recreate the colour evolution of galaxies in simulations, we also test how well theoretical models reproduce the inferred colour evolution.
While predictions have been made from simulation for the colour evolution and colour transition timescales \citep[e.g.,][]{trayford2016,nelson2018a,wright2019}, direct comparisons with equivalent observational results have not been attempted.

The rest of this work is ordered as follows.
We describe the data used in this work in Section \ref{sec:galcat}.
Using this data, we present our method to study the colour evolution in Section \ref{sec:evol_tracks}.
We discuss the results presented in this work in Section \ref{sec:disc}.
Finally, we summarise the findings of this work in Section \ref{sec:sum}.
In this work, we adopt the \citet{planck2016xiii} $\Lambda$CDM cosmology, with values of matter, baryon, and dark energy densities of $\Omega_b=0.0488$, and $\Omega_\Lambda=0.6879$, respectively, and a Hubble parameter of H$_0=67.51$ km s$^{-1}$ Mpc$^{-1}$.

\section{Galaxy catalogues}\label{sec:galcat}

In this work, we use the same galaxy sample presented in \citet[, hereafter B20b]{bellstedt2020b}, together with a similarly selected sample of galaxies from \shark.

\subsection{Observed catalogues (GAMA)}\label{subsec:GAMA}

GAMA is a spectroscopic redshift survey that targeted $\sim300,000$ galaxies in five fields (total of $\sim285\ \deg^2$), selecting galaxies with $r_\mathrm{ap}<19.8$ (save one field, G23, selected with $i_\mathrm{ap}<19.0$.), for which achieved $\sim98\%$ completeness.

From this survey, we combine the most recent version of the GAMA redshift catalogue \citep{liske2015} with the galaxy property catalogue derived from SED fitting by \citetalias{bellstedt2020b}, which used the latest photometry available from \citet{bellstedt2020a}.
The use of the \citetalias{bellstedt2020b} sample limits the available fields from five to the three equatorial (G09, G12 and G15, $\sim180\ \deg^2$).
From this point, we will refer to this data set as GAMA.

Briefly, the majority ($\sim85\%$) of the GAMA redshifts were measured using the AAOmega spectrograph \citep{saunders2004,sharp2006} and the Two-degree Field \citep[2dF;][]{lewis2002} fibre plate on the Anglo-Australian Telescope (AAT).
These redshifts were obtained by cross-correlating the observed spectra with spectral templates using the \textsc{autoz} software \citep{baldry2014}.
The remaining redshifts were collected from previous surveys covering the GAMA footprint \citep[see][ for a detailed description]{liske2015}.
All redshifts were then assigned a quality flag ($nQ$), ranging from 0 (worst) to best (4), with $\sim90\%$ of the galaxies with $nQ\ge3$. 

The new photometry catalogue by \citet{bellstedt2020a} was built from GALEX+VST+VISTA+WISE+Herschel \citep{martin2005,arnaboldi2007,wright2010,pilbratt2010,sutherland2015} imaging, using the software \textsc{ProFound} \citep{robotham2018}.
A combination of $r$ and $Z$ filters were used for source finding, with the photometry measured using the segments defined during source finding for bands FUV to W2, and the positions of these segments for PSF photometry from W3 to S500.
\citetalias{bellstedt2020b} then combined this new photometry catalogue with the existing redshifts to perform SED fitting on a sub-sample of the survey (galaxies with $nQ\ge3$, $z<0.06$ and $r<19.5$\footnote{The new photometry catalogue moved the $95\%$ completeness from $r<19.8$ to $r<19.5$.}) to recover galaxy properties and star formation/metallicity histories (SFH/$Z$H).
This was done using \prospect\footnote{\url{https://github.com/asgr/ProSpect}} \citep{robotham2020}, a high-level SED generator, whose design has influences from existing spectral fitting codes like \textsc{MAGPHYS} \citep{dacunha2008} and \textsc{CIGALE} \citep{noll2009,boquien2019}.
\prospect\ uses a combination of either the \textsc{GALAXEV} \citep{bruzual2003} or \textsc{E-MILES} \citep{vazdekis2016} Stellar Population Synthesis (SSP) libraries with the \citet{charlot2000} multi-component dust attenuation model and the \citet{dale2014} dust re-emission templates, under the assumption of a \citet{chabrier2003} Initial Mass Function (IMF).

From the wide variety of choices of functional forms for the characterisation of the star formation and metallicity histories (SFH and $Z$H, respectively) that \prospect\ offers, \citetalias{bellstedt2020b} used:
\begin{itemize}
    \item the \textsc{GALAXEV} SSP library.
    \item \texttt{massfunc\_snorm\_trunc}, a parametric description of the SFH using a skewed and truncated Gaussian distribution, as the functional form for the SFH, with $m_\mathrm{SFR}$, $m_\mathrm{mpeak}$, $m_\mathrm{mperiod}$, and $m_\mathrm{mskew}$ as free parameters to fit.
    \item \texttt{Zfunc\_massmap\_lin}, a linear map between the metallicity increase and the stellar mass growth, to parameterise the $Z$H,
    fitting $Z\mathrm{final}$, the final metallicity.
    \item $\tau^{}_\mathrm{birth}$, $\tau^{}_\mathrm{screen}$, $\alpha^{}_\mathrm{birth}$ and $\alpha^{}_\mathrm{screen}$ dust parameters as free parameters within the fitting.
    \item pow$_\mathrm{birth}=\mathrm{pow}_\mathrm{screen}=-0.7$, the default value in \prospect.
    \item 13.4 Gyr as the maximum age for star formation, demanding that stars form after $z\sim11$.
\end{itemize}

To fit the parameters, \citetalias{bellstedt2020b} used a Covariance Matrix Adaptation genetic algorithm to make an initial estimate of the parameter.
Then a Component-wise Hit And Run Metropolis Markov-Chain Monte Carlo algorithm was used with $10,000$ steps to determine the best-fitting SFH, $Z$H and dust parameters.
While the observed colours from GAMA are affected by noise, this noise is taken into account in the SED fitting process (i.e., all intrinsic colours are noiseless by construction).
This means that the intrinsic colours recovered with \prospect\ are robust against this noise and that it is fair to compare them to the results directly from our simulations.
For the rest of this work, we will refer to the sample of galaxies from \citetalias{bellstedt2020b} as simply GAMA.

\subsection{Simulated catalogues (\shark)}\label{subsec:SHARK}

In this work, we use the semi-analytic model (SAM) \shark\ (\citetalias{lagos2018}), which has been shown to reproduce a wide variety of observations \citep[][]{amarantidis2019,davies2019a,lagos2019,lagos2020,chauhan2019,chauhan2020,chauhan2021,bravo2020}.
Most critical for this work are the excellent predictions of observed number counts, luminosity functions \citep{lagos2019,lagos2020} and low-redshift colour distributions \citep{bravo2020}.
Briefly, a SAM generates and evolve galaxies using dark matter (DM) only N-body simulations, by following a set of equations that described the relevant physical processes to galaxy evolution.
What follows is a brief description of the base DM-only simulation, \shark, and how we generate in post-process the SEDs for these simulated galaxies using \prospect.

For this work we use the SURFS suite of DM-only simulations \citep{elahi2018a}, which adopts a $\Lambda$CDM \citep{planck2016xiii} cosmology and span a range of box length of $40-210$ h$^{-1}$cMpc (cMpc being comoving megaparsec) and particle mass of $4.13\times10^7$ to $5.90\times10^9$, reaching up to 8.5 billion particles.
This simulation suite was run with a memory lean version of the \textsc{GADGET2} code on the Magnus supercomputer at the Pawsey Supercomputing Centre.
We use the same simulation as in \citetalias{lagos2018}, L210N1536, with a box size of 210 h$^{-1}$cMpc, $1,536^3$ DM particles, a particle mass of $2.21\times10^8$ h$^{-1}$M$_\odot$, and a softening length of $4.5$ h$^{-1}$ckpc.
SURFS produced 200 snapshots for each simulation, with a typical time-span between snapshots in the range of $\sim6-80$ Myr.
The halo catalogues for the SURFS suite were constructed using the 6D FoF finder \textsc{VELOCIraptor} \citep{poulton2018,canas2019,elahi2019a}, and for the halo merger trees \textsc{TreeFrog} \citep{elahi2019b} was used.
The design of L210N1536 ensures that the merger histories for haloes of $\sim10^{11}$ M$_\odot$ are strongly numerically converged.
We refer the reader to \citetalias{lagos2018} for more details on the construction of the merger trees and halo catalogues used in this work.

These catalogues are the input for \shark, which populates these halo catalogues with galaxies and evolves them following prescriptions for key physical processes that shape the formation and evolution of galaxies.
Among these processes are the collapse and merging of DM haloes, gas accretion to both halo and galaxy, star formation, black hole growth, feedback by stellar and AGN, galaxy mergers, disc instabilities and environmental processes affecting the gas supply of satellite galaxies.
Most of these processes can be modelled by a choice of different prescriptions built into \shark, in this work we use the prescription and parameter choices from \citetalias{lagos2018} (see their table 2).
Important for this work is that \shark\ adopts the same universal \citet{chabrier2003} IMF as used in \prospect.
The model presented in \citetalias{lagos2018} has been calibrated only to reproduce the $z=0,1,2$ stellar mass functions (SMFs), the $z=0$ black hole–bulge mass relation and the mass–size relation.

To construct the GAMA-like sample, we first need to generate synthetic SEDs for the \shark\ galaxies to replicate the $r_\mathrm{ap}<19.8$ selection.
For this, we start with the SFH and $Z$H of the galaxies, contained in the \texttt{star\_formation\_histories} output file from \shark\footnote{This is done by setting \texttt{output\_sf\_histories = true} on the \shark\ configuration file.}.
These files contain the information for each of the three channels for star formation (in disc, in bulge due to mergers, and in bulge due to disc instabilities).

We use \textsc{Viperfish}\footnote{\url{https://github.com/asgr/Viperfish}}, a light wrapper around \prospect, to generate the SEDs.
From the discretely-valued SFH and $Z$H at the observation snapshot, it first calculates the stellar light emission for each galaxy.
Next, it accounts for dust screening and re-emission, where young stars (age less than 10 Myr) are attenuated by the dust in birth clouds, and then all stars are attenuated by the diffuse interstellar medium (ISM).
While \prospect\ currently includes several models to account for the AGN contribution to the SEDs, with the \citet{fritz2006,feltre2012} model, in particular, providing a good match to existing observations \citep{thorne2022}, we do not attempt to include this effect in \shark.
We note also that this is a recent addition, and therefore \citetalias{bellstedt2020b} did not include an AGN component in their analysis (also justified by the negligible occurrence of AGN at such low redshift).
We leave how to connect the central black hole properties in \shark\ to the AGN SEDs in \prospect\ for future work.

Following \citet{lagos2019,bravo2020}, we use their parameterisation model for dust\footnote{Called EAGLE-$\tau$ RR14 in \citet{lagos2019}, and T20-RR14 in \citet{bravo2020}.}, which uses the best-fit dust fraction-to-gas metallicity ratio from \citet{remy-ruyer2014} to calculate $\Sigma_\mathrm{dust}$, and then apply the \citet{charlot2000} parameters $\Sigma_\mathrm{dust}$-dependency found in \citet{trayford2020}.
For a more detailed description, we refer the reader to Sections 2.1 and 3.1 of \citet{lagos2019}.

The GAMA sample from \citetalias{bellstedt2020b} was chosen to be a volume-limited sample, which means that creating a synthetic lightcone is not required to reproduce this sample.
Instead, to expand the number of simulated galaxies available we choose our galaxies directly from one of the simulation snapshots, specifically snapshot 195 ($z=0.0668$).
We select galaxies by the GAMA magnitude selection of $r<19.8$, for which all galaxies in the box are assumed to be at the redshift of the snapshot.

As we will also fit these galaxies with \prospect\ (see next subsection), due to the computational cost of SED generation and fitting we are not able to use the full simulation box to draw a galaxy sample.
For parallelisation purposes, the merger trees from the simulation are provided divided into 64 "subvolumes", each containing a similar number of random selection from across the simulation box (i.e., a "subvolume" is not selected based on halo position in the box).
We choose to use five random subvolumes in this work, to balance between a large sample and the computational cost of SED generation and fitting.
This yields a total sample of $\sim30,000$ GAMA-like galaxies, five times more than our observations and $\sim50$ times more than that used by \citet{robotham2020}.

\subsection{Fitted simulations (\sharkfit)}\label{subsec:SHARKfit}

Of special interest for this work is how the reconstructed colour evolution from observational data is informed by the SED modelling choices built into \prospect.
In particular, the assumption of a skewed Normal functional form for the SFH and of a linear map from the mass growth to the $Z$H.
For this purpose, we also fitted the synthetic SEDs generated for \shark\ with \prospect, in a comparable (but not equal) manner to \citetalias{bellstedt2020b}.

\begin{figure}
    \centering
    \includegraphics[width=0.99\linewidth]{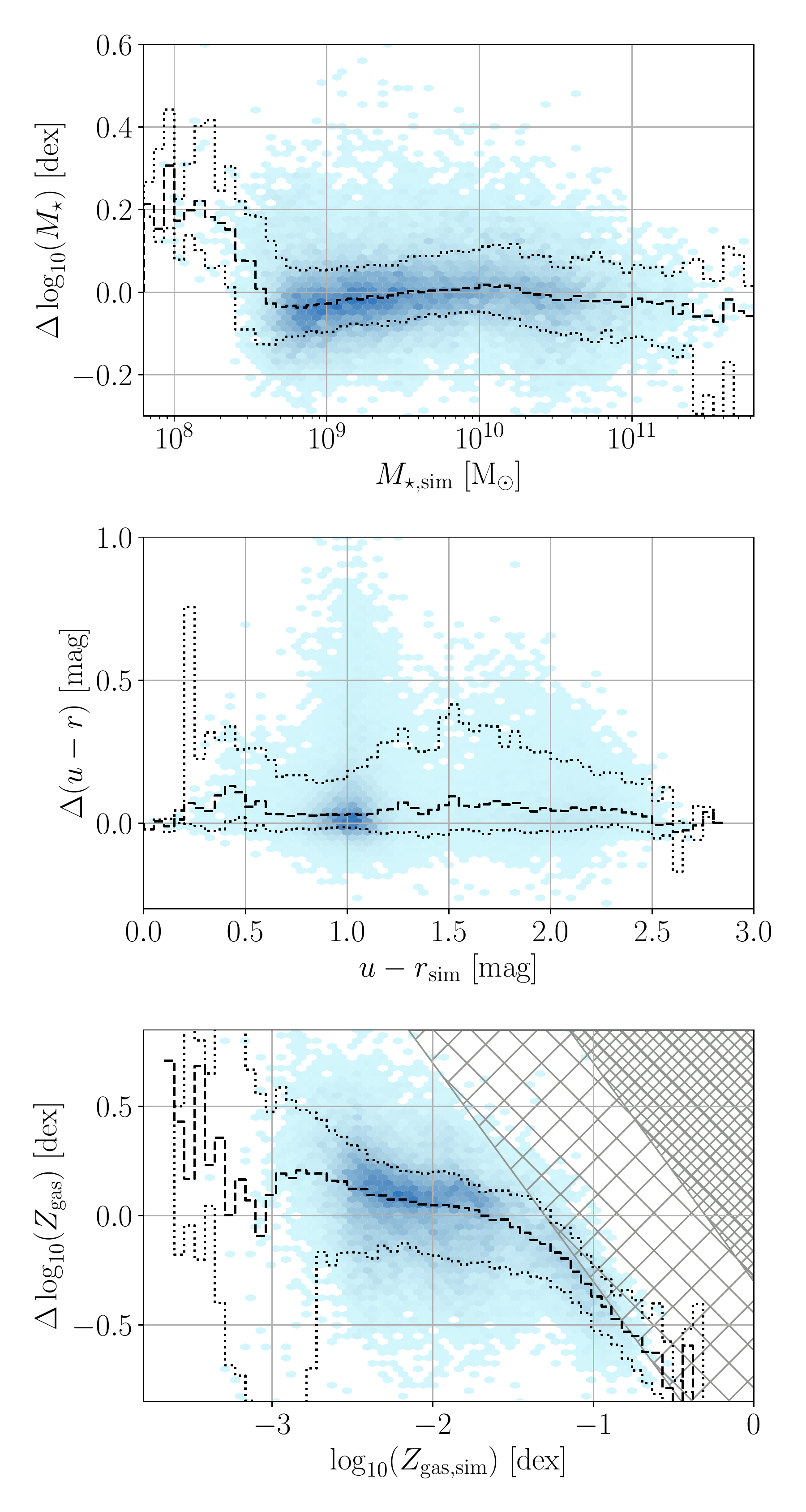}
    \caption{Comparison of the scatter between the fitted and intrinsic values to the intrinsic galaxy properties of \shark.
    The upper panel shows this for the stellar masses, the middle for rest-frame intrinsic colours, and the lower for gas-phase metallicities.
    The colour of the bins represents their galaxy density, larger densities in darker shades.
    The dashed and dotted lines show the running medians and 16$^\mathrm{th}$-84$^\mathrm{th}$ percentiles, respectively.
    The sparsely hatched area in the bottom panel indicates the region where the fitted metallicity is above the \textsc{galaxev} template limits, and the densely hatched region values above the prior range used for the SED fit.
    $6.67\%$ of the galaxies have $|\Delta\log_{10}(M_\star)|>0.2$, $4.08\%$ of the galaxies have $|\Delta\log_{10}(u-r)|>0.5$ mag (equivalent to a flux ratio difference of 0.2 dex), and $40.8\%$ of the galaxies have $|\Delta\log_{10}(Z)|>0.2$ ($9.20\%$ with $|\Delta\log_{10}(Z)|>0.5$).}
    \label{fig:SHARK_fit_diff}
\end{figure}

We assume the same functional forms for the SFH and $Z$H (\texttt{massfunc\_snorm\_trunc} and \texttt{Zfun\_massmap\_lin}) but modify the priors for metallicity and dust parameters.
We set a log-Uniform prior for both opacities, with $-6<\{\log_{10}(\tau^{}_\mathrm{BC}),\log_{10}(\tau^{}_\mathrm{ISM})\}<\log_{10}(5)$, which matches the range of opacities used to generate the SEDs for \shark\ (instead of the narrower -2.5 to 1 range used by \citetalias{bellstedt2020b} and \citealt{thorne2021}, the latter hereafter T21).
The dust temperature parameters ($\alpha^{}_\mathrm{BC}$ and $\alpha^{}_\mathrm{ISM}$) are fixed to 1 and 3, respectively, as per the generative \shark\ SEDs (\citetalias{bellstedt2020b} and \citetalias{thorne2021} used a Gaussian prior of mean 2 and standard deviation of 1).
We expand the log-Uniform prior for $Z_\mathrm{final}$ used by \citetalias{bellstedt2020b} and \citetalias{thorne2021} to cover the gas-phase metallicities in \shark, setting the prior range to $-4<\log_{10}(Z_\mathrm{final})<-0.3$\footnote{This is an order of magnitude higher than the highest metallicity in the \textsc{galaxev} templates, which is the reason for the use of this value as a limit by \citetalias{bellstedt2020b} and \citetalias{thorne2021}. \prospect\ defaults to the highest metallicity template in such cases, which means that this will not impact the fitting for young stellar populations. The main reason for this increase is allowing the use of high-metallicity templates for older stellar populations, as some \shark\ galaxies do reach the template limit long before they have built all of their current stellar mass.}.
For the photometry errors, for each \shark\ galaxy we match to the 5 GAMA galaxies with the most similar observed flux, and then randomly choose one from which we draw its photometry error and assign it to the \shark\ galaxy.
This is done in a filter-by-filter basis, i.e., different filters of a single \shark\ galaxy are matched to different GAMA galaxies.
Finally, we generate a perturbed photometry by drawing fluxes from a Normal distribution centred on the true flux and standard deviation equal to the GAMA-drawn error.
We note that we required a much larger number of fitting steps compared to \citetalias{bellstedt2020b} and \citetalias{thorne2021} for these fits ($\sim30$ times the total number of steps).

Figure \ref{fig:SHARK_fit_diff} compares the stellar mass, intrinsic colour and gas-phase metallicity of \shark\ galaxies to the best fit results, at the redshift of observation ($z=0.0668$).
In line with the results from \citet{robotham2020} (see their figure 30), we see a superb recovery of the stellar masses, with a negligible global bias ($\sim-0.01$ dex) and very small global scatter ($\sim0.08$ dex).
The running percentiles also show that this remains true across all stellar masses.
Globally, the colour recovery also shows a very small bias ($\sim0.04$ mag), with a small scatter ($\sim0.11$ mag).

The recovery of the gas-phase metallicity shows different behaviours above and below the maximum metallicity available in the \textsc{galaxev} templates ($Z=0.05$).
Below this limit, the metallicities are well-recovered, with only a weak trend with true metallicity and a reasonable scatter ($\sim0.25$ dex).
For metal-rich galaxies, running median approaches the template limit, with few galaxies reaching the 1:1 line.
We also use this template set to generate the \shark\ photometry, so this is not the result of any handicap set upon \prospect\ but of old stellar populations being intrinsically harder to fit \citep{conroy2013}.
Despite these results, we should note that the SFH is not as well-recovered.
We further discuss this in Section \ref{app:SHARKfit_qual}, which also include a more in-depth exploration of the consequences of our choice of SFH/$Z$H for the SED fits.
While this complicates the interpretation of some of our results, these fits still provide valuable insight into the dependency of our results on our modelling choices.

\section{Modelling the colour evolution of galaxies}\label{sec:evol_tracks}

The first step in studying galaxy colours is to define the colour we wish to explore.
The results from \citet{strateva2001} and \citet{martin2007} lead to both NUV and $u$ being common choices across the literature for the "blue" band, as they provide the best separation between the colour populations.
As we are using intrinsic colours derived from either SED fitting or SAM generation, we are technically free to choose any band, as long as it is within the spectral range of the \prospect\ stellar templates.
However, fitted SEDs depend on the quality of the photometry for each band, which means that not all parts of the intrinsic SED will be equally reliable.
(For GAMA, the UV photometry from GALEX has larger uncertainties compared to the optical photometry from VST, due to both fewer photon counts and larger point spread function.
This translates to poorer fit constraints in NUV compared to $u$, and as such, we choose the latter as our "blue" band.
We then pair this with $r$ as our "red" band, generating colours in $u-r$ \citep[a choice shared with][]{strateva2001,baldry2004,schawinski2014,trayford2016,bremer2018}.
Our tests comparing specific star formation rates (sSFR) and $u-r$ with all three samples show a narrow linear relation down to $\sim10^{-11}$ yr$^{-1}$ (though \shark\ does exhibit more scatter), at which point $u-r$ saturates.

At the core of this work then lies the choice of how to define a blue and red population in our $u-r$ colour space.
There is a wide variety of literature choices on how to separate these populations (e.g., \citealt{schawinski2014}; \citetalias{taylor2015}; \citealt{trayford2016}; \citealt{bremer2018}; \citealt{nelson2018a}; \citealt{wright2019}), most of which are simple selection functions drawn by eye (i.e., defining a minimum colour for a galaxy to be blue).
Since we will use our presented population model in future work to explore the colour transition timescales of galaxies (Bravo et al., in preparation), defining a single hard cut in colour \citep[like in][]{bell2003,baldry2004,peng2010} to separate between blue and red populations is not appropriate.
Such a classification is binary, with galaxies being either blue or red, with an effectively instantaneous transition timescale.
The measurement of a timescale then requires a boundary region where galaxies are tagged as neither blue nor red.
To overcome this, we devise a  new method for tracking the blue and red populations over time, which can then be used to probabilistically assign galaxies to either population at a given epoch.

\subsection{Model overview}\label{subsec:overview}

\begin{table*}
    \centering
    \begin{tabular}{c|c|c|p{4cm}}
        GMM parameter & $\log_{10}(M_\star)$ dependence & $t^{}_\mathrm{LB}$ dependence & Relevant equations \\
        \hline
        \hline
        \multirow{8}{*}{$f^{}_\mathrm{\{B,R\}}$} & \multirow{4}{*}{$M^{}_\mathrm{T}$} & $M^{}_\mathrm{3T}$ & \multirow{8}{4cm}{\ref{e:GMM}, \ref{e:wR_mstar}, \ref{e:wB_mstar}, \ref{e:wMt_t}, \ref{e:wk_t}} \\
         &  & $M^{}_\mathrm{2T}$ &  \\
         &  & $M^{}_\mathrm{1T}$ &  \\
         &  & $M^{}_\mathrm{0T}$ &  \\
        \cline{2-3}
         & \multirow{3}{*}{$k$} & $k^{}_3$ &  \\
         &  & $k^{}_2$ &  \\
         &  & $k^{}_1$ &  \\
         &  & $k^{}_0$ &  \\
        \hline
        \multirow{8}{*}{$\mu^{}_\mathrm{\{B,R\}}$} & \multirow{4}{*}{$\alpha^{}_{\mu\mathrm{\{B,R\}}}$} & $\alpha^{}_{3\mu\mathrm{B}}$ & \multirow{8}{4cm}{\ref{e:GMM_blue}, \ref{e:GMM_red}, \ref{e:muB_mstar}, \ref{e:muR_mstar}, \ref{e:muB1_t}, \ref{e:muB0_t}, \ref{e:muR1_t}, \ref{e:muR0_t}} \\
         &  & $\alpha^{}_{2\mu\mathrm{B}}$ &  \\
         &  & $\alpha^{}_{1\mu\mathrm{\{B,R\}}}$ &  \\
         &  & $\alpha^{}_{0\mu\mathrm{\{B,R\}}}$ &  \\
        \cline{2-3}
         & \multirow{4}{*}{$\beta^{}_{\mu\mathrm{\{B,R\}}}$} & $\beta^{}_{3\mu\mathrm{\{B,R\}}}$ &  \\
         &  & $\beta^{}_{2\mu\mathrm{\{B,R\}}}$ &  \\
         &  & $\beta^{}_{1\mu\mathrm{\{B,R\}}}$ &  \\
         &  & $\beta^{}_{0\mu\mathrm{\{B,R\}}}$ &  \\
        \hline
        \multirow{8}{*}{$\sigma^{}_\mathrm{\{B,R\}}$} & \multirow{4}{*}{$\alpha^{}_{\sigma\mathrm{\{B,R\}}}$} & $\alpha^{}_{3\sigma\mathrm{B}}$ & \multirow{8}{4cm}{\ref{e:GMM_blue}, \ref{e:GMM_red}, \ref{e:sigmaB_mstar}, \ref{e:sigmaR_mstar}, \ref{e:sigmaB1_t}, \ref{e:sigmaB0_t}, \ref{e:sigmaR1_t}, \ref{e:sigmaR0_t}} \\
         &  & $\alpha^{}_{2\sigma\mathrm{\{B,R\}}}$ &  \\
         &  & $\alpha^{}_{1\sigma\mathrm{\{B,R\}}}$ &  \\
         &  & $\alpha^{}_{0\sigma\mathrm{\{B,R\}}}$ &  \\
        \cline{2-3}
         & \multirow{4}{*}{$\beta^{}_{\sigma\mathrm{\{B,R\}}}$} & $\beta^{}_{3\sigma\mathrm{\{B,R\}}}$ &  \\
         &  & $\beta^{}_{2\sigma\mathrm{\{B,R\}}}$ &  \\
         &  & $\beta^{}_{1\sigma\mathrm{\{B,R\}}}$ &  \\
         &  & $\beta^{}_{0\sigma\mathrm{\{B,R\}}}$ &  \\
        \hline
    \end{tabular}
    \caption{Summary of the variables that describe our galaxy colour population model.
    This is a Gaussian Mixture Model, which are described by a set of 6 parameters as shown in the left column: the fraction that the blue and red populations contribute to the combined PDF, the means of each population, and the standard deviations of each population.
    We model these parameters as a function of stellar mass, with a logistic curve for the fractions and linear relations for the means and standard deviations, which are shown in the middle column.
    For the fractions, $M^{}_\mathrm{T}$ is the stellar mass at which $f^{}_\mathrm{B}=f^{}_\mathrm{R}$, and $k$ is the sharpness of the transition from blue- to red-dominated (as a function of stellar mass).
    For the means and standard deviations, $\alpha$ is the slope and $\beta$ the normalisation.
    We further model these 10 parameters as a function of lookback time, where the number index represents the order of the term, e.g., $\alpha^{}_{0\sigma\mathrm{R}}$ is the zeroth-order term of the time evolution of the slope of the red population standard deviation as a function of stellar mass.
    These are shown in the right column.}
    \label{t:param_sum}
\end{table*}

To study the colour evolution of galaxies from both GAMA and \shark\ we use a method inspired by \citetalias{taylor2015}, who studied the observed $z\sim0$ colour population in GAMA.
Briefly, their starting point was the assumption, corroborated in their analysis, that the colour distribution is well-represented by two Gaussians at any stellar mass.
Then, they assumed that the means and standard deviations are a function of stellar mass, using a fairly flexible model (two smoothly-joined line segments).
As their interest was in the stellar mass functions, they implicitly parameterised the fractions of each Gaussian by assuming a double Schechter function \citep{schechter1976} for each population.
Finally, they also assumed a model to bias against "bad" data\footnote{Quotation marks as in \citetalias{taylor2015}.}.
In total, the \citetalias{taylor2015} model contains 40 free parameters.
To obtain the best fitting values of these parameters they used a Markov chain Monte Carlo sampling to explore the parameter space, which for their sample of $\sim26,000$ galaxies required $\sim90,000$ CPU hours of computation.

The challenge for this work is that we aim to also study the time evolution of these parameterisations, which makes for a more complex fitting process.
Combined with our total sample of 67,000 galaxies to fit at 91 evolutionary time steps, which is a total of $\sim6,100,000$ implied data points, means that the method described by \citetalias{taylor2015} is too computationally expensive for our work.
We simplify this model by ignoring both stellar mass function, which we do not intend to explore, and the "bad" data modelling.
Since these choices are based on some of our results, we will defer our justifications for ignoring stellar mass function and "bad" data modelling to Sections \ref{subsubsec:w} and \ref{subsec:GMM_fit}, respectively.
A simple two-parameter description for the fractions, the minimum required to include a stellar mass dependency, would bring the number of free parameters of the \citetalias{taylor2015} model down to 27.
However, even with this simplification, using a simple linear model for the time evolution of each parameter would double this number, and this may not be enough to model the true measured colour evolution.
This is exacerbated by our desire to also test fitting three populations instead of just two (to evaluate the existence of a third "green" population).

For this reason, we further reduce the computation cost by splitting our fitting into multiple simple steps.
This greatly reduces computation time at the cost of requiring careful control of every step, to avoid fitting errors cascading throughout our method.
Given the significant number of steps in our method, and the necessity to introduce 21 equations with 51 variables involved, here we first provide an explanation of our nomenclature system, followed by a simple overview of this method.
We provide Table \ref{t:param_sum} as a quick reference for the nomenclature scheme we adopt.

\begin{figure*}
    \centering
    \includegraphics[width=\linewidth]{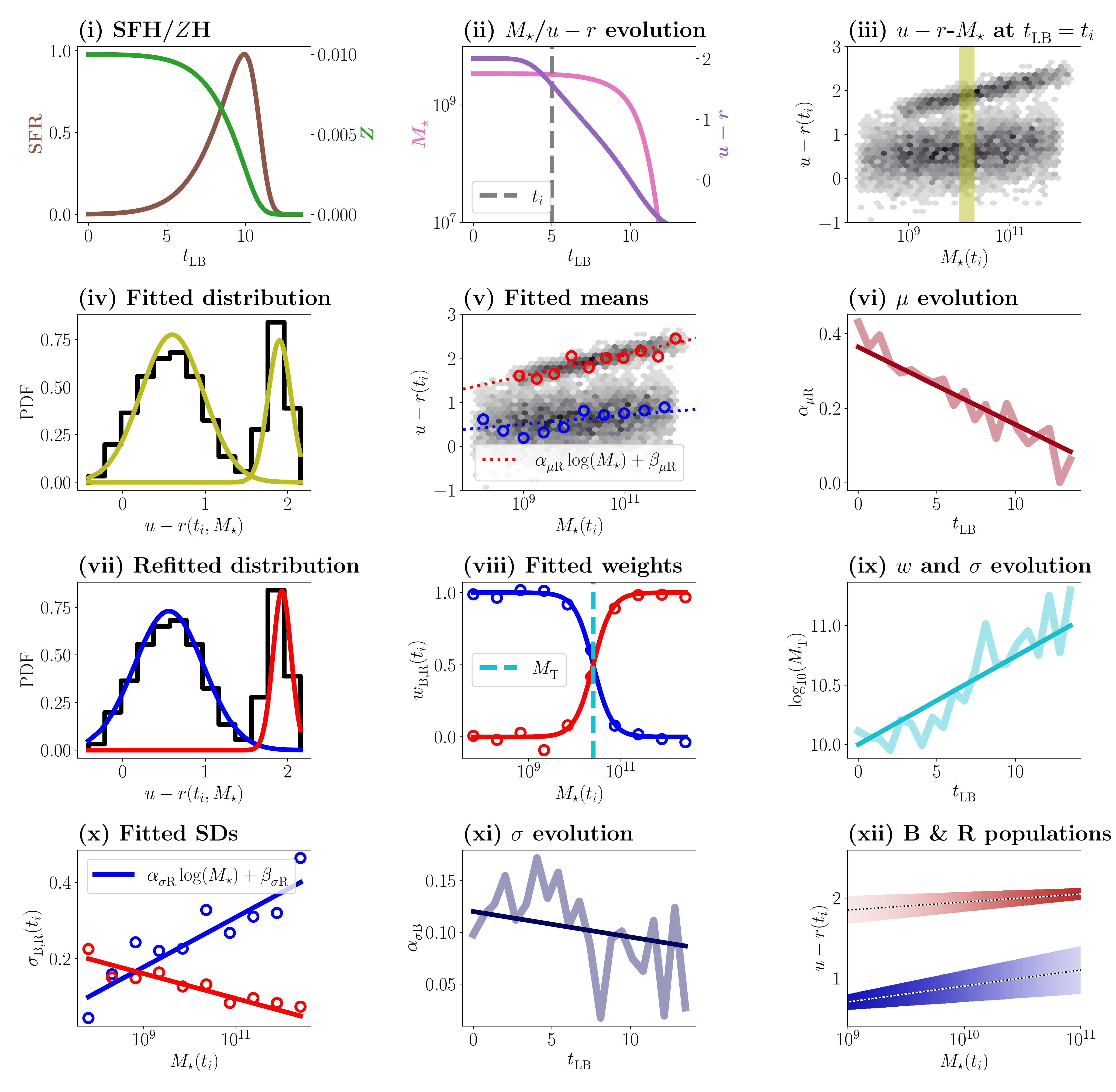}
    \caption{Schematic representation of the methodology employed in this work, using our two-component fit as an example.
    Our starting point (i, described in Section \ref{sec:galcat}) are the SFHs and $Z$Hs of the galaxies in each sample, retrieved either directly from \shark\ (Section \ref{subsec:SHARK}) or through SED fitting with \prospect\ (GAMA and \sharkfit, Sections \ref{subsec:GAMA} and \ref{subsec:SHARKfit}).
    From those, we generate the stellar mass and colour histories for these galaxies (ii, Section \ref{subsec:Mstar_col}), for every 100 Myr from 1 to 10 Gyr of lookback time, using the generative mode of \prospect.
    At each of these time steps, we bin the galaxies in stellar mass bins of a width 0.3 dex, and use a two- or three-component GMM to fit the colour distribution in each mass bin (iii and iv, Section \ref{subsec:GMM_fit}).
    We then fit the blue and red populations with linear relations as a function of mass, and then we fit the slope and value at $M_\star=$\mstar{9.5} ($M_\star=$\mstar{10.5}) for the blue (red) population as a function of time (v and vi, Section \ref{subsubsec:mu}).
    We show as an example a representation of the time evolution of the slope of the red means ($\alpha^{}_{\mu\mathrm{R}}$).
    Next, we use these fits to fix the means of the components in our GMM and refit the populations (vii, Section \ref{subsec:GMM_refit}).
    From these refits, we then first parameterise the dependency of the fractions and standard deviations on stellar mass (viii and x, Section \ref{subsubsec:w} for the fractions and \ref{subsubsec:sigma} for the standard deviations), followed by fits to the time evolution of these parameters (ix and xi, Sections \ref{subsubsec:w} and \ref{subsubsec:sigma}).
    As example, we display a representation of the time evolution of two parameters: the mass where both populations have equal fractions ($M^{}_\mathrm{T}$), and the slope of the standard deviation as a function of stellar mass for the blue population ($\alpha^{}_{\sigma\mathrm{B}}$).
    With these parameterisations, we have a full description of the populations as a function of time (xii, Section \ref{sec:disc}), where the dotted lines show the means, the width show the standard deviations and the shading the fractions.}
    \label{fig:explanation}
\end{figure*}

For the parameterisation as a function of stellar mass, we use first-order polynomials for the means and standard deviations, where $\alpha$ indicates the first-order term and $\beta$ the zeroth-order.
As an example, $\alpha^{}_{\mu\mathrm{B}}$ then represents the slope as a function of stellar mass of the means of the blue population ($\mu^{}_\mathrm{B}$).
The parameterisation of the fractions as a function of stellar mass does not follow this convention, as we use a noticeably different parameterisation, for which the parameters have a different physical interpretation.
For the temporal parameterisation, we use a numbered subscript to indicate the order, though note that it is not necessarily indicative of a polynomial expression, as we use a greater variety of functional forms for these fits.
I.e., $\alpha^{}_{0\mu\mathrm{B}}$ is the zeroth order of the time evolution of the slope as a function of stellar mass ($\alpha^{}_{\mu\mathrm{B}}$) of the mean of the blue population ($\mu^{}_\mathrm{B}$).

Our method can be summarised by the following steps:
\begin{enumerate}
    \item The foundation of this work are the SFHs and $Z$Hs of the galaxies in our three samples: GAMA, \shark\ and \sharkfit.
    \item Given these SFHs and $Z$Hs, we use \prospect\ to reconstruct the stellar mass and model the SED (to derive intrinsic colour) of each low-redshift galaxy at a range of prior lookback times (from 1 to 10 Gyr, in 100 Myr steps).
    \item At every time step we divide the galaxies in stellar mass bins with $0.3$ dex width, chosen as the best balance between a robust measurement of the colour distribution for $\log_{10}(M_\star/M_\odot)\in[9,11]$ and a good number of bins sampling the red population.
    \item We fit the colour distribution in each stellar mass bins using a two- or three-component Gaussian Mixture Model (GMM).
    \item[(v), (vi)] We assign the Gaussians as being blue or red (or green for the three-component fit).
    We then fit the means, as a function of stellar mass, and parameterise the time evolution of these fits.
    \setcounter{enumi}{6}
    \item As the number of galaxies in each bin rapidly decays as one moves away from the median stellar mass, the fits do not show a smooth evolution between time steps (an expected outcome of individually fitting each mass bin).
    To obtain a smoother evolution for the standard deviations and fractions we refit our GMM, this time fixing the means to values calculated with the evolution fit in the previous step.
    \item[(viii), (ix), (x), (xi)] We fit the standard deviations and fractions in the same manner as the means.
\end{enumerate}
Figure \ref{fig:explanation} also provides a visual representation of these steps.

We need to justify our choice of parameterising the time evolution linearly with lookback time.
The challenge is that SED fitting favours a (roughly) logarithmic scale in lookback time, as galaxy SEDs are more sensitive to recent star formation than to older stellar populations, while most cosmological simulation use a logarithmic scale in cosmic time that becomes sparser at recent times (including SURFS, as described in Section \ref{subsec:SHARK}).
A linear scale in lookback time will emphasise the poorly constrained reconstruction at  early times from SED fitting, but this is done to avoid overly interpolating the galaxy evolution at recent times from our simulations.
For both parameterisations of the dependence with lookback time and stellar mass, we choose the functions that with the fewest free parameters ensures both a good representation of the data and a stable parameter evolution with time.

Panel (xii) of Figure \ref{fig:explanation} provides a schematic representation of the resulting parameterisation of the colour populations.
The dotted lines indicate the means of each population, the widths show the standard deviations and the shading shows the fractions.
What follows for the remainder of this section is a detailed presentation of how we implement this method, together with the results from which we construct the representation of the colour evolution in our galaxy samples shown in Figure \ref{fig:summary}, the main summary of our results.

\subsection{Converting SFH and $Z$H into $M_\star$ and $u-r$ histories}\label{subsec:Mstar_col}

From the SFHs it is straightforward to generate stellar mass histories.
In this work, we use the remaining stellar mass, not the formed stellar mass \citep[i.e., removing the recycled stellar mass, see][ for details]{robotham2020}.
For \shark, since recycling is instantaneous, this is calculated simply by integrating the SFH of the galaxy and scaling it by $1-f_\mathrm{recycled}$\footnote{$f_\mathrm{recycled}=0.46$ for the \citet{chabrier2003} IMF used in \shark, as described in section 4.4.6 of \citetalias{lagos2018}.}.
For GAMA and \sharkfit\ we calculate the remaining stellar mass with \prospect, which uses the lookup tables from the \textsc{GALAXEV} templates.
For the colours, we use \prospect\ in its generative mode, in a similar manner to how we generated the synthetic SEDs described in Section \ref{subsec:SHARK}, the big distinction being that we restrict the SFH and $Z$H to the lookback time of each time step.
We remark that with this method we are following the colour evolution across cosmic time of the same set of low-redshift galaxies, and not selecting different galaxies at different lookback times.

\begin{figure}
    \centering
    \includegraphics[width=\linewidth]{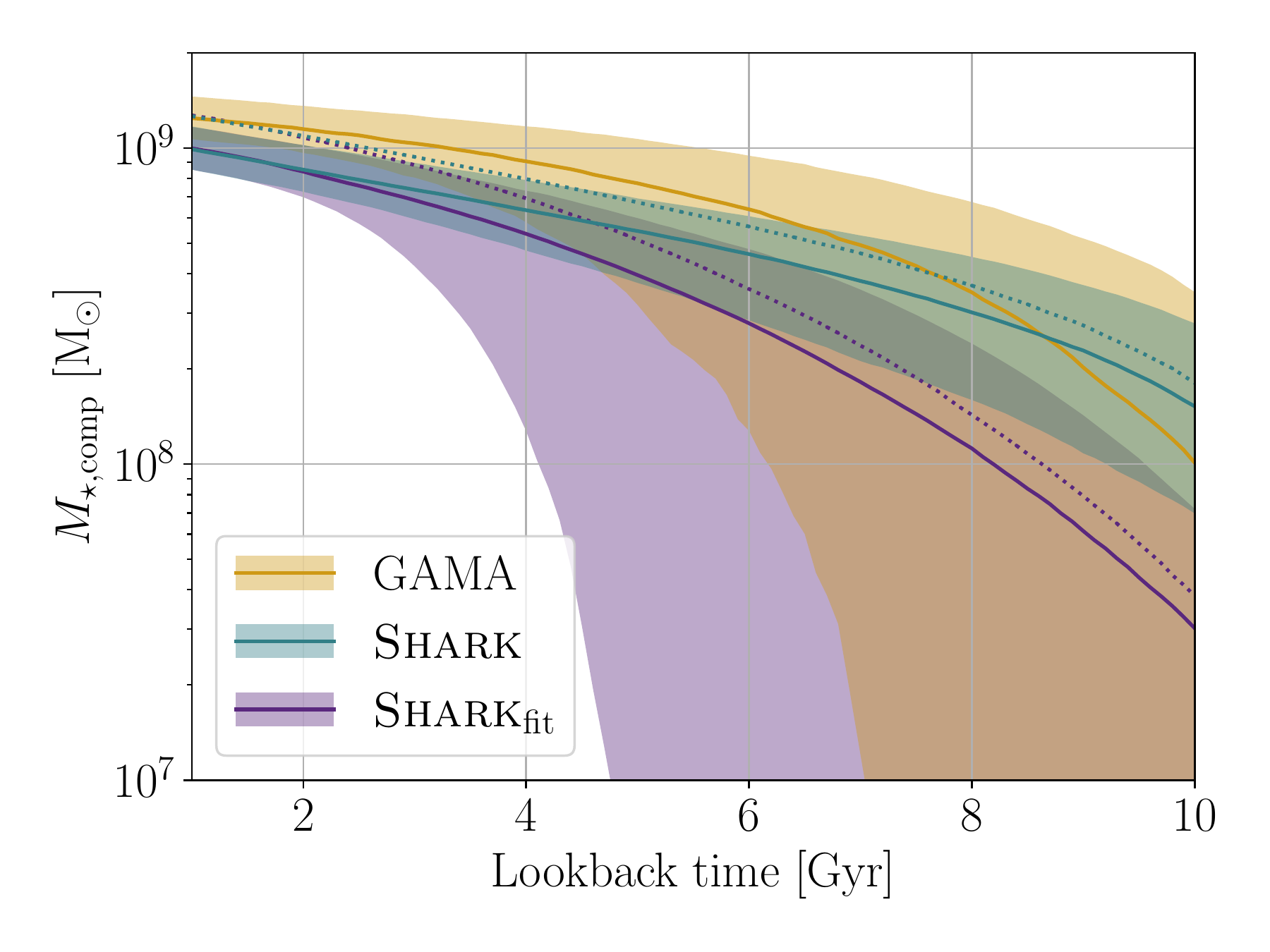}
    \caption{Evolution of the stellar masses of the galaxies $0.2$ dex around the mass completeness of each sample, $10^{9.0}$-\mstar{9.2} for GAMA and $10^{8.9}$-\mstar{9.1} for \shark/\sharkfit.
    GAMA is shown in orange, \shark\ in cyan and \sharkfit\ in purple.
    The solid lines show the median stellar mass at any given lookback time, with the shaded areas showing the 16$^\mathrm{th}$-84$^\mathrm{th}$ percentile range.
    The dotted lines indicate the median evolution for \shark/\sharkfit\ galaxies in the same stellar mass range as GAMA.}
    \label{fig:mass_comp}
\end{figure}

While \citetalias{bellstedt2020b} selected the GAMA sample such that it is volume-limited, this does not ensure mass completeness.
The challenge that reconstructing the evolution of this sample presents is how to model the evolution of the mass completeness.
Since at any time step we have the same galaxies, we define mass completeness by tracking the galaxies around the mass completeness limit at observation time.
The samples from GAMA and \shark/\sharkfit\ differ slightly in their mass completeness, with the former having a higher completeness limit.
We then select all galaxies in a 0.2 dex range around the mass completeness limit for each sample ($9.0<\log_{10}(M_\star/\mathrm{M}_\odot)<9.2$ for GAMA and $8.9<\log_{10}(M_\star/\mathrm{M}_\odot)<9.1$ for \shark/\sharkfit), and trace their evolution with time.

We show the evolution of these galaxies in Figure \ref{fig:mass_comp}.
We find that \shark\ and \sharkfit\ are in excellent agreement for lookback times below $\sim3$ Gyr, but start to diverge considerably by 4 Gyr.
This is driven by the slower mass build-up in our \prospect\ fits compared to \shark\ galaxies, which we explore in further detail in Section \ref{app:SHARKfit_qual}.
For the reader interested in the technical aspect of SED fitting, discussed in detail in Section \ref{subsec:SHARKfit}, we also show the median evolution of \shark/\sharkfit\ galaxies in the same mass range of GAMA.
GAMA and \sharkfit\ are in remarkable agreement in the evolution of the dispersion, while dispersion is markedly smaller in \shark, which shows that not all differences stem from the quality of our SED fits in \sharkfit.
This points to the modelling choices we make (following \citetalias{bellstedt2020b} and \citetalias{thorne2021}), the parameterisation of the SFHs specifically, as the source of this increased scatter.
As the medians are in reasonable agreement (when accounting for SFH differences) this is evidence that they are the more reliable tracer.
For this reason, we use the medians to define the mass completeness for each sample as a function of time.

\subsection{Unconstrained GMM and means parameterisation}\label{subsec:GMM_fit}

We start with an unconstrained GMM fit to all stellar mass bins with at least 30 galaxies.
These fits were carried out using the \texttt{normalmixEM} function from the \textsc{mixtools} R package, and are be described by:
\begin{align}
    P(u-r)&=f^{}_\mathrm{B}P^{}_\mathrm{B}(u-r)+f^{}_\mathrm{R}P^{}_\mathrm{R}(u-r)\label{e:GMM}\\
    P^{}_\mathrm{B}(u-r)&=\frac{1}{2\pi\sigma^{}_\mathrm{B}}\exp\left(-\frac{1}{2}\left(\frac{(u-r)-\mu^{}_\mathrm{B}}{\sigma^{}_\mathrm{B}}\right)^2\right)\label{e:GMM_blue}\\
    P^{}_\mathrm{R}(u-r)&=\frac{1}{2\pi\sigma^{}_\mathrm{R}}\exp\left(-\frac{1}{2}\left(\frac{(u-r)-\mu^{}_\mathrm{R}}{\sigma^{}_\mathrm{R}}\right)^2\right)\label{e:GMM_red}
\end{align}
\noindent where $P(u-r)$ is the fitted PDF describing the colour distribution, composed by the weighted addition of the Gaussian distributions $P^{}_\mathrm{B}(u-r)$ and $P^{}_\mathrm{R}(u-r)$, representing the blue and red populations.
These are weighted by $f^{}_\mathrm{B}$ and $f^{}_\mathrm{R}$, which are defined such that $f^{}_\mathrm{B}+f^{}_\mathrm{R}=1$.
The sets $\{\mu^{}_\mathrm{B},\sigma^{}_\mathrm{B}\}$ and $\{\mu^{}_\mathrm{R},\sigma^{}_\mathrm{R}\}$ are the means and standard deviations of $P^{}_\mathrm{B}(u-r)$ and $P^{}_\mathrm{R}(u-r)$, respectively.
In this step, the set of free parameters is then $\{f^{}_\mathrm{B},\mu^{}_\mathrm{B}, \sigma^{}_\mathrm{B},f^{}_\mathrm{R},\mu^{}_\mathrm{R},\sigma^{}_\mathrm{R}\}$.
It is worth noting that, in the definitions presented, for a cleaner notation we have ignored the fact that all of these quantities are a function of both stellar mass and lookback time, but we fit them at each stellar mass bin and time step.

We also tested an equivalent three-component model:
\begin{equation*}
    P(u-r)=f^{}_\mathrm{B}P^{}_\mathrm{B}(u-r)+f^{}_\mathrm{x}P^{}_\mathrm{x}(u-r)+f^{}_\mathrm{R}P^{}_\mathrm{R}(u-r)
\end{equation*}
\noindent where $P^{}_\mathrm{x}(u-r)$ follows the same equation as $P^{}_\mathrm{B}(u-r)$ and $P^{}_\mathrm{R}(u-r)$.
We found that, like \citetalias{taylor2015}, two was sufficient, with small fractions ($f^{}_\mathrm{x}<0.1$) and large standard deviations ($\sigma^{}_\mathrm{x}\gtrsim1$ mag) fitted for the third component.
This reflects the fact that this third component was not capturing a unique (continuous) population, but small residuals of the blue and red (hence the use of $x$ subscripts instead of $G$).

Interestingly, this seems to roughly reproduce the "bad" data modelling of \citetalias{taylor2015}, who find that it captures $\sim3\%$ of their sample.
This posses the question of why we exclude such population, as it could still help capturing the other two populations.
The logic behind this model in \citetalias{taylor2015} is to not only to reduce the effect of strong outliers in the colour-mass plane in the parameterisation of the population, but also to capture features that are not easily represented with a two-component GMM.
This means that it will capture features that we would classify as part of the blue or red populations, which is particularly noticeably in the low-mass end ($<$\mstar{9.5}) of the blue population (see Figure 4\footnote{This can also be seen in figure 7 of \citetalias{taylor2015}, which is discussed in their sections 6.3 and 9.3.4.}).

We are further justified by the fact that we impose a mass completeness limit, which removes low mass galaxies from our sample, which given our near volume-complete sample are also the ones with a noisier photometry.
This contrasts with citet{taylor2015}, who correct for incompleteness in the low-mass regime, which enhances the influence of low SNR data.
This, combined with the small population being assigned to the third population ($<10\%$), are the reasons why we will only present and discuss results for the two-component GMM fits.

\begin{figure*}
    \centering
    \includegraphics[width=\linewidth]{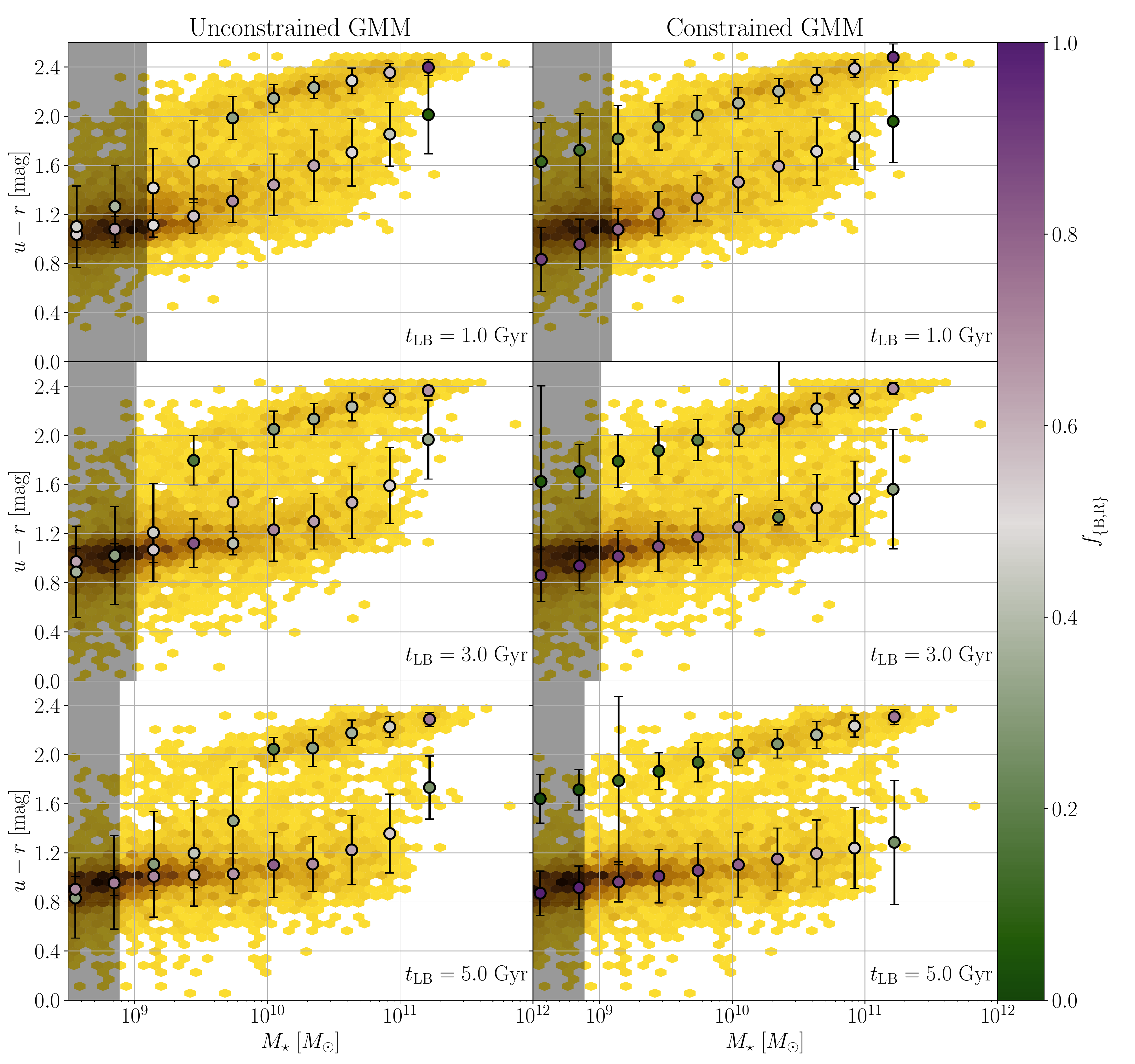}
    \caption{Example of the results from the unconstrained (left column) and constrained (right column) GMM fits to GAMA data at three time steps (1, 3 and 5 Gyr, from top to bottom).
    Each panel shows the distribution of rest-frame intrinsic $u-r$ colours and stellar mass, together with a display of the resulting fits from the GMM.
    Histogram bins show the 2D PDF, with colour range optimised for each panel to display the relative distribution of galaxies.
    The position of the markers shows the mean ($\mu_{\{\mathrm{B},\mathrm{R}\}}$) of each Gaussian component.
    The inner colour of the markers shows the fractions ($f_{\{\mathrm{B},\mathrm{R}\}}$), following the colour bar shown on the right.
    The vertical bars show the standard deviation ($\sigma_{\{\mathrm{B},\mathrm{R}\}}$).
    The grey shaded regions show the stellar masses below the mass completeness limit.}
    \label{fig:GAMA_GMM_fit}
\end{figure*}

The left column of Figure \ref{fig:GAMA_GMM_fit} shows three examples of these fits for GAMA.
For the bluer component, the slope as a function of stellar mass ($\alpha^{}_{\mu\mathrm{B}}$) increases at lower lookback times, which is driven by the high-mass end becoming redder, while near the mass completeness limit the locus remains nearly constant.
By contrast, the redder components (not merged with the bluer) show a weaker evolution, showing a shallower slope at recent times, driven by the reddening of the low-mass end.
Near the completeness limit, both components show significant overlap, though it is clear that the redder Gaussian is becoming redder with time.

To parameterise the means of both populations, we first need to choose which stellar mass bins and GMM components to use.
Firstly, to remove poor fits we discard all fits for which the reduced chi-square ($\chi^2_\nu$) is outside the range [0.85,1.1], which removes a median of one mass bin per time step.
Our choice of this non-symmetric range comes from a detailed inspection of the GMM [unconstrained] fits.
Fits with $\chi^2_\nu$ values of 0.85-0.90 tend to properly capture the red population at low stellar masses, while those in the 1.10-1.15 range are driven by one of the two components being a poor fit, leaving a significant part of the true distribution unaccounted for\footnote{We remark that, as stated before, this is not evidence of a third ("green") population, as a third component does not improve these fits.
Instead, this is a consequence of our approach of fitting each time step and mass bin individually, which leads to poorer fits where one of the populations is strongly dominant.}.

We then apply further selection criteria to the GMM fits that pass our $\chi^2_\nu$ selection.
First, we have the second significant difference between our approach and that of \citetalias{taylor2015}, as we consider the existence of a red population only when 1) there is a local minimum in the combined PDF between the means of both Gaussians and 2) the separation between means is at least 0.5 mag.
Where this is not the case we consider the dominant component as tracing the blue population, with the secondary accounting for non-Gaussianities.
It may seem that requiring the presence of a local minimum between the means of the two GMM components and a minimum separation between the means is redundant, but they complement each other.
The former ensures a strong distinction between populations, while the latter removes the rare occurrences of spurious narrow fits.
This could be roughly replicated by setting a minimum value for the standard deviation or fractions of the components, but either would also more strongly impact the red population.
We set the limit to 0.5 mag based on the results from the unconstrained GMM fits.

For the GMM fits that pass our criterion of having two distinct populations, we impose further checks on the individual components.
We set a maximum for the standard deviations, which can be no greater than 0.3 mag, to ensure a clean selection for both populations.
Small variations to this limit have no strong effect, but changes larger than $\sim0.1$ mag lead to significantly larger uncertainties in our parameterisation of means\footnote{Smaller values lead to the inclusion of poorly-constrained means, while larger values lead to only a few data points to fit.}.
Finally, we remove the fits at very high masses, which have a small number of galaxies ($\lesssim0.5\%$ of the full sample), only considering means from fits below $10^{10.5}$ and \mstar{11} for the blue and red populations, respectively.
The result from this selection is shown in Figure \ref{fig:GAMA_mean_fit} with open (removed) and solid (included) markers.

\subsubsection{Means parameterisation}\label{subsubsec:mu}

\begin{figure}
    \centering
    \includegraphics[width=\linewidth]{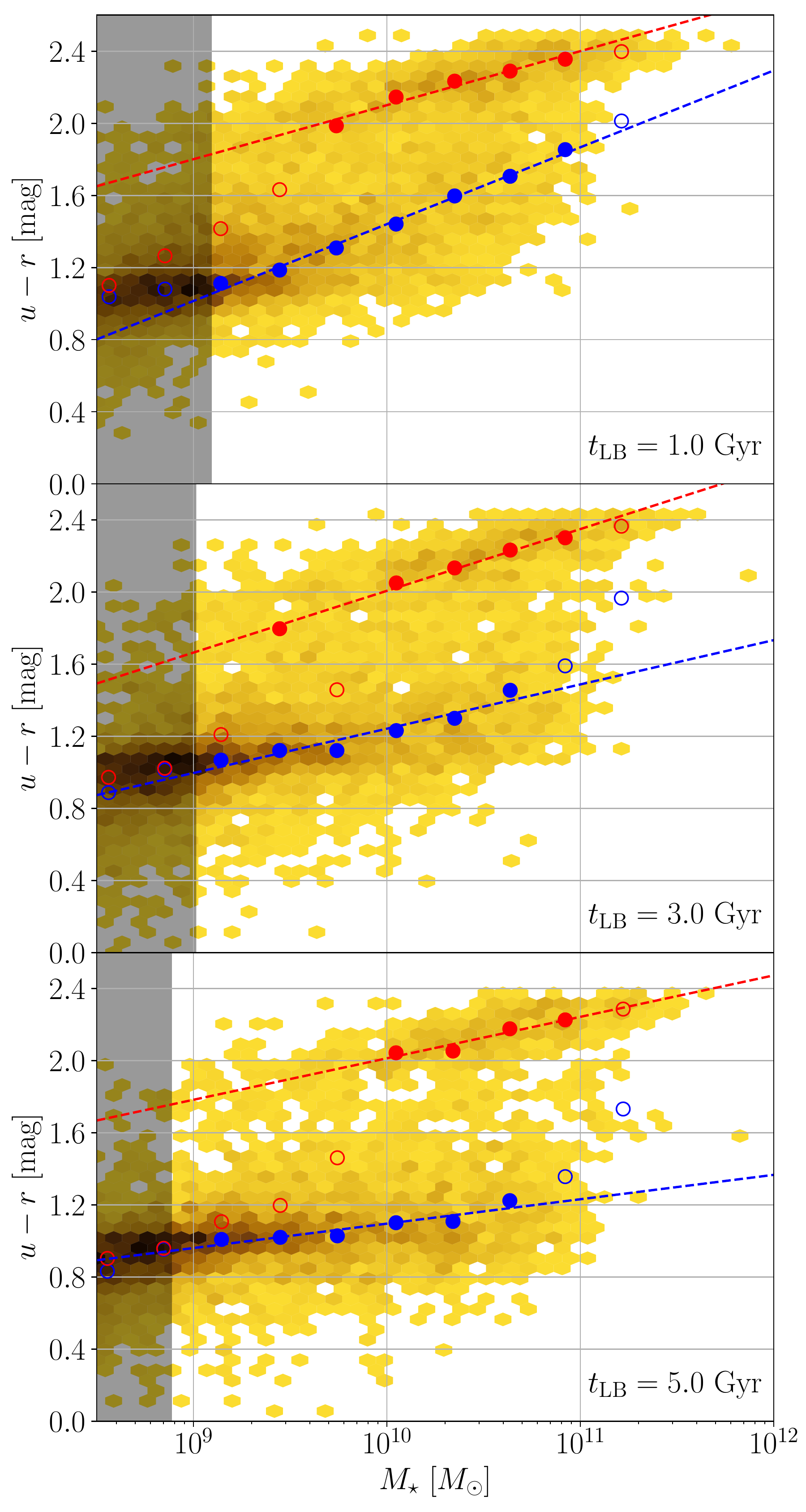}
    \caption{Example of the parameterisation of the means of the Gaussian components $(\mu^{}_\mathrm{\{B,R\}})$, as a function of stellar mass.
    Panels and histograms as in the left column of Figure \ref{fig:GAMA_GMM_fit}.
    The blue and red points show the selected means assigned to either the blue or red population, respectively.
    The blue and red curves show the fitted parameterisation to each, defined by Equations \ref{e:muB_mstar} and \ref{e:muR_mstar}.
    Open markers denote the points not used to fit the curves.
    Histograms and shaded regions as in Figure \ref{fig:GAMA_GMM_fit}.}
    \label{fig:GAMA_mean_fit}
\end{figure}

We use the following equations to fit the means as a function of stellar mass for both populations, at every time step:
\begin{align}
    \mu^{}_\mathrm{B}&=\alpha^{}_{\mu\mathrm{B}}\log_{10}(M_\star/10^{9.5}\mathrm{M}_\odot)+\beta^{}_{\mu\mathrm{B}}\label{e:muB_mstar}\\
    \mu^{}_\mathrm{R}&=\alpha^{}_{\mu\mathrm{R}}\log_{10}(M_\star/10^{10.5}\mathrm{M}_\odot)+\beta^{}_{\mu\mathrm{R}}\label{e:muR_mstar}
\end{align}
\noindent where $\{\alpha^{}_{\mu\mathrm{B}},\alpha^{}_{\mu\mathrm{R}}\}$ are the slope of the means as a function of stellar mass, and $\{\beta^{}_{\mu\mathrm{B}},\beta^{}_{\mu\mathrm{R}}\}$ are the value of the means at $M_\star=\{$\mstar{9.5},\mstar{10.5}$\}$\footnote{The main reason for this choice is that the intercepts with the $y$-axis ($\log(M_\star/\mathrm{M}_\odot)=0$) can be dominated by small fluctuations in $\{\alpha^{}_{\mu\mathrm{B}},\alpha^{}_{\mu\mathrm{R}}\}$, leading to strongly correlated values. Our choices are not optimal, i.e., they do not minimise the off-diagonal terms of the covariance matrices for each blue and red populations, but they are a simple and good approximation to reduce correlation. Proper minimisation would require having these masses as free parameters, which would likely evolve with time, requiring even more free parameters.}.
Figure \ref{fig:GAMA_mean_fit} shows that, given our choices to select points as accurately representing either population, these parameterisations provide a good representation of the data.

\begin{figure*}
    \centering
    \includegraphics[width=\linewidth]{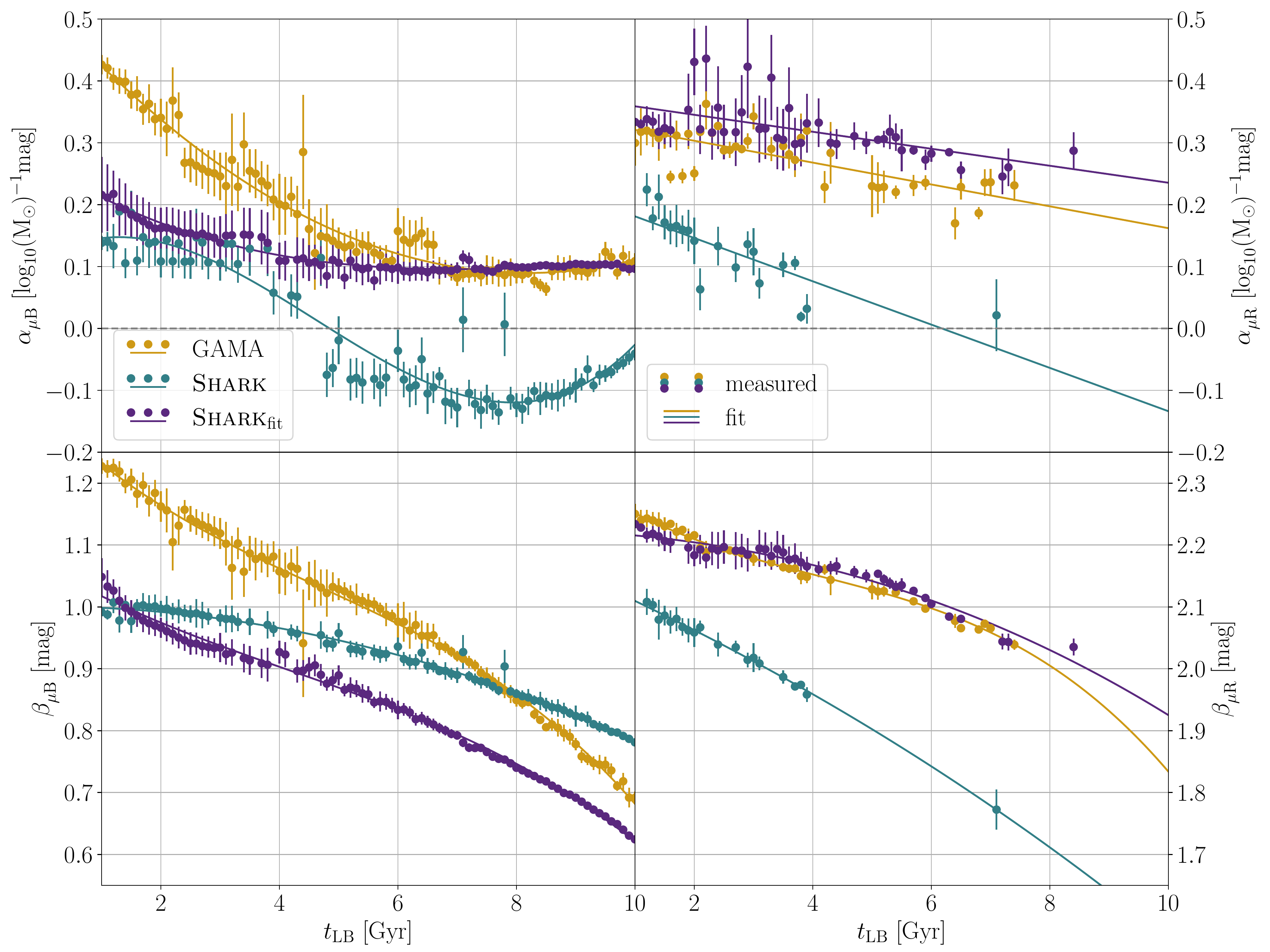}
    \caption{Time evolution of the blue and red population mean parameterisations.
    Colours as in Figure \ref{fig:mass_comp}.
    The left column shows the results for the blue population and the right for the red population.
    In the top row, the evolution of the slope of the means as a function of stellar mass ($\alpha^{}_{\mu\mathrm{\{B,R\}}}$, Equations \ref{e:muB_mstar} and \ref{e:muR_mstar}) is shown with the round markers, and the fit to these (Equations \ref{e:muB1_t} and \ref{e:muR1_t}) with solid lines.
    The bottom shows the value of the means at $M_\star=$\mstar{\{9.5,10.5\}} ($\beta^{}_{\mu\mathrm{\{B,R\}}}$, Equations \ref{e:muB_mstar} and \ref{e:muR_mstar}) and fits (Equations \ref{e:muB0_t} and \ref{e:muR0_t}) in the same manner.
    The error bars indicate the uncertainty in the fitted parameters.}
    \label{fig:evol_mu}
\end{figure*}

Based on the observed evolution of the mean parameters seen in Figure \ref{fig:evol_mu} we decide to fit $\beta^{}_{\mu\mathrm{R}}$ with a first-order polynomial as a function of lookback time, with the others parameters fit with a third-order polynomial ($\alpha^{}_{\mu\mathrm{B}}$, $\beta^{}_{\mu\mathrm{B}}$ and $\beta^{}_{\mu\mathrm{R}}$).
These fits were carried out with the \texttt{curve\_fit} function from the \textsc{scipy} Python package, as with all other non-GMM fits in this work.
The equations used for the fits are:
\begin{align}
    \alpha^{}_{\mu\mathrm{B}}&=\alpha^{}_{3\mu\mathrm{B}}t^{3}_\mathrm{LB}+\alpha^{}_{2\mu\mathrm{B}}t^{2}_\mathrm{LB}+\alpha^{}_{1\mu\mathrm{B}}t^{}_\mathrm{LB}+\alpha^{}_{0\mu\mathrm{B}}\label{e:muB1_t}\\
    \beta^{}_{\mu\mathrm{B}}&=\beta^{}_{3\mu\mathrm{B}}t^{3}_\mathrm{LB}+=\beta^{}_{2\mu\mathrm{B}}t^{2}_\mathrm{LB}+\beta^{}_{1\mu\mathrm{B}}t^{}_\mathrm{LB}+\beta^{}_{0\mu\mathrm{B}}\label{e:muB0_t}\\
    \alpha^{}_{\mu\mathrm{R}}&=\alpha^{}_{1\mu\mathrm{R}}t^{}_\mathrm{LB}+\alpha^{}_{0\mu\mathrm{R}}\label{e:muR1_t}\\
    \beta^{}_{\mu\mathrm{R}}&=\beta^{}_{3\mu\mathrm{R}}t^{3}_\mathrm{LB}+\beta^{}_{2\mu\mathrm{R}}t^{2}_\mathrm{LB}+\beta^{}_{1\mu\mathrm{R}}t^{}_\mathrm{LB}+\beta^{}_{0\mu\mathrm{R}}.\label{e:muR0_t}
\end{align}

Starting with the blue population, while both \shark\ and \sharkfit\ show matching slopes at low lookback times, the same is not true above $\sim5$ Gyr, where they are in tension in the sign of the slope.
Note that \sharkfit\ and GAMA are well-matched with GAMA.
This strongly suggests that the slope of the blue population in GAMA is likely set by our chosen SFH model, as due our use of a skewed Normal does not allow a large degree of variation on the SFH/ZH of early blue galaxies.
In theory, a more flexible SFH model could recover this, but recovering the evolution of $\sim$8-12 Gyr old stellar populations is intrinsically hard, so we do not expect this to significantly improve with other SFH parameterisations.
We can conclude that \shark\ produces a blue population whose colour is not as strongly dependent on stellar mass as in observations at low lookback times, but no firm conclusion can be drawn at older times.

The normalisation of the blue population means is systematically lower for \sharkfit\ than \shark, with the discrepancy increasing with lookback time, albeit with overall small differences (reaching $\sim0.1$ mag at $\sim8$ Gyr).
This reinforces the interpretation that our modelling choices play a role at early times, but the more recent evolution of the blue population is well-captured.
More significant is the difference between GAMA and \shark/\sharkfit\ at recent times, with our observations suggesting a redder blue population than our simulations.
Since figure 15 of \citetalias{lagos2018} shows that the stellar metallicities in \shark\ are in good agreement with observations, this difference suggest a difference in the stellar ages between the blue galaxies in \shark\ and GAMA.

In contrast to the blue population, there is a good agreement between GAMA and \sharkfit\ for the red population mean parameters at all lookback times.
This is a strong indication that the position of this colour population is dictated by our modelling choices for the SED fits.
Since $u-r$ is not sensitive to sSFR below $\sim10^{-11}$ yr, this is mostly indicative of possible limitations to the $Z$H modelling we adopt.

\subsection{Constrained GMM and standard deviation/fractions parameterisation}\label{subsec:GMM_refit}

With $\mu^{}_\mathrm{\{B,R\}}$ fully parameterised, we then repeat the GMM fits using these parameterisations to fix the means.
The right column of Figure \ref{fig:GAMA_GMM_fit} displays the results of this refit, which in comparison with the left column shows that our parameterisation is well behaved.
Not shown in the Figure is that this refit does affect the fits in a statistical sense, as the $\chi^2_\nu$ values increase in spread (from $\sim0.9$-$1.1$ to $\sim0.8$-$1.2$), with the values now showing a clear trend in stellar mass (higher $\chi^2_\nu$ with higher $M_\star$).
This trend should not come as a surprise, as it is clear that the red population has greatly decreased contribution to the overall population at low masses, so by forcing only one of the Gaussians to account for most of the population (as seen by comparing the fractions between columns) a smaller value is expected.
At higher masses, our assumption of linearity is not ideal, as small offsets can be seen in the means between columns, which leads to our refits to account for most, but not all, of the colour distribution.
We remark that this change is small, so this does not provide a strong argument against this parameterisation of the fractions.
Furthermore, our definition of blue and red is not purely phenomenological, as we are not asking which two Gaussian components best describe the colour distribution, but which two "distinct" components do.

From Figure \ref{fig:GAMA_GMM_fit} it can be seen that, as with the means, not all fits provide meaningful information.
As an example, the fractions and standard deviations of the red components below $\sim10^{9.5}$ M$_\odot$ in the bottom-right panel indicate that they are being used to fit a small residual population from the main (blue) population, with no identifiable red population.
As with the means, we make several selection cuts.
We retain the same selection criteria of distinguishable populations (local minimum present between means), quality of fit ($\chi^2_\nu$) and maximum stellar mass as before, only mildly increasing the maximum standard deviation allowed to 0.4 mag.

\subsubsection{Fractions parameterisation}\label{subsubsec:w}

\begin{figure}
    \centering
    \includegraphics[width=\linewidth]{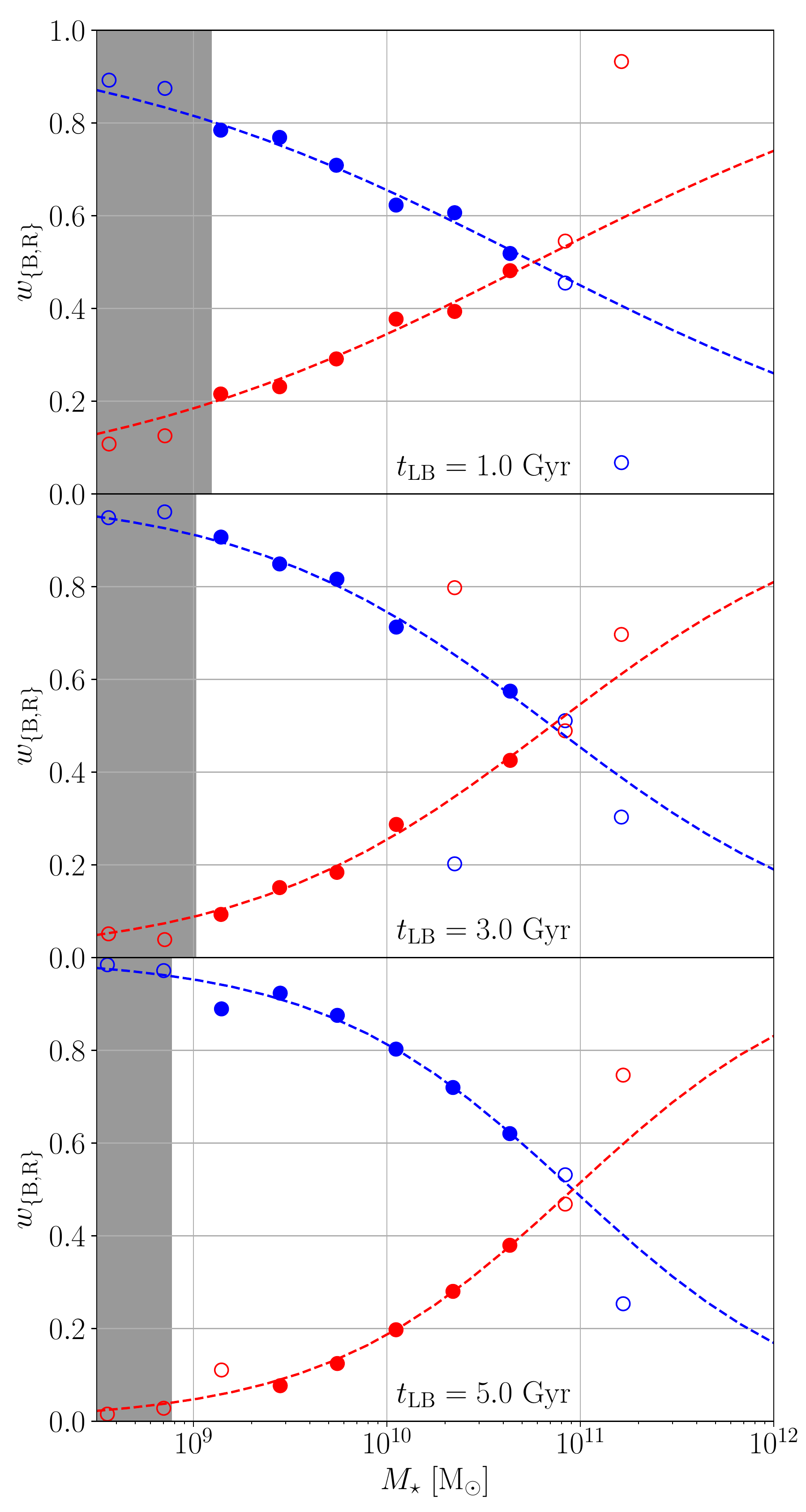}
    \caption{Example of the parameterisation of the fractions of the Gaussian components $(f^{}_\mathrm{\{B,R\}})$, as a function of stellar mass.
    Lookback times, shaded areas and points and curve colours as in Figure \ref{fig:GAMA_mean_fit}.
    Curves defined by Equations \ref{e:wR_mstar} and \ref{e:wB_mstar}.}
    \label{fig:GAMA_weight_fit}
\end{figure}

From these fits we first proceed to parameterise the fractions ($\{f^{}_\mathrm{B},f^{}_\mathrm{R}\}$) and standard deviations ($\{\sigma^{}_\mathrm{B},\sigma^{}_\mathrm{R}\}$) of the distributions.
For the fractions, we first fit them logistic curves as a function of stellar mass:
\begin{align}
    f^{}_\mathrm{R}&=\frac{1}{1+e^{-k\left(\log_{10}(M_\star)-M^{}_\mathrm{T}\right)}}\label{e:wR_mstar}\\
    f^{}_\mathrm{B}&=1-f^{}_\mathrm{R}\label{e:wB_mstar}
\end{align}
\noindent where $M^{}_\mathrm{T}$ is the stellar mass where both populations have equal fractions and $k$ defines the sharpness of the transition\footnote{The slope of the curve at $M^{}_\mathrm{T}$ is one quarter of $k$, $f^{'}_\mathrm{R}(M^\mathrm{T})=k/4$.}.
This parameterisation naturally models two populations where one dominates above $M^{}_\mathrm{T}$ and the other below, with the benefit of only requiring two free parameters for a description of both.
Figure \ref{fig:GAMA_weight_fit} shows three examples of these fits for GAMA, which shows that this model is well justified by the data, even for points that have not been used to fit the free parameters.
As with the mean parameterisation, we find only a weak change in the $\chi^2_\nu$ values when using this parameterisation.
This is the reason why we do not adopt the more complex stellar mass function modelling used by \citetalias{taylor2015}, two parameters are enough to describe our data.

\begin{figure}
    \centering
    \includegraphics[width=\linewidth]{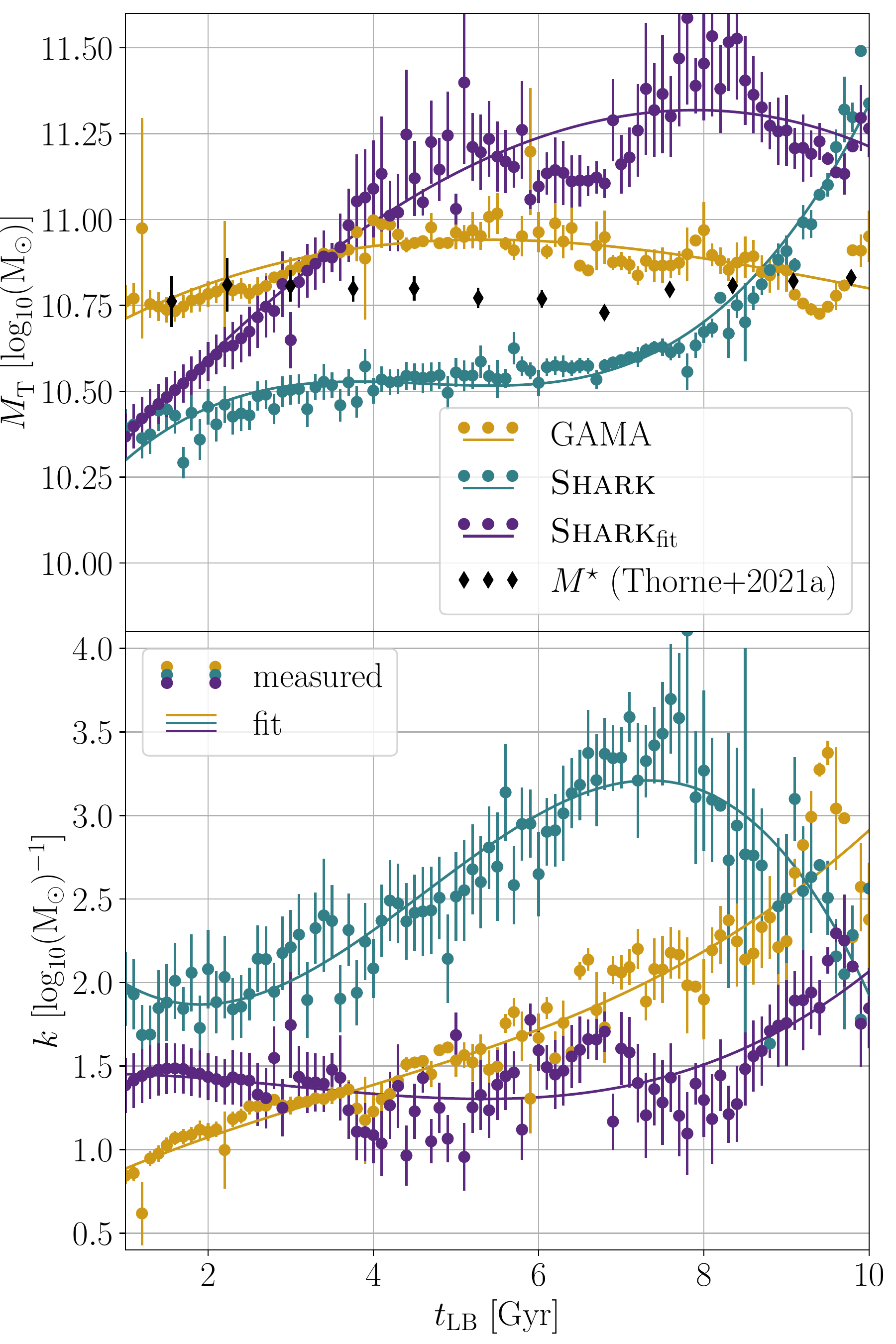}
    \caption{Time evolution of the blue/red population fractions parameterisation.
    Curves and markers as in Figure \ref{fig:evol_mu}.
    The top panel shows the evolution of the transition mass  and time evolution fit ($M^{}_\mathrm{T}$, Equations \ref{e:wR_mstar} and \ref{e:wMt_t}), together with the measured evolution of the stellar mass function characteristic mass ($M^\star$) from table D1 of \citetalias{thorne2021}.
    The bottom shows the evolution of the sharpness of the transition and the time evolution fit ($k$, Equations \ref{e:wR_mstar} and \ref{e:wk_t}) in the same manner.
    The error bars indicate the uncertainty in the fitted parameters.}
    \label{fig:evol_w}
\end{figure}

Based on evolution of $M^{}_\mathrm{T}$ and $k$ seen in Figure \ref{fig:evol_w} we fit the time evolution of these parameters using the third-order polynomials:
\begin{align}
    M^{}_\mathrm{T}&=M^{}_{3\mathrm{T}}t^3_\mathrm{LB}+M^{}_{2\mathrm{T}}t^2_\mathrm{LB}+M^{}_{1\mathrm{T}}t^{}_\mathrm{LB}+M^{}_{0\mathrm{T}}\label{e:wMt_t},\\
    k&=k_3t^3_\mathrm{LB}+k_2t^2_\mathrm{LB}+k_1t^{}_\mathrm{LB}+k_0.\label{e:wk_t}
\end{align}
\noindent \shark\ and \sharkfit\ exhibit consistent transition masses ($M^{}_\mathrm{T}$) at recent ($\lesssim2$ Gyr) and early ($\gtrsim9$ Gyr) lookback times but diverge in between, showing that the oldest red galaxies are being modelled correctly but the rest of the population build-up is delayed.
Compared to our simulations, for which $M^{}_\mathrm{T}$ decreases by $\sim1$ dex, we find a comparatively weak evolution of this parameter in GAMA.
Interestingly, our finding that $M^{}_\mathrm{T}$ in \shark\ is lower by $\sim0.3$ dex than in GAMA at recent times ($t^{}_\mathrm{LB}\leq2$ Gyr) is in tension with the results from \citet{bravo2020}, where with a more qualitative assessment we found [in dust-attenuated $g-i$] the opposite to be true (\shark\ transitioning $\sim0.3$ dex higher).
While there are caveats in this comparison, mainly that here we use a subset of one simulation and intrinsic $u-r$ colour instead of the GAMA lightcone and attenuated $g-i$\footnote{Our use of absolute magnitude and apparent magnitudes in \citet{bravo2020} is of secondary concern, as in both cases we are dealing with low redshifts, which are also roughly comparable, $z=0.0668$ compared to $0.003<z<0.12$ in \citet{bravo2020}.}, this suggests that the driver of this difference is our dust model in \shark.

Figures 5 and 13 of \citetalias{taylor2015} suggest that we should expect this transition mass
to be in good agreement with the characteristic mass of the stellar mass function ($M^\star$).
For this purpose, we also include the measured evolution of $M^\star$ found by \citetalias{thorne2021}, measured from the Deep Extragalatic VIsible Legacy Survey \citep[DEVILS][]{davies2018}.
We find these values to be in remarkable agreement with our measured evolution of $M^{}_\mathrm{T}$, given that we are reconstructing the evolution of $M^{}_\mathrm{T}$ from low-redshift data, while \citetalias{thorne2021} directly measured $M^\star$ at every lookback time indicated by the data in the Figure.
While there are no available fits of Schechter functions to the stellar mass function from \shark, figure 5 from \citetalias{lagos2018} suggest that $M^\star\sim$\mstar{10.5} for $z\leq1$, in good agreement with our measured $M^{}_\mathrm{T}$ for \shark. 
This is strong evidence that we are justified in our simpler model for the relative fraction of blue/red galaxies compared to \citetalias{taylor2015}.

The sharpness of the transition in \sharkfit\ shows a weak evolution with lookback time, indicating that in this sample the assembly of the red population is fully captured only with the change in the transition mass.
This is in contrast to GAMA and \shark, which exhibit comparatively little evolution in the transition mass (below $\sim8$ Gyr for \shark), with the assembly of the red population being captured by a strong decrease in the sharpness of the transition.
This difference suggest that.
As with the transition mas, \shark\ and \sharkfit\ are in reasonable agreement at recent/early lookback times, but diverge strongly in the $\sim4$-$8$ Gyr range.
The poor agreement between GAMA and \sharkfit\ suggest that this tension is not the result of the assumed SFH/$Z$H for our SED fits, but that this is a consequence of the dust parameters we assume for our fits (see Section \ref{app:SHARKfit_qual}).

\subsubsection{Standard deviations parameterisation}\label{subsubsec:sigma}

\begin{figure}
    \centering
    \includegraphics[width=\linewidth]{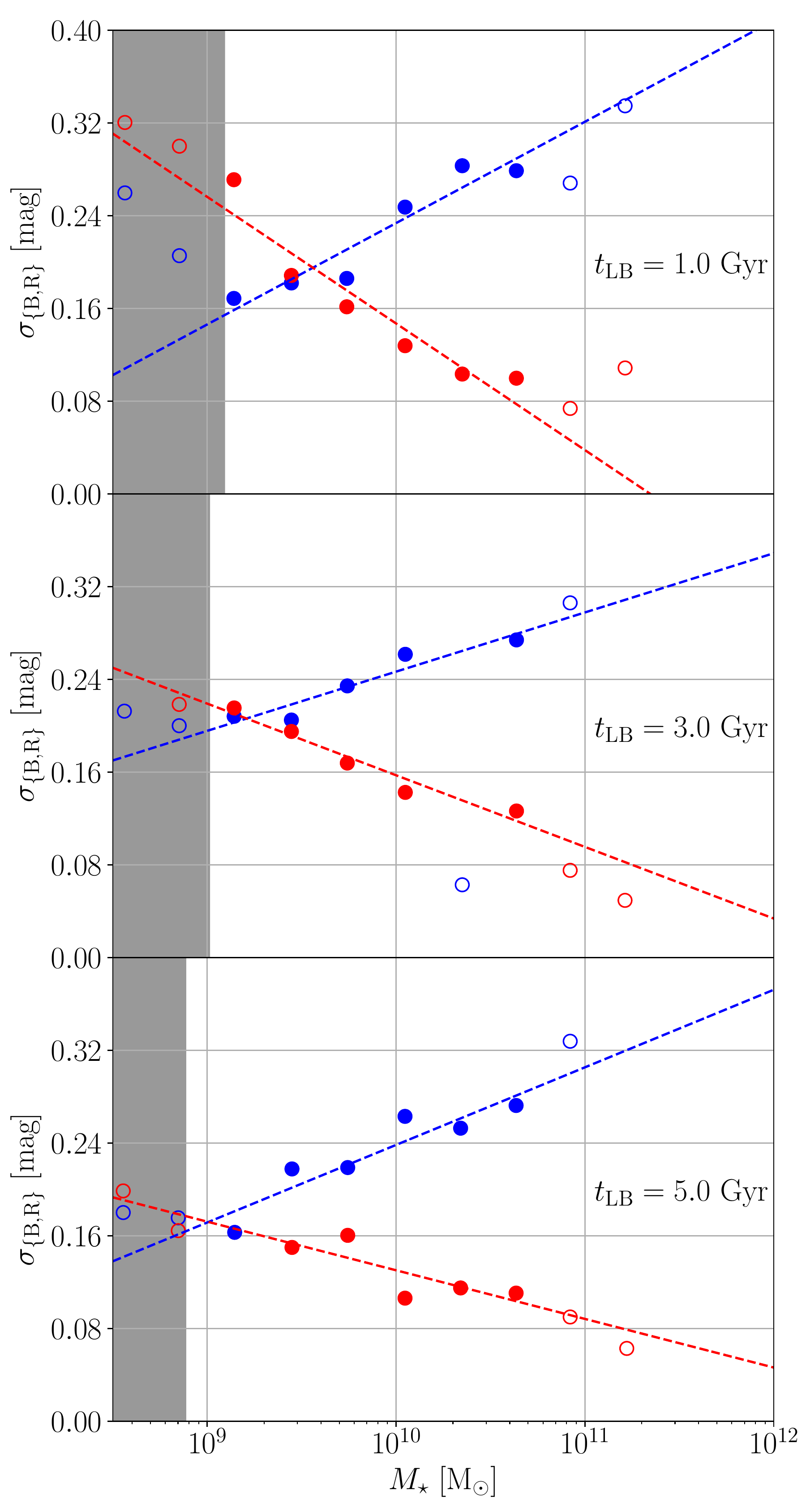}
    \caption{Example of the parameterisation of the standard deviations of the Gaussian components ($\sigma^{}_\mathrm{\{B,R\}}$), as a function of stellar mass.
    Lookback times, shaded areas and points and curve colours as in Figure \ref{fig:GAMA_mean_fit}.
    Curves defined by Equations \ref{e:sigmaB_mstar} and \ref{e:sigmaR_mstar}.}
    \label{fig:GAMA_std_fit}
\end{figure}

\citet{katsianis2019,davies2019a,davies2022} found that the galaxy star-forming main sequence displays a local minimum dispersion at $\sim10^9$ M$_\odot$, which can be well described by a second-order polynomial.
We considered a similar choice for the parameterisation of the standard deviation of the blue population, but we find no evidence of a similar behaviour above our chosen mass limit, as shown in Figure \ref{fig:GAMA_std_fit}.

\begin{figure*}
    \centering
    \includegraphics[width=\linewidth]{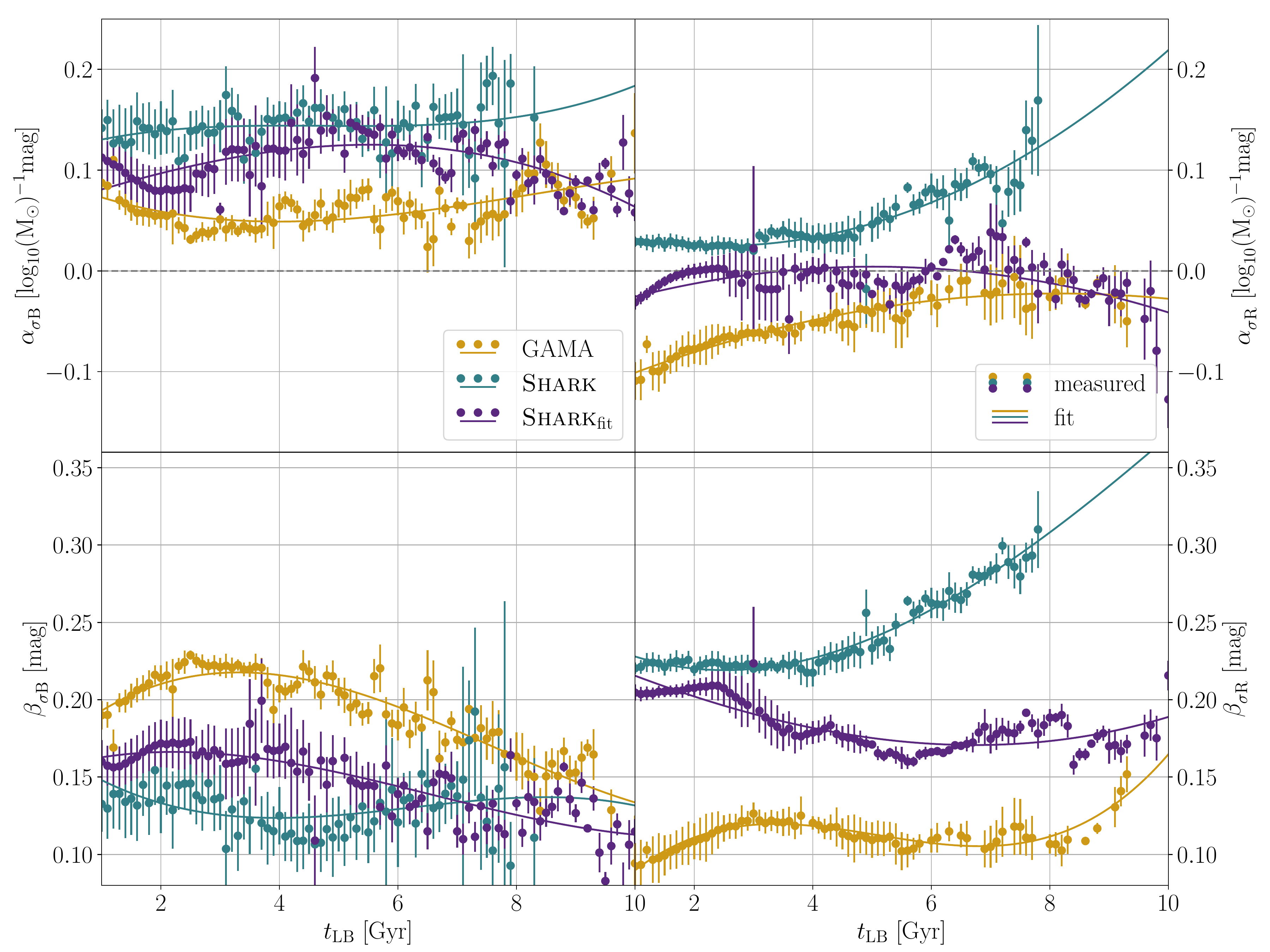}
    \caption{Time evolution of the blue and red population standard deviation parameterisations.
    Curves and markers as in Figure \ref{fig:evol_mu}.
    The left column shows the results for the blue population and the right for the red population.
    In the top row, the evolution of the slope of the standard deviation as a function of stellar mass ($\alpha^{}_{\sigma\mathrm{\{B,R\}}}$, Equations \ref{e:sigmaB_mstar} and \ref{e:sigmaR_mstar}) is shown with the round markers, and the fit to these (Equations \ref{e:sigmaB1_t} and \ref{e:sigmaR1_t}) with solid lines.
    The bottom shows the value of the standard deviations at $M_\star=$\mstar{\{9.5,10.5\}} ($\beta^{}_{\sigma\mathrm{\{B,R\}}}$, Equations \ref{e:sigmaB_mstar} and \ref{e:sigmaR_mstar}) and fits (Equations \ref{e:sigmaB0_t} and \ref{e:sigmaR0_t}) in the same manner.
    The error bars indicate the uncertainty in the fitted parameters.}
    \label{fig:evol_sigma}
\end{figure*}

For this reason, we have fitted both standard deviations using linear relations as a function of mass:
\begin{align}
    \sigma^{}_\mathrm{B}&=\alpha^{}_{\sigma\mathrm{B}}\log_{10}(M_\star/10^{9.5}\mathrm{M}_\odot)+\beta^{}_{\sigma\mathrm{B}}\label{e:sigmaB_mstar}\\
    \sigma^{}_\mathrm{R}&=\alpha^{}_{\sigma\mathrm{R}}\log_{10}(M_\star/10^{10.5}\mathrm{M}_\odot)+\beta^{}_{\sigma\mathrm{R}}\label{e:sigmaR_mstar}
\end{align}
\noindent where $\{\alpha^{}_{\sigma\mathrm{B}},\alpha^{}_{\sigma\mathrm{R}}\}$ are the slopes of the relation and $\{\beta^{}_{\sigma\mathrm{B}},\beta^{}_{\sigma\mathrm{R}}\}$ the values of the standard deviations at $M_\star=\{$\mstar{9.5},\mstar{10.5}$\}$.
One limitation of this approach, as seen in Figure \ref{fig:GAMA_std_fit}, is that we run the risk of under-predicting the standard deviations at either stellar mass end.
Examining all three samples we find that sensible lower limits are $0.12$ and $0.08$ mag for the blue and red populations respectively\footnote{We tested using second-order polynomials, but the second-order term were markedly unstable, strongly fluctuating around zero as a function of lookback time (for all samples).
This suggests that our data is not constraining enough for such parameterisation. Still, just a simple linear model is not enough, particularly for Shark, which exhibits a flat dependency of the standard deviation on stellar mass in the lowest 2-3 stellar mass bins.
Not being able to constrain second-order fits, and with linear fits underpredicting the dispersion in some cases, we decided to add this floor.
Since the floor is essentially a third free parameter (that we choose to fix), this model is equivalent in degrees of freedom to a second-order polynomial.}, as at no time step the values from the selected GMM fits go below the mentioned limits for any of our samples.
These values are between $\sim20\%$ of the lower limits we establish, which allows some flexibility for the minimum in our parameterised evolution, while still avoiding artificially low parameterised values.

Similar to the means, based on the results shown Figure \ref{fig:evol_sigma} we decide to fit with third-order polynomials for all parameters save the slope as function of stellar mass for the red population ($\alpha^{}_{\sigma\mathrm{R}}$), for which we use a polynomial of lower order (second-order in this case).
These equations are:
\begin{align}
    \alpha^{}_{\sigma\mathrm{B}}&=\alpha^{}_{3\sigma\mathrm{B}}t^3_\mathrm{LB}+\alpha^{}_{2\sigma\mathrm{B}}t^2_\mathrm{LB}+\alpha^{}_{1\sigma\mathrm{B}}t^{}_\mathrm{LB}+\alpha^{}_{0\sigma\mathrm{B}}\label{e:sigmaB1_t},\\
    \beta^{}_{\sigma\mathrm{B}}&=\beta^{}_{3\sigma\mathrm{B}}t^3_\mathrm{LB}+\beta^{}_{2\sigma\mathrm{B}}t^2_\mathrm{LB}+\beta^{}_{1\sigma\mathrm{B}}t^{}_\mathrm{LB}+\beta^{}_{0\sigma\mathrm{B}}\label{e:sigmaB0_t},\\
    \alpha^{}_{\sigma\mathrm{R}}&=\alpha^{}_{2\sigma\mathrm{R}}t^2_\mathrm{LB}+\alpha^{}_{1\sigma\mathrm{R}}t^{}_\mathrm{LB}+\alpha^{}_{0\sigma\mathrm{R}},\label{e:sigmaR1_t}\\
    \beta^{}_{\sigma\mathrm{R}}&=\beta^{}_{3\sigma\mathrm{R}}t^2_\mathrm{LB}+\beta^{}_{3\sigma\mathrm{R}}t^2_\mathrm{LB}+\beta^{}_{1\sigma\mathrm{R}}t^{}_\mathrm{LB}+\beta^{}_{0\sigma\mathrm{R}}.\label{e:sigmaR0_t}
\end{align}
\noindent The similar slopes of the blue standard deviation between \shark and \sharkfit\ suggest that this is not strongly sensitive to the modelling choices in \prospect, though we cannot fully rule this out as \sharkfit\ straddles between GAMA and \shark\ at recent times ($t^{}_\mathrm{LB}\lesssim3$ Gyr).
The strongest evidence for any effect is at early times ($t^{}_\mathrm{LB}\gtrsim8$ Gyr), where GAMA and \sharkfit\ come into good agreement.
The normalisation of the standard deviation of the blue population shows similar results, with a better agreement between \shark\ and \sharkfit\ than either to GAMA.
Both parameters of the standard deviation of the red population ($\alpha^{}_{\sigma\mathrm{R}}$ and $\beta^{}_{\sigma\mathrm{R}}$) show a similar trend to what we find for the slope of means of the blue population ($\alpha^{}_{\mu\mathrm{B}}$), where at recent times they are in better agreement with \shark\ but at early times with GAMA, pointing to another aspect encoded in the modelling choices in \prospect.

\section{Discussion}\label{sec:disc}

\begin{figure*}
    \centering
    \includegraphics[width=\linewidth]{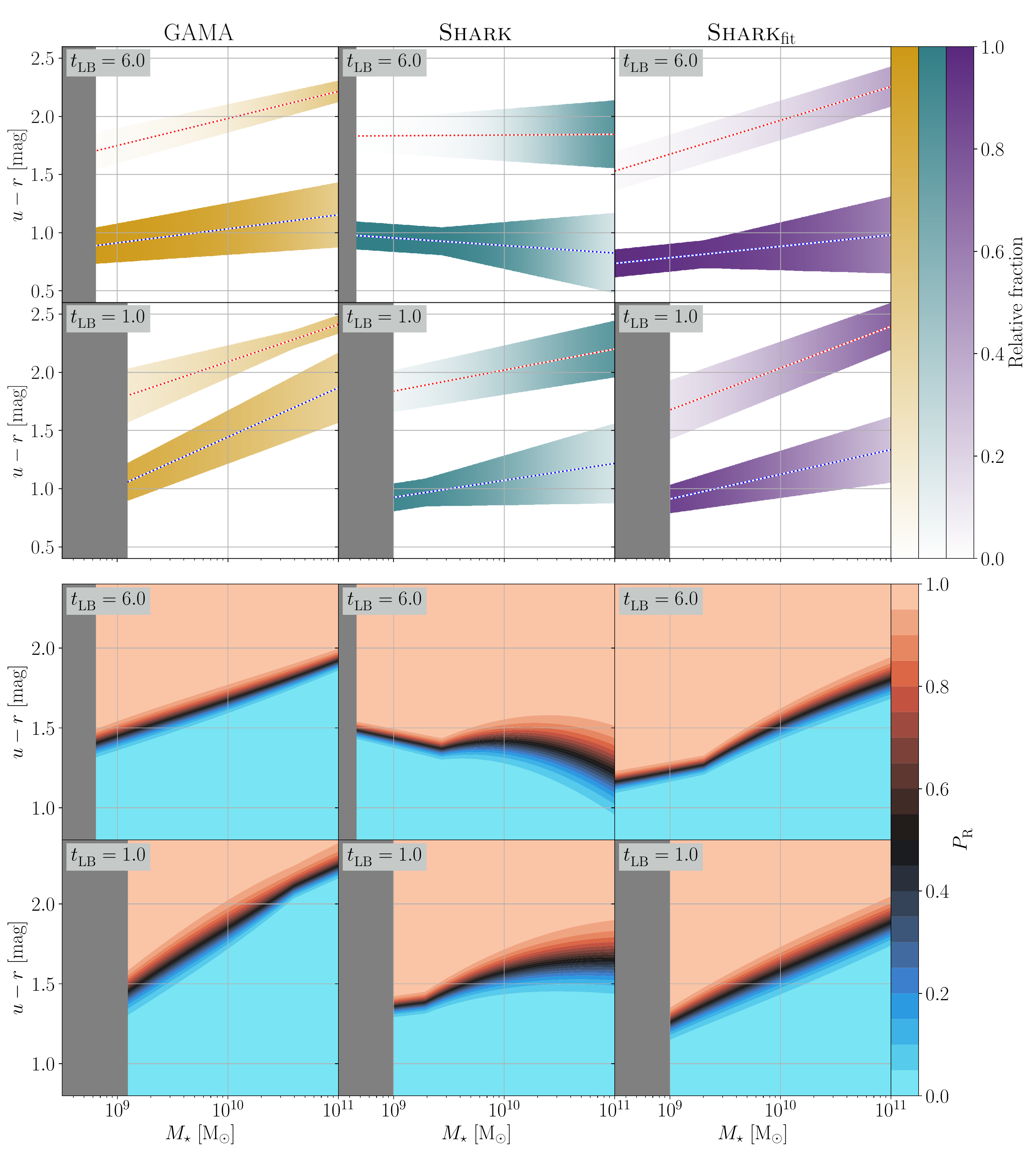}
    \caption{Top set of panels: Schematic representation of the evolution, drawn using the parameterisations presented in Section \ref{sec:evol_tracks} (\ref{subsec:GMM_fit} and \ref{subsec:GMM_refit} in particular).
    The top row shows the populations at a lookback time of 6 Gyr, the bottom at 1 Gyr.
    Each column corresponds to one of our galaxy samples: GAMA (left, orange), \shark\ (middle, cyan) and \sharkfit\ (right, purple).
    The lines indicate the mean of each population as a function of stellar mass (colour-coded by the population they trace, blue or red), the shaded areas the 1-sigma width of the population as a function of mass, and the opacity shows the relative contribution of each population to the full population at a given stellar mass (lighter colours indicating a smaller contribution).
    Bottom set of panels: Contour map of the probability for a galaxy of being red ($P^{}_\mathrm{R}$), as a function of stellar mass, colour and time.
    Note the smaller range in $u-r$ compared to the top group of panels, which is to highlight the transition zone.
    The grey-shaded areas in all panels indicate the mass completeness limit for each sample and lookback time.
    An animated version of this Figure is available in the supplemental material.}
    \label{fig:summary}
\end{figure*}

As mentioned earlier, before we can address the main question (i), there are three related questions (ii-iv) we must first consider.
To answer these questions, Figure \ref{fig:summary} presents a schematic view of the evolution of both blue and red populations for our samples and how they translate into probability maps, and Figure \ref{fig:PRex} shows an example of these probability maps to each sample in both colour-mass and sSFR-mass planes.

\subsection{(ii) How well we can reconstruct the colour evolution of galaxies from their panchromatic SEDs?}

\begin{figure*}
    \centering
    \includegraphics[width=\linewidth]{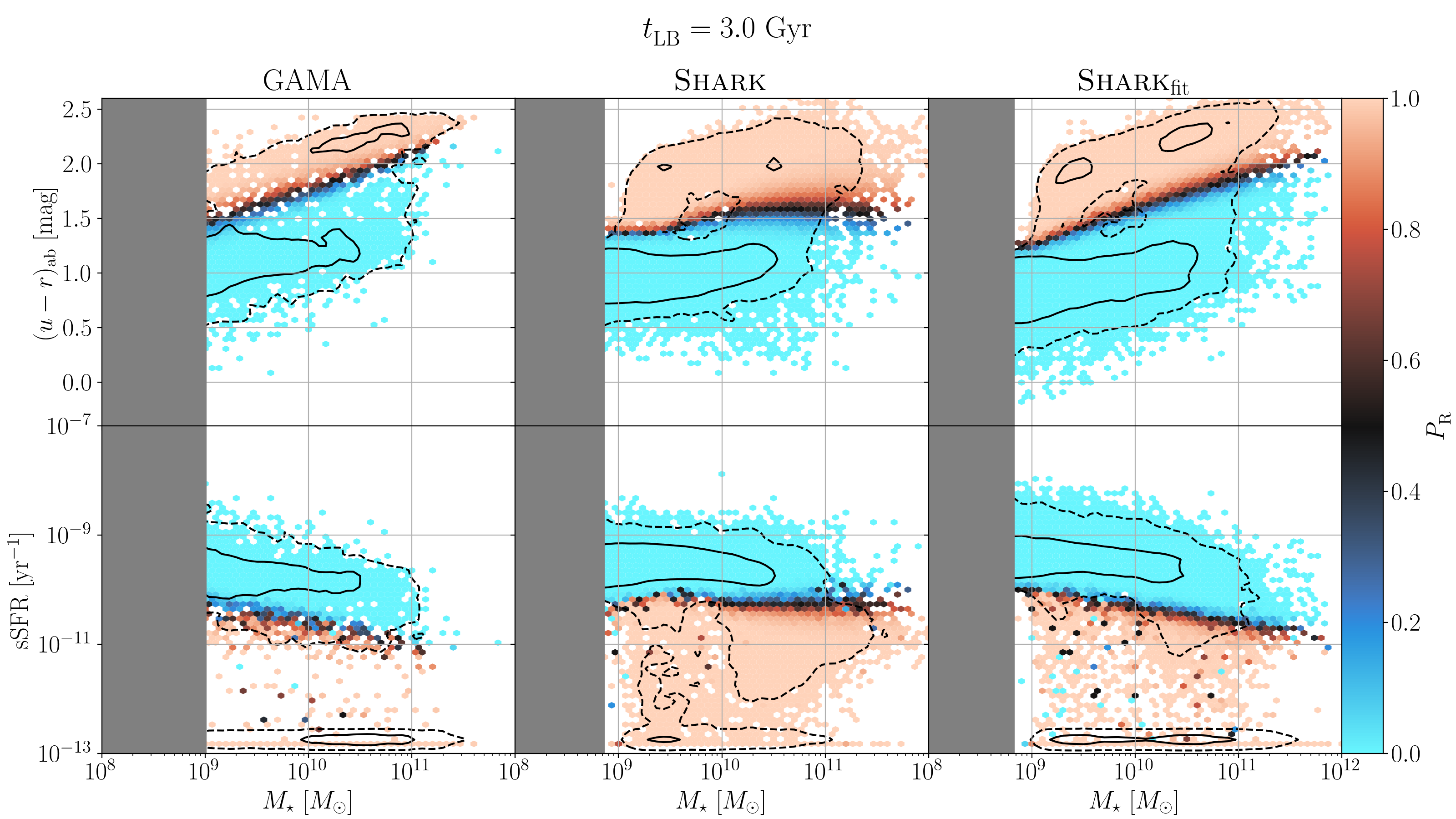}
    \caption{Example of our population modelling applied to our samples to quantify the probability of any given galaxy of being red.
    The top row show this classification in colour-mass space, the bottom in sSFR-mass.
    Each row corresponds to each different sample.
    For visualisation purposes we have set the sSFR of all galaxies below $10^{12.8}$ Gyr$^{-1}$ at that value. 
    Black lines indicate contours encircling the highest-density region that contains $68.3\%$ (solid) and $95.5\%$ (dashed) of the galaxies
    An animated version of this Figure is available in the supplemental material.}
    \label{fig:PRex}
\end{figure*}

We can utilise the results from \sharkfit\ to discern which aspects of the colour evolution we can recover with \prospect\ (where \sharkfit\ and \shark\ match) from those that we cannot (where \sharkfit\ matches GAMA).
The one challenge to such analysis is the evolution of the fraction parameterisation, where outside of narrow time windows, \sharkfit\ is in tension with both GAMA and \shark.
While clear in our measurements, we note that the effect of this tension in the populations is subtler than suggested by Figure \ref{fig:evol_w}, as shown by the top set of panels in Figure \ref{fig:summary}.
The match between \sharkfit\ and \shark\ at recent and early times suggests that the difference is not likely to be a consequence of the modelling in \prospect, instead likely an outcome of our particular SED fits to \shark, which we explore in more detail in Section \ref{app:SHARKfit_qual}.

Regarding the means, we see clear evidence that the position of the red population is set by the modelling choices in \prospect, and to some degree for the blue population.
Starting with the latter, \sharkfit\ is in good agreement with \shark\ at recent times ($\lesssim4$ Gyr), but at early times ($\gtrsim8$ Gyr) it is closer to GAMA, particularly for the slope of the means.
This suggests that we are able to reconstruct the position of the blue population a few Gyr into the past, but beyond that the reconstruction is biased.
The latter is likely a result of combining the reduced constraining power of galaxy SEDs as a function of stellar age with the skewed Normal SFH we use in \prospect.
Regarding the means of the red population, Figures \ref{fig:summary} and \ref{fig:PRex} show that \sharkfit\ is well-matched to \shark\ only for low-mass galaxies, with more massive ones being recovered as too red.
This mismatch is the outcome of assuming a fixed value of $\eta^{}_\mathrm{ISM}$ and $\eta^{}_\mathrm{ISM}$, which is not only is a poor representation of these values in \shark\ \citep[figure 4 from][]{lagos2019}, but also sets the slope of the red population.

As with the means, we see different behaviours for the match between \sharkfit\ and GAMA/\shark\ for the blue and red populations.
The standard deviation of the blue population in \sharkfit\ is consistently in better agreement with \shark\ than GAMA, suggesting, at best, small biases in the recovery of this aspect of the colour population evolution.
For the red population, we see a similar behaviour to the slope of the means of the blue population, where \sharkfit\ broadly agrees well with \shark\ at recent times but with GAMA at longer lookback times.
In practice, this sets an upper limit for the trustworthiness of our reconstruction, which we estimate at $t^{}_\mathrm{LB}\sim6$ Gyr, given that \shark\ and \sharkfit\ are broadly similar up to that lookback time.
We will use this time limit in Bravo et al. (in preparation).

\subsection{(iii) How can we best define the blue and red populations across cosmic time?}

We can fully describe the galaxy colour distribution in our three galaxy samples (GAMA, \shark\ and \sharkfit) as a combination of two Gaussians.
At a given stellar mass and time, the parameters of these Gaussians (means, fractions and standard deviations) can be accurately described by simple functional forms with only two free parameters.
For the means, this is not only in line with the well-established colour-magnitude relation for both early- and late-type galaxies \citep[e.g.,][]{baum1959,faber1973,visvanathan1977,bower1992}, but also with the results shown for GAMA by \citetalias{taylor2015} in the overlapping mass range.
Also broadly aligned with \citetalias{taylor2015} is our finding of linear relations between the width of the Gaussians and stellar mass.
We, however, expand over previous work in showing that these results hold at earlier cosmic times as well as at $z\sim0$.

Together the means and standard deviations produce a fairly simple description of the blue and red populations, as can be seen in the upper panels of Figure \ref{fig:summary}.
The inclusion of fractions in our modelling unlocks the more complex distribution of the probability of a galaxy being red that is seen in the bottom (top) panels of Figure \ref{fig:summary} (\ref{fig:PRex}).
These show that a statistically-driven selection of blue and red galaxies can not be replicated by the more simple selection criteria that are common in literature \citep[e.g.,][]{goncalves2012,smethurst2015,bremer2018,phillipps2019}.
Figure \ref{fig:PRex} shows that this classification will not lead to a clean separation in sSFR-mass, but still remains fairly well-defined.

\subsection{(iv) Is the green value a superposition of the blue and red populations, or a population on its own?}

In agreement with the results from \citet{schawinski2014} and \citetalias{taylor2015}, we do not find evidence for a separate green population at any cosmic time.
Hence, the green valley is only a product of the overlap of the blue and red populations.
The presence of this overlap between the blue and red populations indicates that galaxies become red before they reach a colour close to the mean of the red population, suggesting that the transformation from blue to red happens on shorter timescales than the transition from being confidently-classified as blue to confidently classified as red.

While small, the overlap between the two populations also indicates that there are blue (red) galaxies that are redder (bluer) than the bluest (reddest) galaxies of the red (blue) population.
While that statement may seem contradictory, it points to an intrinsic difference between both populations that it is not fully captured just by colour and stellar mass.
This reinforces what was already suggested by \citet{schawinski2014}, that the processes responsible for transforming a blue galaxy into a red galaxy are not the same, or at least operate differently, to those that make a blue galaxy a comparatively red member of the blue population.

Figure \ref{fig:PRex} shows that the overlap between the two populations is remarkably narrow, as indicated by the regions where the shading in the Figure transitions from blue to red.
This narrow transition is similar to the results seen in figure 11 of \citetalias{taylor2015} but unlike classifications used to measure the timescale on which galaxies cross the green valley by e.g., \citet{goncalves2012,smethurst2015,bremer2018}.
The expectation from this is that we will measure shorter transition timescales than previous literature using similar methods, which we will explore in Bravo et al. (in preparation).

\subsection{(i) How have the colours of the local blue and red populations evolve with time?}

We find that both blue and red massive galaxies become redder with time, while smaller galaxies remain a similar colour through cosmic time, in agreement with the cosmic SFH results in \citetalias{bellstedt2020b}.
These two results combined are consistent with the idea of downsizing \citep[e.g.,][]{cowie1996,kodama2004,neistein2006,fontanot2009,goncalves2012}.
This suggests that small galaxies follow evolutionary paths that are not directly affected by the age of the Universe, while massive galaxies become noticeably more metal-rich, less star-forming, or a mix of both, with cosmic time.
Another almost constant quantity with time is the width of the colour distribution of each population, which suggests that, for a given mass, the variety of evolutionary paths that galaxies follow to reach said mass is invariant with cosmic time.
The sharpness of the transition from blue- to red-dominated decreases with time, which implies that the quenching timescale for galaxies near the transition mass becomes larger with time\newtext.
At the low-mass end, the near static nature of both populations suggests the timescales are near time-invariant for these galaxies.

\section{Conclusions}\label{sec:sum}

In this work we have introduced a novel method to reconstruct the colour evolution of low-redshift galaxies, using the SED fitting software \prospect.
We then present a method to classify galaxies into colour populations and characterising their evolution in time.
We test these methods using simulated galaxies from \shark, testing both the predicted evolution from the simulation and how our reconstructing method matches the true evolution when the answer is known beforehand.

Our main findings can be summarised as:
\begin{itemize}
    \item We can to reconstruct the evolution of galaxies up to a lookback time of $\sim6$ Gyr, from where our results become driven by the modelling choices we adopt for the SED fitting.
    \item We find no evidence of a green galaxy population, with the green valley being a mix of blue and red galaxies.
    \item While in good qualitative agreement, small but measurable tensions in the colour evolution of galaxies are apparent at the quantitative level between simulations and observations.
    \item At a fixed stellar mass, we observe a strong colour evolution for massive galaxies, both blue and red, while low-mass galaxies remain of a similar colour.
    \item We find that galaxies reaching a given stellar mass display a variety of evolutionary paths that is invariant with time, as encoded in the almost complete lack of evolution of the width of the populations.
    \item We find further evidence for the red population assembling from the high-mass end down.
\end{itemize}

These results will serve as the foundation for Bravo et al. (in preparation), where we will use this statistical model of the colour population to select present-day red galaxies and study the timescales for their transition from blue to red galaxies.


\section*{Acknowledgements}

We thank Chris Power and Pascal Elahi for their role in completing the SURFS $N$-body DM-only simulations suite, Rodrigo Tobar for his contributions to \shark, and the anonymous referee for their constructive report.
MB acknowledges the support of the University of Western Australia through a Scholarship for International Research Fees and Ad Hoc Postgraduate Scholarship.
LJMD and ASGR acknowledge support from the Australian Research Councils Future Fellowship scheme (FT200100055 and FT200100375, respectively)
CdPL is funded by the ARC Centre of Excellence for All Sky Astrophysics in 3 Dimensions (ASTRO 3D), through project number CE170100013.
CdPL also thanks the MERAC Foundation for a Postdoctoral Research Award.
SB acknowledges support by the Australian Research Council’s funding scheme DP180103740.
JET is supported by the Australian Government Research Training Program (RTP) Scholarship.

This work was supported by resources provided by the Pawsey Supercomputing Centre with funding from the Australian Government and the Government of Western Australia.
We gratefully acknowledge DUG Technology for their support and HPC services.

GAMA is a joint European-Australasian project based around a spectroscopic campaign using the Anglo-Australian Telescope.
The GAMA input catalogue is based on data taken from the Sloan Digital Sky Survey and the UKIRT Infrared Deep Sky Survey.
Complementary imaging of the GAMA regions is being obtained by a number of independent survey programmes including GALEX MIS, VST KiDS, VISTA VIKING, WISE, Herschel-ATLAS, GMRT and ASKAP providing UV to radio coverage.
GAMA is funded by the STFC (UK), the ARC (Australia), the AAO, and the participating institutions.
The GAMA website is \url{http://www.gama-survey.org/}.
Based on observations made with ESO Telescopes at the La Silla Paranal Observatory under programme ID 179.A-2004.
Based on observations made with ESO Telescopes at the La Silla Paranal Observatory under programme ID 177.A-3016.

The analysis on this work was performed using the programming languages \textsc{Python} v3.8 (\url{https://www.python.org}) and R v4.0 (\url{https://www.r-project.org}), with the open source libraries \textsc{celestial} \citep{celestial}, \textsc{data.table} (\url{https://github.com/Rdatatable/data.table}), \textsc{foreach} (\url{https://github.com/RevolutionAnalytics/foreach}), \textsc{matplotlib} \citep{matplotlib}, \textsc{mixtools} \citep{mixtools}, \textsc{NumPy} \citep{numpy}, \textsc{pandas} \citep{pandas}, \textsc{scicm} (\url{https://github.com/MBravoS/scicm}), and \textsc{SciPy} \citep{scipy}, in addition of the software previously described.


\section*{Data availability}

The SED fitting data from GAMA was provided by Sabine Bellstedt by permission, and will be shared on request to the corresponding author with permission of Sabine Bellstedt.
The \shark\ simulated SEDs and SFH/$Z$H were produced by the \shark\ team for this work, and will be shared on reasonable request to the corresponding author.
The \prospect\ fits to \shark\ galaxies and all reconstructed colour evolution tracks generated for this work will be shared on reasonable request to the corresponding author.

\bibliographystyle{mnras}
\bibliography{papers} 

\begin{thebibliography}{}
\makeatletter
\relax
\def\mn@urlcharsother{\let\do\@makeother \do\$\do\&\do\#\do\^\do\_\do\%\do\~}
\def\mn@doi{\begingroup\mn@urlcharsother \@ifnextchar [ {\mn@doi@}
  {\mn@doi@[]}}
\def\mn@doi@[#1]#2{\def\@tempa{#1}\ifx\@tempa\@empty \href
  {http://dx.doi.org/#2} {doi:#2}\else \href {http://dx.doi.org/#2} {#1}\fi
  \endgroup}
\def\mn@eprint#1#2{\mn@eprint@#1:#2::\@nil}
\def\mn@eprint@arXiv#1{\href {http://arxiv.org/abs/#1} {{\tt arXiv:#1}}}
\def\mn@eprint@dblp#1{\href {http://dblp.uni-trier.de/rec/bibtex/#1.xml}
  {dblp:#1}}
\def\mn@eprint@#1:#2:#3:#4\@nil{\def\@tempa {#1}\def\@tempb {#2}\def\@tempc
  {#3}\ifx \@tempc \@empty \let \@tempc \@tempb \let \@tempb \@tempa \fi \ifx
  \@tempb \@empty \def\@tempb {arXiv}\fi \@ifundefined
  {mn@eprint@\@tempb}{\@tempb:\@tempc}{\expandafter \expandafter \csname
  mn@eprint@\@tempb\endcsname \expandafter{\@tempc}}}

\bibitem[\protect\citeauthoryear{{Amarantidis} et~al.,}{{Amarantidis}
  et~al.}{2019}]{amarantidis2019}
{Amarantidis} S.,  et~al., 2019, \mn@doi [\mnras] {10.1093/mnras/stz551}, \href
  {https://ui.adsabs.harvard.edu/abs/2019MNRAS.485.2694A} {485, 2694}

\bibitem[\protect\citeauthoryear{{Arnaboldi}, {Neeser}, {Parker}, {Rosati},
  {Lombardi}, {Dietrich}  \& {Hummel}}{{Arnaboldi}
  et~al.}{2007}]{arnaboldi2007}
{Arnaboldi} M.,  {Neeser} M.~J.,  {Parker} L.~C.,  {Rosati} P.,  {Lombardi} M.,
   {Dietrich} J.~P.,   {Hummel} W.,  2007, The Messenger, \href
  {https://ui.adsabs.harvard.edu/abs/2007Msngr.127...28A} {127, 28}

\bibitem[\protect\citeauthoryear{{Baldry}, {Glazebrook}, {Brinkmann},
  {Ivezi{\'c}}, {Lupton}, {Nichol}  \& {Szalay}}{{Baldry}
  et~al.}{2004}]{baldry2004}
{Baldry} I.~K.,  {Glazebrook} K.,  {Brinkmann} J.,  {Ivezi{\'c}} {\v{Z}}.,
  {Lupton} R.~H.,  {Nichol} R.~C.,   {Szalay} A.~S.,  2004, \mn@doi [\apj]
  {10.1086/380092}, \href
  {https://ui.adsabs.harvard.edu/abs/2004ApJ...600..681B} {600, 681}

\bibitem[\protect\citeauthoryear{{Baldry}, {Balogh}, {Bower}, {Glazebrook},
  {Nichol}, {Bamford}  \& {Budavari}}{{Baldry} et~al.}{2006}]{baldry2006}
{Baldry} I.~K.,  {Balogh} M.~L.,  {Bower} R.~G.,  {Glazebrook} K.,  {Nichol}
  R.~C.,  {Bamford} S.~P.,   {Budavari} T.,  2006, \mn@doi [\mnras]
  {10.1111/j.1365-2966.2006.11081.x}, \href
  {https://ui.adsabs.harvard.edu/abs/2006MNRAS.373..469B} {373, 469}

\bibitem[\protect\citeauthoryear{{Baldry} et~al.,}{{Baldry}
  et~al.}{2014}]{baldry2014}
{Baldry} I.~K.,  et~al., 2014, \mn@doi [\mnras] {10.1093/mnras/stu727}, \href
  {https://ui.adsabs.harvard.edu/abs/2014MNRAS.441.2440B} {441, 2440}

\bibitem[\protect\citeauthoryear{{Balogh}, {Navarro}  \& {Morris}}{{Balogh}
  et~al.}{2000}]{balogh2000}
{Balogh} M.~L.,  {Navarro} J.~F.,   {Morris} S.~L.,  2000, \mn@doi [\apj]
  {10.1086/309323}, \href
  {https://ui.adsabs.harvard.edu/abs/2000ApJ...540..113B} {540, 113}

\bibitem[\protect\citeauthoryear{{Baum}, {Hiltner}, {Johnson}  \&
  {Sandage}}{{Baum} et~al.}{1959}]{baum1959}
{Baum} W.~A.,  {Hiltner} W.~A.,  {Johnson} H.~L.,   {Sandage} A.~R.,  1959,
  \mn@doi [\apj] {10.1086/146766}, \href
  {https://ui.adsabs.harvard.edu/abs/1959ApJ...130..749B} {130, 749}

\bibitem[\protect\citeauthoryear{{Bell}, {McIntosh}, {Katz}  \&
  {Weinberg}}{{Bell} et~al.}{2003}]{bell2003}
{Bell} E.~F.,  {McIntosh} D.~H.,  {Katz} N.,   {Weinberg} M.~D.,  2003, \mn@doi
  [\apjs] {10.1086/378847}, \href
  {https://ui.adsabs.harvard.edu/abs/2003ApJS..149..289B} {149, 289}

\bibitem[\protect\citeauthoryear{{Bell} et~al.,}{{Bell}
  et~al.}{2004}]{bell2004}
{Bell} E.~F.,  et~al., 2004, \mn@doi [\apj] {10.1086/420778}, \href
  {https://ui.adsabs.harvard.edu/abs/2004ApJ...608..752B} {608, 752}

\bibitem[\protect\citeauthoryear{{Belli}, {Newman}  \& {Ellis}}{{Belli}
  et~al.}{2015}]{belli2015}
{Belli} S.,  {Newman} A.~B.,   {Ellis} R.~S.,  2015, \mn@doi [\apj]
  {10.1088/0004-637X/799/2/206}, \href
  {https://ui.adsabs.harvard.edu/abs/2015ApJ...799..206B} {799, 206}

\bibitem[\protect\citeauthoryear{{Belli}, {Newman}  \& {Ellis}}{{Belli}
  et~al.}{2019}]{belli2019}
{Belli} S.,  {Newman} A.~B.,   {Ellis} R.~S.,  2019, \mn@doi [\apj]
  {10.3847/1538-4357/ab07af}, \href
  {https://ui.adsabs.harvard.edu/abs/2019ApJ...874...17B} {874, 17}

\bibitem[\protect\citeauthoryear{{Belli} et~al.,}{{Belli}
  et~al.}{2021}]{belli2021}
{Belli} S.,  et~al., 2021, \mn@doi [\apjl] {10.3847/2041-8213/abe6a6}, \href
  {https://ui.adsabs.harvard.edu/abs/2021ApJ...909L..11B} {909, L11}

\bibitem[\protect\citeauthoryear{{Bellstedt} et~al.,}{{Bellstedt}
  et~al.}{2020a}]{bellstedt2020a}
{Bellstedt} S.,  et~al., 2020a, \mn@doi [\mnras] {10.1093/mnras/staa1466},
  \href {https://ui.adsabs.harvard.edu/abs/2020MNRAS.496.3235B} {496, 3235}

\bibitem[\protect\citeauthoryear{{Bellstedt} et~al.,}{{Bellstedt}
  et~al.}{2020b}]{bellstedt2020b}
{Bellstedt} S.,  et~al., 2020b, \mn@doi [\mnras] {10.1093/mnras/staa2620},
  \href {https://ui.adsabs.harvard.edu/abs/2020MNRAS.498.5581B} {498, 5581}

\bibitem[\protect\citeauthoryear{{Bellstedt} et~al.,}{{Bellstedt}
  et~al.}{2021}]{bellstedt2021}
{Bellstedt} S.,  et~al., 2021, \mn@doi [\mnras] {10.1093/mnras/stab550}, \href
  {https://ui.adsabs.harvard.edu/abs/2021MNRAS.503.3309B} {503, 3309}

\bibitem[\protect\citeauthoryear{Benaglia, Chauveau, Hunter  \& Young}{Benaglia
  et~al.}{2009}]{mixtools}
Benaglia T.,  Chauveau D.,  Hunter D.~R.,   Young D.,  2009, Journal of
  Statistical Software, 32, 1

\bibitem[\protect\citeauthoryear{{Blanton}}{{Blanton}}{2006}]{blanton2006}
{Blanton} M.~R.,  2006, \mn@doi [\apj] {10.1086/505628}, \href
  {https://ui.adsabs.harvard.edu/abs/2006ApJ...648..268B} {648, 268}

\bibitem[\protect\citeauthoryear{{Blanton} et~al.,}{{Blanton}
  et~al.}{2003}]{blanton2003}
{Blanton} M.~R.,  et~al., 2003, \mn@doi [\apj] {10.1086/375528}, \href
  {https://ui.adsabs.harvard.edu/abs/2003ApJ...594..186B} {594, 186}

\bibitem[\protect\citeauthoryear{{Boquien}, {Burgarella}, {Roehlly}, {Buat},
  {Ciesla}, {Corre}, {Inoue}  \& {Salas}}{{Boquien} et~al.}{2019}]{boquien2019}
{Boquien} M.,  {Burgarella} D.,  {Roehlly} Y.,  {Buat} V.,  {Ciesla} L.,
  {Corre} D.,  {Inoue} A.~K.,   {Salas} H.,  2019, \mn@doi [\aap]
  {10.1051/0004-6361/201834156}, \href
  {https://ui.adsabs.harvard.edu/abs/2019A&A...622A.103B} {622, A103}

\bibitem[\protect\citeauthoryear{{Borch} et~al.,}{{Borch}
  et~al.}{2006}]{borch2006}
{Borch} A.,  et~al., 2006, \mn@doi [\aap] {10.1051/0004-6361:20054376}, \href
  {https://ui.adsabs.harvard.edu/abs/2006A&A...453..869B} {453, 869}

\bibitem[\protect\citeauthoryear{{Bower}, {Lucey}  \& {Ellis}}{{Bower}
  et~al.}{1992}]{bower1992}
{Bower} R.~G.,  {Lucey} J.~R.,   {Ellis} R.~S.,  1992, \mn@doi [\mnras]
  {10.1093/mnras/254.4.601}, \href
  {https://ui.adsabs.harvard.edu/abs/1992MNRAS.254..601B} {254, 601}

\bibitem[\protect\citeauthoryear{{Bower}, {Benson}, {Malbon}, {Helly}, {Frenk},
  {Baugh}, {Cole}  \& {Lacey}}{{Bower} et~al.}{2006}]{bower2006}
{Bower} R.~G.,  {Benson} A.~J.,  {Malbon} R.,  {Helly} J.~C.,  {Frenk} C.~S.,
  {Baugh} C.~M.,  {Cole} S.,   {Lacey} C.~G.,  2006, \mn@doi [\mnras]
  {10.1111/j.1365-2966.2006.10519.x}, \href
  {http://adsabs.harvard.edu/abs/2006MNRAS.370..645B} {370, 645}

\bibitem[\protect\citeauthoryear{{Bravo}, {Lagos}, {Robotham}, {Bellstedt}  \&
  {Obreschkow}}{{Bravo} et~al.}{2020}]{bravo2020}
{Bravo} M.,  {Lagos} C. d.~P.,  {Robotham} A. S.~G.,  {Bellstedt} S.,
  {Obreschkow} D.,  2020, \mn@doi [\mnras] {10.1093/mnras/staa2027}, \href
  {https://ui.adsabs.harvard.edu/abs/2020MNRAS.497.3026B} {497, 3026}

\bibitem[\protect\citeauthoryear{{Bremer} et~al.,}{{Bremer}
  et~al.}{2018}]{bremer2018}
{Bremer} M.~N.,  et~al., 2018, \mn@doi [\mnras] {10.1093/mnras/sty124}, \href
  {https://ui.adsabs.harvard.edu/abs/2018MNRAS.476...12B} {476, 12}

\bibitem[\protect\citeauthoryear{{Brown}, {Martini}  \& {Andrews}}{{Brown}
  et~al.}{2016}]{brown2016}
{Brown} J.~S.,  {Martini} P.,   {Andrews} B.~H.,  2016, \mn@doi [\mnras]
  {10.1093/mnras/stw392}, \href
  {https://ui.adsabs.harvard.edu/abs/2016MNRAS.458.1529B} {458, 1529}

\bibitem[\protect\citeauthoryear{{Brown}, {Cortese}, {Catinella}  \&
  {Kilborn}}{{Brown} et~al.}{2018}]{brown2018}
{Brown} T.,  {Cortese} L.,  {Catinella} B.,   {Kilborn} V.,  2018, \mn@doi
  [\mnras] {10.1093/mnras/stx2452}, \href
  {https://ui.adsabs.harvard.edu/abs/2018MNRAS.473.1868B} {473, 1868}

\bibitem[\protect\citeauthoryear{{Bruzual} \& {Charlot}}{{Bruzual} \&
  {Charlot}}{2003}]{bruzual2003}
{Bruzual} G.,  {Charlot} S.,  2003, \mn@doi [\mnras]
  {10.1046/j.1365-8711.2003.06897.x}, \href
  {https://ui.adsabs.harvard.edu/abs/2003MNRAS.344.1000B} {344, 1000}

\bibitem[\protect\citeauthoryear{{Ca{\~n}as}, {Elahi}, {Welker}, {del P Lagos},
  {Power}, {Dubois}  \& {Pichon}}{{Ca{\~n}as} et~al.}{2019}]{canas2019}
{Ca{\~n}as} R.,  {Elahi} P.~J.,  {Welker} C.,  {del P Lagos} C.,  {Power} C.,
  {Dubois} Y.,   {Pichon} C.,  2019, \mn@doi [\mnras] {10.1093/mnras/sty2725},
  \href {https://ui.adsabs.harvard.edu/abs/2019MNRAS.482.2039C} {482, 2039}

\bibitem[\protect\citeauthoryear{{Carnall}, {Leja}, {Johnson}, {McLure},
  {Dunlop}  \& {Conroy}}{{Carnall} et~al.}{2019}]{carnall2019}
{Carnall} A.~C.,  {Leja} J.,  {Johnson} B.~D.,  {McLure} R.~J.,  {Dunlop}
  J.~S.,   {Conroy} C.,  2019, \mn@doi [\apj] {10.3847/1538-4357/ab04a2}, \href
  {https://ui.adsabs.harvard.edu/abs/2019ApJ...873...44C} {873, 44}

\bibitem[\protect\citeauthoryear{{Chabrier}}{{Chabrier}}{2003}]{chabrier2003}
{Chabrier} G.,  2003, \mn@doi [\pasp] {10.1086/376392}, \href
  {https://ui.adsabs.harvard.edu/abs/2003PASP..115..763C} {115, 763}

\bibitem[\protect\citeauthoryear{{Charlot} \& {Fall}}{{Charlot} \&
  {Fall}}{2000}]{charlot2000}
{Charlot} S.,  {Fall} S.~M.,  2000, \mn@doi [\apj] {10.1086/309250}, \href
  {https://ui.adsabs.harvard.edu/abs/2000ApJ...539..718C} {539, 718}

\bibitem[\protect\citeauthoryear{{Chauhan}, {Lagos}, {Obreschkow}, {Power},
  {Oman}  \& {Elahi}}{{Chauhan} et~al.}{2019}]{chauhan2019}
{Chauhan} G.,  {Lagos} C. d.~P.,  {Obreschkow} D.,  {Power} C.,  {Oman} K.,
  {Elahi} P.~J.,  2019, \mn@doi [\mnras] {10.1093/mnras/stz2069}, \href
  {https://ui.adsabs.harvard.edu/abs/2019MNRAS.488.5898C} {488, 5898}

\bibitem[\protect\citeauthoryear{{Chauhan}, {Lagos}, {Stevens}, {Obreschkow},
  {Power}  \& {Meyer}}{{Chauhan} et~al.}{2020}]{chauhan2020}
{Chauhan} G.,  {Lagos} C. d.~P.,  {Stevens} A. R.~H.,  {Obreschkow} D.,
  {Power} C.,   {Meyer} M.,  2020, \mn@doi [\mnras] {10.1093/mnras/staa2251},
  \href {https://ui.adsabs.harvard.edu/abs/2020MNRAS.498...44C} {498, 44}

\bibitem[\protect\citeauthoryear{{Chauhan}, {Lagos}, {Stevens}, {Bravo},
  {Rhee}, {Power}, {Obreschkow}  \& {Meyer}}{{Chauhan}
  et~al.}{2021}]{chauhan2021}
{Chauhan} G.,  {Lagos} C. d.~P.,  {Stevens} A. R.~H.,  {Bravo} M.,  {Rhee} J.,
  {Power} C.,  {Obreschkow} D.,   {Meyer} M.,  2021, arXiv e-prints, \href
  {https://ui.adsabs.harvard.edu/abs/2021arXiv210212203C} {p. arXiv:2102.12203}

\bibitem[\protect\citeauthoryear{{Chevallard} \& {Charlot}}{{Chevallard} \&
  {Charlot}}{2016}]{chevallard2016}
{Chevallard} J.,  {Charlot} S.,  2016, \mn@doi [\mnras]
  {10.1093/mnras/stw1756}, \href
  {https://ui.adsabs.harvard.edu/abs/2016MNRAS.462.1415C} {462, 1415}

\bibitem[\protect\citeauthoryear{{Conroy}}{{Conroy}}{2013}]{conroy2013}
{Conroy} C.,  2013, \mn@doi [\araa] {10.1146/annurev-astro-082812-141017},
  \href {https://ui.adsabs.harvard.edu/abs/2013ARA&A..51..393C} {51, 393}

\bibitem[\protect\citeauthoryear{{Cowie}, {Songaila}, {Hu}  \& {Cohen}}{{Cowie}
  et~al.}{1996}]{cowie1996}
{Cowie} L.~L.,  {Songaila} A.,  {Hu} E.~M.,   {Cohen} J.~G.,  1996, \mn@doi
  [\aj] {10.1086/118058}, \href
  {https://ui.adsabs.harvard.edu/abs/1996AJ....112..839C} {112, 839}

\bibitem[\protect\citeauthoryear{{Croton} et~al.,}{{Croton}
  et~al.}{2006}]{croton2006}
{Croton} D.~J.,  et~al., 2006, \mn@doi [\mnras]
  {10.1111/j.1365-2966.2005.09675.x}, \href
  {https://ui.adsabs.harvard.edu/abs/2006MNRAS.365...11C} {365, 11}

\bibitem[\protect\citeauthoryear{{Crowl}, {Kenney}, {van Gorkom}  \&
  {Vollmer}}{{Crowl} et~al.}{2005}]{crowl2005}
{Crowl} H.~H.,  {Kenney} J. D.~P.,  {van Gorkom} J.~H.,   {Vollmer} B.,  2005,
  \mn@doi [\aj] {10.1086/430526}, \href
  {https://ui.adsabs.harvard.edu/abs/2005AJ....130...65C} {130, 65}

\bibitem[\protect\citeauthoryear{{Dale}, {Helou}, {Magdis}, {Armus},
  {D{\'\i}az-Santos}  \& {Shi}}{{Dale} et~al.}{2014}]{dale2014}
{Dale} D.~A.,  {Helou} G.,  {Magdis} G.~E.,  {Armus} L.,  {D{\'\i}az-Santos}
  T.,   {Shi} Y.,  2014, \mn@doi [\apj] {10.1088/0004-637X/784/1/83}, \href
  {https://ui.adsabs.harvard.edu/abs/2014ApJ...784...83D} {784, 83}

\bibitem[\protect\citeauthoryear{{Dalla Vecchia} \& {Schaye}}{{Dalla Vecchia}
  \& {Schaye}}{2012}]{dallavecchia2012}
{Dalla Vecchia} C.,  {Schaye} J.,  2012, \mn@doi [\mnras]
  {10.1111/j.1365-2966.2012.21704.x}, \href
  {https://ui.adsabs.harvard.edu/abs/2012MNRAS.426..140D} {426, 140}

\bibitem[\protect\citeauthoryear{{Davies} et~al.,}{{Davies}
  et~al.}{2018}]{davies2018}
{Davies} L.~J.~M.,  et~al., 2018, \mn@doi [\mnras] {10.1093/mnras/sty1553},
  \href {https://ui.adsabs.harvard.edu/#abs/2018MNRAS.480..768D} {480, 768}

\bibitem[\protect\citeauthoryear{{Davies} et~al.,}{{Davies}
  et~al.}{2019a}]{davies2019a}
{Davies} L.~J.~M.,  et~al., 2019a, \mn@doi [\mnras] {10.1093/mnras/sty2957},
  \href {https://ui.adsabs.harvard.edu/\#abs/2019MNRAS.483.1881D} {483, 1881}

\bibitem[\protect\citeauthoryear{{Davies} et~al.,}{{Davies}
  et~al.}{2019b}]{davies2019b}
{Davies} L.~J.~M.,  et~al., 2019b, \mn@doi [\mnras] {10.1093/mnras/sty3393},
  \href {https://ui.adsabs.harvard.edu/\#abs/2019MNRAS.483.5444D} {483, 5444}

\bibitem[\protect\citeauthoryear{{Davies} et~al.,}{{Davies}
  et~al.}{2022}]{davies2022}
{Davies} L.~J.~M.,  et~al., 2022, \mn@doi [\mnras] {10.1093/mnras/stab3145},
  \href {https://ui.adsabs.harvard.edu/abs/2022MNRAS.509.4392D} {509, 4392}

\bibitem[\protect\citeauthoryear{{Dekel} \& {Birnboim}}{{Dekel} \&
  {Birnboim}}{2006}]{dekel2006}
{Dekel} A.,  {Birnboim} Y.,  2006, \mn@doi [\mnras]
  {10.1111/j.1365-2966.2006.10145.x}, \href
  {https://ui.adsabs.harvard.edu/abs/2006MNRAS.368....2D} {368, 2}

\bibitem[\protect\citeauthoryear{{Driver} et~al.,}{{Driver}
  et~al.}{2006}]{driver2006}
{Driver} S.~P.,  et~al., 2006, \mn@doi [\mnras]
  {10.1111/j.1365-2966.2006.10126.x}, \href
  {https://ui.adsabs.harvard.edu/abs/2006MNRAS.368..414D} {368, 414}

\bibitem[\protect\citeauthoryear{{Driver} et~al.,}{{Driver}
  et~al.}{2011}]{driver2011}
{Driver} S.~P.,  et~al., 2011, \mn@doi [\mnras]
  {10.1111/j.1365-2966.2010.18188.x}, \href
  {https://ui.adsabs.harvard.edu/#abs/2011MNRAS.413..971D} {413, 971}

\bibitem[\protect\citeauthoryear{{Elahi}, {Welker}, {Power}, {Lagos},
  {Robotham}, {Ca{\~n}as}  \& {Poulton}}{{Elahi} et~al.}{2018}]{elahi2018a}
{Elahi} P.~J.,  {Welker} C.,  {Power} C.,  {Lagos} C. d.~P.,  {Robotham} A.
  S.~G.,  {Ca{\~n}as} R.,   {Poulton} R.,  2018, \mn@doi [\mnras]
  {10.1093/mnras/sty061}, \href
  {https://ui.adsabs.harvard.edu/#abs/2018MNRAS.475.5338E} {475, 5338}

\bibitem[\protect\citeauthoryear{{Elahi}, {Ca{\~n}as}, {Poulton}, {Tobar},
  {Willis}, {Lagos}, {Power}  \& {Robotham}}{{Elahi}
  et~al.}{2019a}]{elahi2019a}
{Elahi} P.~J.,  {Ca{\~n}as} R.,  {Poulton} R. J.~J.,  {Tobar} R.~J.,  {Willis}
  J.~S.,  {Lagos} C. d.~P.,  {Power} C.,   {Robotham} A. S.~G.,  2019a, \mn@doi
  [\pasa] {10.1017/pasa.2019.12}, \href
  {https://ui.adsabs.harvard.edu/abs/2019PASA...36...21E} {36, e021}

\bibitem[\protect\citeauthoryear{{Elahi}, {Poulton}, {Tobar}, {Ca{\~n}as},
  {Lagos}, {Power}  \& {Robotham}}{{Elahi} et~al.}{2019b}]{elahi2019b}
{Elahi} P.~J.,  {Poulton} R. J.~J.,  {Tobar} R.~J.,  {Ca{\~n}as} R.,  {Lagos}
  C. d.~P.,  {Power} C.,   {Robotham} A. S.~G.,  2019b, \mn@doi [\pasa]
  {10.1017/pasa.2019.18}, \href
  {https://ui.adsabs.harvard.edu/abs/2019PASA...36...28E} {36, e028}

\bibitem[\protect\citeauthoryear{{Faber}}{{Faber}}{1973}]{faber1973}
{Faber} S.~M.,  1973, \mn@doi [\apj] {10.1086/151912}, \href
  {https://ui.adsabs.harvard.edu/abs/1973ApJ...179..731F} {179, 731}

\bibitem[\protect\citeauthoryear{{Faber} et~al.,}{{Faber}
  et~al.}{2007}]{faber2007}
{Faber} S.~M.,  et~al., 2007, \mn@doi [\apj] {10.1086/519294}, \href
  {https://ui.adsabs.harvard.edu/abs/2007ApJ...665..265F} {665, 265}

\bibitem[\protect\citeauthoryear{{Fang}, {Faber}, {Koo}  \& {Dekel}}{{Fang}
  et~al.}{2013}]{fang2013}
{Fang} J.~J.,  {Faber} S.~M.,  {Koo} D.~C.,   {Dekel} A.,  2013, \mn@doi [\apj]
  {10.1088/0004-637X/776/1/63}, \href
  {https://ui.adsabs.harvard.edu/abs/2013ApJ...776...63F} {776, 63}

\bibitem[\protect\citeauthoryear{{Feltre}, {Hatziminaoglou}, {Fritz}  \&
  {Franceschini}}{{Feltre} et~al.}{2012}]{feltre2012}
{Feltre} A.,  {Hatziminaoglou} E.,  {Fritz} J.,   {Franceschini} A.,  2012,
  \mn@doi [\mnras] {10.1111/j.1365-2966.2012.21695.x}, \href
  {https://ui.adsabs.harvard.edu/abs/2012MNRAS.426..120F} {426, 120}

\bibitem[\protect\citeauthoryear{{Fontanot}, {De Lucia}, {Monaco}, {Somerville}
   \& {Santini}}{{Fontanot} et~al.}{2009}]{fontanot2009}
{Fontanot} F.,  {De Lucia} G.,  {Monaco} P.,  {Somerville} R.~S.,   {Santini}
  P.,  2009, \mn@doi [\mnras] {10.1111/j.1365-2966.2009.15058.x}, \href
  {https://ui.adsabs.harvard.edu/abs/2009MNRAS.397.1776F} {397, 1776}

\bibitem[\protect\citeauthoryear{{Fritz}, {Franceschini}  \&
  {Hatziminaoglou}}{{Fritz} et~al.}{2006}]{fritz2006}
{Fritz} J.,  {Franceschini} A.,   {Hatziminaoglou} E.,  2006, \mn@doi [\mnras]
  {10.1111/j.1365-2966.2006.09866.x}, \href
  {https://ui.adsabs.harvard.edu/abs/2006MNRAS.366..767F} {366, 767}

\bibitem[\protect\citeauthoryear{{Gon{\c{c}}alves}, {Martin},
  {Men{\'e}ndez-Delmestre}, {Wyder}  \& {Koekemoer}}{{Gon{\c{c}}alves}
  et~al.}{2012}]{goncalves2012}
{Gon{\c{c}}alves} T.~S.,  {Martin} D.~C.,  {Men{\'e}ndez-Delmestre} K.,
  {Wyder} T.~K.,   {Koekemoer} A.,  2012, \mn@doi [\apj]
  {10.1088/0004-637X/759/1/67}, \href
  {https://ui.adsabs.harvard.edu/abs/2012ApJ...759...67G} {759, 67}

\bibitem[\protect\citeauthoryear{{Hahn}, {Tinker}  \& {Wetzel}}{{Hahn}
  et~al.}{2017}]{hahn2017}
{Hahn} C.,  {Tinker} J.~L.,   {Wetzel} A.,  2017, \mn@doi [\apj]
  {10.3847/1538-4357/aa6d6b}, \href
  {https://ui.adsabs.harvard.edu/abs/2017ApJ...841....6H} {841, 6}

\bibitem[\protect\citeauthoryear{Harris et~al.,}{Harris et~al.}{2020}]{numpy}
Harris C.~R.,  et~al., 2020, \nat, 585, 357

\bibitem[\protect\citeauthoryear{{Hogg} et~al.,}{{Hogg}
  et~al.}{2002}]{hogg2002}
{Hogg} D.~W.,  et~al., 2002, \mn@doi [\aj] {10.1086/341392}, \href
  {https://ui.adsabs.harvard.edu/abs/2002AJ....124..646H} {124, 646}

\bibitem[\protect\citeauthoryear{{Hopkins}, {Hernquist}, {Cox}, {Di Matteo},
  {Robertson}  \& {Springel}}{{Hopkins} et~al.}{2006}]{hopkins2006}
{Hopkins} P.~F.,  {Hernquist} L.,  {Cox} T.~J.,  {Di Matteo} T.,  {Robertson}
  B.,   {Springel} V.,  2006, \mn@doi [\apjs] {10.1086/499298}, \href
  {https://ui.adsabs.harvard.edu/abs/2006ApJS..163....1H} {163, 1}

\bibitem[\protect\citeauthoryear{Hunter}{Hunter}{2007}]{matplotlib}
Hunter J.~D.,  2007, \mn@doi [Computing In Science \& Engineering]
  {10.1109/MCSE.2007.55}, 9, 90

\bibitem[\protect\citeauthoryear{{Iyer} \& {Gawiser}}{{Iyer} \&
  {Gawiser}}{2017}]{iyer2017}
{Iyer} K.,  {Gawiser} E.,  2017, \mn@doi [\apj] {10.3847/1538-4357/aa63f0},
  \href {https://ui.adsabs.harvard.edu/abs/2017ApJ...838..127I} {838, 127}

\bibitem[\protect\citeauthoryear{{Johnson}, {Leja}, {Conroy}  \&
  {Speagle}}{{Johnson} et~al.}{2021}]{johnson2021}
{Johnson} B.~D.,  {Leja} J.,  {Conroy} C.,   {Speagle} J.~S.,  2021, \mn@doi
  [\apjs] {10.3847/1538-4365/abef67}, \href
  {https://ui.adsabs.harvard.edu/abs/2021ApJS..254...22J} {254, 22}

\bibitem[\protect\citeauthoryear{{Katsianis} et~al.,}{{Katsianis}
  et~al.}{2019}]{katsianis2019}
{Katsianis} A.,  et~al., 2019, \mn@doi [\apj] {10.3847/1538-4357/ab1f8d}, \href
  {https://ui.adsabs.harvard.edu/abs/2019ApJ...879...11K} {879, 11}

\bibitem[\protect\citeauthoryear{{Kauffmann}, {White}, {Heckman}, {M{\'e}nard},
  {Brinchmann}, {Charlot}, {Tremonti}  \& {Brinkmann}}{{Kauffmann}
  et~al.}{2004}]{kauffmann2004}
{Kauffmann} G.,  {White} S. D.~M.,  {Heckman} T.~M.,  {M{\'e}nard} B.,
  {Brinchmann} J.,  {Charlot} S.,  {Tremonti} C.,   {Brinkmann} J.,  2004,
  \mn@doi [\mnras] {10.1111/j.1365-2966.2004.08117.x}, \href
  {https://ui.adsabs.harvard.edu/abs/2004MNRAS.353..713K} {353, 713}

\bibitem[\protect\citeauthoryear{{Kaviraj}, {Devriendt}, {Ferreras}, {Yi}  \&
  {Silk}}{{Kaviraj} et~al.}{2009}]{kaviraj2009}
{Kaviraj} S.,  {Devriendt} J.~E.~G.,  {Ferreras} I.,  {Yi} S.~K.,   {Silk} J.,
  2009, \mn@doi [\aap] {10.1051/0004-6361/200810483}, \href
  {https://ui.adsabs.harvard.edu/abs/2009A&A...503..445K} {503, 445}

\bibitem[\protect\citeauthoryear{{Kawata} \& {Mulchaey}}{{Kawata} \&
  {Mulchaey}}{2008}]{kawata2008}
{Kawata} D.,  {Mulchaey} J.~S.,  2008, \mn@doi [\apjl] {10.1086/526544}, \href
  {https://ui.adsabs.harvard.edu/abs/2008ApJ...672L.103K} {672, L103}

\bibitem[\protect\citeauthoryear{{Kere{\v{s}}}, {Katz}, {Weinberg}  \&
  {Dav{\'e}}}{{Kere{\v{s}}} et~al.}{2005}]{keres2005}
{Kere{\v{s}}} D.,  {Katz} N.,  {Weinberg} D.~H.,   {Dav{\'e}} R.,  2005,
  \mn@doi [\mnras] {10.1111/j.1365-2966.2005.09451.x}, \href
  {https://ui.adsabs.harvard.edu/abs/2005MNRAS.363....2K} {363, 2}

\bibitem[\protect\citeauthoryear{{Kodama} et~al.,}{{Kodama}
  et~al.}{2004}]{kodama2004}
{Kodama} T.,  et~al., 2004, \mn@doi [\mnras]
  {10.1111/j.1365-2966.2004.07711.x}, \href
  {https://ui.adsabs.harvard.edu/abs/2004MNRAS.350.1005K} {350, 1005}

\bibitem[\protect\citeauthoryear{{Lagos}, {Lacey}  \& {Baugh}}{{Lagos}
  et~al.}{2013}]{lagos2013}
{Lagos} C. d.~P.,  {Lacey} C.~G.,   {Baugh} C.~M.,  2013, \mn@doi [\mnras]
  {10.1093/mnras/stt1696}, \href
  {https://ui.adsabs.harvard.edu/abs/2013MNRAS.436.1787L} {436, 1787}

\bibitem[\protect\citeauthoryear{{Lagos}, {Tobar}, {Robotham}, {Obreschkow},
  {Mitchell}, {Power}  \& {Elahi}}{{Lagos} et~al.}{2018}]{lagos2018}
{Lagos} C. d.~P.,  {Tobar} R.~J.,  {Robotham} A. S.~G.,  {Obreschkow} D.,
  {Mitchell} P.~D.,  {Power} C.,   {Elahi} P.~J.,  2018, \mn@doi [\mnras]
  {10.1093/mnras/sty2440}, \href
  {https://ui.adsabs.harvard.edu/abs/2018MNRAS.481.3573L} {481, 3573}

\bibitem[\protect\citeauthoryear{{Lagos} et~al.,}{{Lagos}
  et~al.}{2019}]{lagos2019}
{Lagos} C. d.~P.,  et~al., 2019, \mn@doi [\mnras] {10.1093/mnras/stz2427},
  \href {https://ui.adsabs.harvard.edu/abs/2019MNRAS.489.4196L} {489, 4196}

\bibitem[\protect\citeauthoryear{{Lagos}, {da Cunha}, {Robotham}, {Obreschkow},
  {Valentino}, {Fujimoto}, {Magdis}  \& {Tobar}}{{Lagos}
  et~al.}{2020}]{lagos2020}
{Lagos} C. d.~P.,  {da Cunha} E.,  {Robotham} A. S.~G.,  {Obreschkow} D.,
  {Valentino} F.,  {Fujimoto} S.,  {Magdis} G.~E.,   {Tobar} R.,  2020, arXiv
  e-prints, \href {https://ui.adsabs.harvard.edu/abs/2020arXiv200709853L} {p.
  arXiv:2007.09853}

\bibitem[\protect\citeauthoryear{{Lara-L{\'o}pez} et~al.,}{{Lara-L{\'o}pez}
  et~al.}{2010}]{lara-lopez2010}
{Lara-L{\'o}pez} M.~A.,  et~al., 2010, \mn@doi [\aap]
  {10.1051/0004-6361/201014803}, \href
  {https://ui.adsabs.harvard.edu/abs/2010A&A...521L..53L} {521, L53}

\bibitem[\protect\citeauthoryear{{Lara-L{\'o}pez}, {L{\'o}pez-S{\'a}nchez}  \&
  {Hopkins}}{{Lara-L{\'o}pez} et~al.}{2013}]{lara-lopez2013}
{Lara-L{\'o}pez} M.~A.,  {L{\'o}pez-S{\'a}nchez} {\'A}.~R.,   {Hopkins} A.~M.,
  2013, \mn@doi [\apj] {10.1088/0004-637X/764/2/178}, \href
  {https://ui.adsabs.harvard.edu/abs/2013ApJ...764..178L} {764, 178}

\bibitem[\protect\citeauthoryear{{Leja}, {Carnall}, {Johnson}, {Conroy}  \&
  {Speagle}}{{Leja} et~al.}{2019a}]{leja2019a}
{Leja} J.,  {Carnall} A.~C.,  {Johnson} B.~D.,  {Conroy} C.,   {Speagle} J.~S.,
   2019a, \mn@doi [\apj] {10.3847/1538-4357/ab133c}, \href
  {https://ui.adsabs.harvard.edu/abs/2019ApJ...876....3L} {876, 3}

\bibitem[\protect\citeauthoryear{{Leja} et~al.,}{{Leja}
  et~al.}{2019b}]{leja2019b}
{Leja} J.,  et~al., 2019b, \mn@doi [\apj] {10.3847/1538-4357/ab1d5a}, \href
  {https://ui.adsabs.harvard.edu/abs/2019ApJ...877..140L} {877, 140}

\bibitem[\protect\citeauthoryear{{Lewis} et~al.,}{{Lewis}
  et~al.}{2002}]{lewis2002}
{Lewis} I.~J.,  et~al., 2002, \mn@doi [\mnras]
  {10.1046/j.1365-8711.2002.05333.x}, \href
  {https://ui.adsabs.harvard.edu/#abs/2002MNRAS.333..279L} {333, 279}

\bibitem[\protect\citeauthoryear{{Liske} et~al.,}{{Liske}
  et~al.}{2015}]{liske2015}
{Liske} J.,  et~al., 2015, \mn@doi [\mnras] {10.1093/mnras/stv1436}, \href
  {https://ui.adsabs.harvard.edu/abs/2015MNRAS.452.2087L} {452, 2087}

\bibitem[\protect\citeauthoryear{{Lower}, {Narayanan}, {Leja}, {Johnson},
  {Conroy}  \& {Dav{\'e}}}{{Lower} et~al.}{2020}]{lower2020}
{Lower} S.,  {Narayanan} D.,  {Leja} J.,  {Johnson} B.~D.,  {Conroy} C.,
  {Dav{\'e}} R.,  2020, \mn@doi [\apj] {10.3847/1538-4357/abbfa7}, \href
  {https://ui.adsabs.harvard.edu/abs/2020ApJ...904...33L} {904, 33}

\bibitem[\protect\citeauthoryear{{Machacek}, {Jones}, {Forman}  \&
  {Nulsen}}{{Machacek} et~al.}{2006}]{machacek2006}
{Machacek} M.,  {Jones} C.,  {Forman} W.~R.,   {Nulsen} P.,  2006, \mn@doi
  [\apj] {10.1086/503350}, \href
  {https://ui.adsabs.harvard.edu/abs/2006ApJ...644..155M} {644, 155}

\bibitem[\protect\citeauthoryear{{Martin} et~al.,}{{Martin}
  et~al.}{2005}]{martin2005}
{Martin} D.~C.,  et~al., 2005, \mn@doi [\apjl] {10.1086/426387}, \href
  {https://ui.adsabs.harvard.edu/abs/2005ApJ...619L...1M} {619, L1}

\bibitem[\protect\citeauthoryear{{Martin} et~al.,}{{Martin}
  et~al.}{2007}]{martin2007}
{Martin} D.~C.,  et~al., 2007, \mn@doi [\apjs] {10.1086/516639}, \href
  {https://ui.adsabs.harvard.edu/abs/2007ApJS..173..342M} {173, 342}

\bibitem[\protect\citeauthoryear{{McCarthy}, {Frenk}, {Font}, {Lacey}, {Bower},
  {Mitchell}, {Balogh}  \& {Theuns}}{{McCarthy} et~al.}{2008}]{mccarthy2008}
{McCarthy} I.~G.,  {Frenk} C.~S.,  {Font} A.~S.,  {Lacey} C.~G.,  {Bower}
  R.~G.,  {Mitchell} N.~L.,  {Balogh} M.~L.,   {Theuns} T.,  2008, \mn@doi
  [\mnras] {10.1111/j.1365-2966.2007.12577.x}, \href
  {https://ui.adsabs.harvard.edu/abs/2008MNRAS.383..593M} {383, 593}

\bibitem[\protect\citeauthoryear{{Moustakas} et~al.,}{{Moustakas}
  et~al.}{2013}]{moustakas2013}
{Moustakas} J.,  et~al., 2013, \mn@doi [\apj] {10.1088/0004-637X/767/1/50},
  \href {https://ui.adsabs.harvard.edu/abs/2013ApJ...767...50M} {767, 50}

\bibitem[\protect\citeauthoryear{{Neistein}, {van den Bosch}  \&
  {Dekel}}{{Neistein} et~al.}{2006}]{neistein2006}
{Neistein} E.,  {van den Bosch} F.~C.,   {Dekel} A.,  2006, \mn@doi [\mnras]
  {10.1111/j.1365-2966.2006.10918.x}, \href
  {https://ui.adsabs.harvard.edu/abs/2006MNRAS.372..933N} {372, 933}

\bibitem[\protect\citeauthoryear{{Nelson} et~al.,}{{Nelson}
  et~al.}{2018}]{nelson2018a}
{Nelson} D.,  et~al., 2018, \mn@doi [\mnras] {10.1093/mnras/stx3040}, \href
  {https://ui.adsabs.harvard.edu/abs/2018MNRAS.475..624N} {475, 624}

\bibitem[\protect\citeauthoryear{{Noll}, {Burgarella}, {Giovannoli}, {Buat},
  {Marcillac}  \& {Mu{\~n}oz-Mateos}}{{Noll} et~al.}{2009}]{noll2009}
{Noll} S.,  {Burgarella} D.,  {Giovannoli} E.,  {Buat} V.,  {Marcillac} D.,
  {Mu{\~n}oz-Mateos} J.~C.,  2009, \mn@doi [\aap]
  {10.1051/0004-6361/200912497}, \href
  {https://ui.adsabs.harvard.edu/abs/2009A&A...507.1793N} {507, 1793}

\bibitem[\protect\citeauthoryear{{Pacifici}, {Charlot}, {Blaizot}  \&
  {Brinchmann}}{{Pacifici} et~al.}{2012}]{pacifici2012}
{Pacifici} C.,  {Charlot} S.,  {Blaizot} J.,   {Brinchmann} J.,  2012, \mn@doi
  [\mnras] {10.1111/j.1365-2966.2012.20431.x}, \href
  {https://ui.adsabs.harvard.edu/abs/2012MNRAS.421.2002P} {421, 2002}

\bibitem[\protect\citeauthoryear{{Pacifici}, {Oh}, {Oh}, {Lee}  \&
  {Yi}}{{Pacifici} et~al.}{2016a}]{pacifici2016a}
{Pacifici} C.,  {Oh} S.,  {Oh} K.,  {Lee} J.,   {Yi} S.~K.,  2016a, \mn@doi
  [\apj] {10.3847/0004-637X/824/1/45}, \href
  {https://ui.adsabs.harvard.edu/abs/2016ApJ...824...45P} {824, 45}

\bibitem[\protect\citeauthoryear{{Pacifici} et~al.,}{{Pacifici}
  et~al.}{2016b}]{pacifici2016b}
{Pacifici} C.,  et~al., 2016b, \mn@doi [\apj] {10.3847/0004-637X/832/1/79},
  \href {https://ui.adsabs.harvard.edu/abs/2016ApJ...832...79P} {832, 79}

\bibitem[\protect\citeauthoryear{{Peng} et~al.,}{{Peng}
  et~al.}{2010}]{peng2010}
{Peng} Y.-j.,  et~al., 2010, \mn@doi [\apj] {10.1088/0004-637X/721/1/193},
  \href {https://ui.adsabs.harvard.edu/#abs/2010ApJ...721..193P} {721, 193}

\bibitem[\protect\citeauthoryear{{Peng}, {Maiolino}  \& {Cochrane}}{{Peng}
  et~al.}{2015}]{peng2015}
{Peng} Y.,  {Maiolino} R.,   {Cochrane} R.,  2015, \mn@doi [\nat]
  {10.1038/nature14439}, \href
  {https://ui.adsabs.harvard.edu/abs/2015Natur.521..192P} {521, 192}

\bibitem[\protect\citeauthoryear{{Phillipps} et~al.,}{{Phillipps}
  et~al.}{2019}]{phillipps2019}
{Phillipps} S.,  et~al., 2019, \mn@doi [\mnras] {10.1093/mnras/stz799}, \href
  {https://ui.adsabs.harvard.edu/abs/2019MNRAS.485.5559P} {485, 5559}

\bibitem[\protect\citeauthoryear{{Pilbratt} et~al.,}{{Pilbratt}
  et~al.}{2010}]{pilbratt2010}
{Pilbratt} G.~L.,  et~al., 2010, \mn@doi [\aap] {10.1051/0004-6361/201014759},
  \href {https://ui.adsabs.harvard.edu/abs/2010A&A...518L...1P} {518, L1}

\bibitem[\protect\citeauthoryear{{Planck Collaboration} et~al.,}{{Planck
  Collaboration} et~al.}{2016}]{planck2016xiii}
{Planck Collaboration} et~al., 2016, \mn@doi [\aap]
  {10.1051/0004-6361/201525830}, \href
  {https://ui.adsabs.harvard.edu/abs/2016A&A...594A..13P} {594, A13}

\bibitem[\protect\citeauthoryear{{Poulton}, {Robotham}, {Power}  \&
  {Elahi}}{{Poulton} et~al.}{2018}]{poulton2018}
{Poulton} R. J.~J.,  {Robotham} A. S.~G.,  {Power} C.,   {Elahi} P.~J.,  2018,
  \mn@doi [\pasa] {10.1017/pasa.2018.34}, \href
  {https://ui.adsabs.harvard.edu/abs/2018PASA...35...42P} {35, 42}

\bibitem[\protect\citeauthoryear{{R{\'e}my-Ruyer} et~al.,}{{R{\'e}my-Ruyer}
  et~al.}{2014}]{remy-ruyer2014}
{R{\'e}my-Ruyer} A.,  et~al., 2014, \mn@doi [\aap]
  {10.1051/0004-6361/201322803}, \href
  {https://ui.adsabs.harvard.edu/abs/2014A&A...563A..31R} {563, A31}

\bibitem[\protect\citeauthoryear{{Robotham}}{{Robotham}}{2016}]{celestial}
{Robotham} A. S.~G.,  2016, {Celestial: Common astronomical conversion routines
  and functions} (\mn@eprint {ascl} {1602.011})

\bibitem[\protect\citeauthoryear{{Robotham}, {Davies}, {Driver}, {Koushan},
  {Taranu}, {Casura}  \& {Liske}}{{Robotham} et~al.}{2018}]{robotham2018}
{Robotham} A.~S.~G.,  {Davies} L.~J.~M.,  {Driver} S.~P.,  {Koushan} S.,
  {Taranu} D.~S.,  {Casura} S.,   {Liske} J.,  2018, \mn@doi [\mnras]
  {10.1093/mnras/sty440}, \href
  {https://ui.adsabs.harvard.edu/abs/2018MNRAS.476.3137R} {476, 3137}

\bibitem[\protect\citeauthoryear{{Robotham}, {Bellstedt}, {Lagos}, {Thorne},
  {Davies}, {Driver}  \& {Bravo}}{{Robotham} et~al.}{2020}]{robotham2020}
{Robotham} A.~S.~G.,  {Bellstedt} S.,  {Lagos} C. d.~P.,  {Thorne} J.~E.,
  {Davies} L.~J.,  {Driver} S.~P.,   {Bravo} M.,  2020, \mn@doi [\mnras]
  {10.1093/mnras/staa1116}, \href
  {https://ui.adsabs.harvard.edu/abs/2020MNRAS.495..905R} {495, 905}

\bibitem[\protect\citeauthoryear{{Saunders} et~al.,}{{Saunders}
  et~al.}{2004}]{saunders2004}
{Saunders} W.,  et~al., 2004, in {Moorwood} A. F.~M.,  {Iye} M.,  eds,  Society
  of Photo-Optical Instrumentation Engineers (SPIE) Conference Series Vol.
  5492, Ground-based Instrumentation for Astronomy. pp 389--400,
  \mn@doi{10.1117/12.550871}

\bibitem[\protect\citeauthoryear{{Schaller}, {Schaerer}, {Meynet}  \&
  {Maeder}}{{Schaller} et~al.}{1992}]{schaller1992}
{Schaller} G.,  {Schaerer} D.,  {Meynet} G.,   {Maeder} A.,  1992, \aaps, \href
  {https://ui.adsabs.harvard.edu/abs/1992A&AS...96..269S} {96, 269}

\bibitem[\protect\citeauthoryear{{Schawinski} et~al.,}{{Schawinski}
  et~al.}{2014}]{schawinski2014}
{Schawinski} K.,  et~al., 2014, \mn@doi [\mnras] {10.1093/mnras/stu327}, \href
  {https://ui.adsabs.harvard.edu/abs/2014MNRAS.440..889S} {440, 889}

\bibitem[\protect\citeauthoryear{{Schaye} et~al.,}{{Schaye}
  et~al.}{2015}]{schaye2015}
{Schaye} J.,  et~al., 2015, \mn@doi [\mnras] {10.1093/mnras/stu2058}, \href
  {http://adsabs.harvard.edu/abs/2015MNRAS.446..521S} {446, 521}

\bibitem[\protect\citeauthoryear{{Schechter}}{{Schechter}}{1976}]{schechter1976}
{Schechter} P.,  1976, \mn@doi [\apj] {10.1086/154079}, \href
  {https://ui.adsabs.harvard.edu/abs/1976ApJ...203..297S} {203, 297}

\bibitem[\protect\citeauthoryear{{Sharp} et~al.,}{{Sharp}
  et~al.}{2006}]{sharp2006}
{Sharp} R.,  et~al., 2006, in Society of Photo-Optical Instrumentation
  Engineers (SPIE) Conference Series. p. 62690G (\mn@eprint {arXiv}
  {astro-ph/0606137}), \mn@doi{10.1117/12.671022}

\bibitem[\protect\citeauthoryear{{Smethurst} et~al.,}{{Smethurst}
  et~al.}{2015}]{smethurst2015}
{Smethurst} R.~J.,  et~al., 2015, \mn@doi [\mnras] {10.1093/mnras/stv161},
  \href {https://ui.adsabs.harvard.edu/abs/2015MNRAS.450..435S} {450, 435}

\bibitem[\protect\citeauthoryear{{Somerville}, {Hopkins}, {Cox}, {Robertson}
  \& {Hernquist}}{{Somerville} et~al.}{2008}]{somerville2008}
{Somerville} R.~S.,  {Hopkins} P.~F.,  {Cox} T.~J.,  {Robertson} B.~E.,
  {Hernquist} L.,  2008, \mn@doi [\mnras] {10.1111/j.1365-2966.2008.13805.x},
  \href {https://ui.adsabs.harvard.edu/abs/2008MNRAS.391..481S} {391, 481}

\bibitem[\protect\citeauthoryear{{Springel}, {Di Matteo}  \&
  {Hernquist}}{{Springel} et~al.}{2005}]{springel2005b}
{Springel} V.,  {Di Matteo} T.,   {Hernquist} L.,  2005, \mn@doi [\mnras]
  {10.1111/j.1365-2966.2005.09238.x}, \href
  {https://ui.adsabs.harvard.edu/abs/2005MNRAS.361..776S} {361, 776}

\bibitem[\protect\citeauthoryear{{Strateva} et~al.,}{{Strateva}
  et~al.}{2001}]{strateva2001}
{Strateva} I.,  et~al., 2001, \mn@doi [\aj] {10.1086/323301}, \href
  {https://ui.adsabs.harvard.edu/abs/2001AJ....122.1861S} {122, 1861}

\bibitem[\protect\citeauthoryear{{Sutherland} et~al.,}{{Sutherland}
  et~al.}{2015}]{sutherland2015}
{Sutherland} W.,  et~al., 2015, \mn@doi [\aap] {10.1051/0004-6361/201424973},
  \href {https://ui.adsabs.harvard.edu/#abs/2015A&A...575A..25S} {575, A25}

\bibitem[\protect\citeauthoryear{{Taylor} et~al.,}{{Taylor}
  et~al.}{2011}]{taylor2011}
{Taylor} E.~N.,  et~al., 2011, \mn@doi [\mnras]
  {10.1111/j.1365-2966.2011.19536.x}, \href
  {https://ui.adsabs.harvard.edu/#abs/2011MNRAS.418.1587T} {418, 1587}

\bibitem[\protect\citeauthoryear{{Taylor} et~al.,}{{Taylor}
  et~al.}{2015}]{taylor2015}
{Taylor} E.~N.,  et~al., 2015, \mn@doi [\mnras] {10.1093/mnras/stu1900}, \href
  {https://ui.adsabs.harvard.edu/abs/2015MNRAS.446.2144T} {446, 2144}

\bibitem[\protect\citeauthoryear{{Thorne} et~al.,}{{Thorne}
  et~al.}{2021}]{thorne2021}
{Thorne} J.~E.,  et~al., 2021, \mn@doi [\mnras] {10.1093/mnras/stab1294}, \href
  {https://ui.adsabs.harvard.edu/abs/2021MNRAS.505..540T} {505, 540}

\bibitem[\protect\citeauthoryear{{Thorne} et~al.,}{{Thorne}
  et~al.}{2022}]{thorne2022}
{Thorne} J.~E.,  et~al., 2022, \mn@doi [\mnras] {10.1093/mnras/stab3208}, \href
  {https://ui.adsabs.harvard.edu/abs/2022MNRAS.509.4940T} {509, 4940}

\bibitem[\protect\citeauthoryear{{Trayford}, {Theuns}, {Bower}, {Crain},
  {Lagos}, {Schaller}  \& {Schaye}}{{Trayford} et~al.}{2016}]{trayford2016}
{Trayford} J.~W.,  {Theuns} T.,  {Bower} R.~G.,  {Crain} R.~A.,  {Lagos} C.
  d.~P.,  {Schaller} M.,   {Schaye} J.,  2016, \mn@doi [\mnras]
  {10.1093/mnras/stw1230}, \href
  {https://ui.adsabs.harvard.edu/abs/2016MNRAS.460.3925T} {460, 3925}

\bibitem[\protect\citeauthoryear{{Trayford}, {Lagos}, {Robotham}  \&
  {Obreschkow}}{{Trayford} et~al.}{2020}]{trayford2020}
{Trayford} J.~W.,  {Lagos} C. d.~P.,  {Robotham} A. S.~G.,   {Obreschkow} D.,
  2020, \mn@doi [\mnras] {10.1093/mnras/stz3234}, \href
  {https://ui.adsabs.harvard.edu/abs/2020MNRAS.491.3937T} {491, 3937}

\bibitem[\protect\citeauthoryear{{Vazdekis}, {Koleva}, {Ricciardelli},
  {R{\"o}ck}  \& {Falc{\'o}n-Barroso}}{{Vazdekis} et~al.}{2016}]{vazdekis2016}
{Vazdekis} A.,  {Koleva} M.,  {Ricciardelli} E.,  {R{\"o}ck} B.,
  {Falc{\'o}n-Barroso} J.,  2016, \mn@doi [\mnras] {10.1093/mnras/stw2231},
  \href {https://ui.adsabs.harvard.edu/abs/2016MNRAS.463.3409V} {463, 3409}

\bibitem[\protect\citeauthoryear{{Virtanen} et~al.,}{{Virtanen}
  et~al.}{2020}]{scipy}
{Virtanen} P.,  et~al., 2020, \mn@doi [Nature Methods]
  {https://doi.org/10.1038/s41592-019-0686-2}, \href {https://rdcu.be/b08Wh}
  {17, 261}

\bibitem[\protect\citeauthoryear{{Visvanathan} \& {Sandage}}{{Visvanathan} \&
  {Sandage}}{1977}]{visvanathan1977}
{Visvanathan} N.,  {Sandage} A.,  1977, \mn@doi [\apj] {10.1086/155464}, \href
  {https://ui.adsabs.harvard.edu/abs/1977ApJ...216..214V} {216, 214}

\bibitem[\protect\citeauthoryear{{Walcher}, {Groves}, {Budav{\'a}ri}  \&
  {Dale}}{{Walcher} et~al.}{2011}]{walcher2011}
{Walcher} J.,  {Groves} B.,  {Budav{\'a}ri} T.,   {Dale} D.,  2011, \mn@doi
  [\apss] {10.1007/s10509-010-0458-z}, \href
  {https://ui.adsabs.harvard.edu/abs/2011Ap&SS.331....1W} {331, 1}

\bibitem[\protect\citeauthoryear{{Wetzel}, {Tinker}, {Conroy}  \& {van den
  Bosch}}{{Wetzel} et~al.}{2013}]{wetzel2013}
{Wetzel} A.~R.,  {Tinker} J.~L.,  {Conroy} C.,   {van den Bosch} F.~C.,  2013,
  \mn@doi [\mnras] {10.1093/mnras/stt469}, \href
  {https://ui.adsabs.harvard.edu/abs/2013MNRAS.432..336W} {432, 336}

\bibitem[\protect\citeauthoryear{{Williams}, {Quadri}, {Franx}, {van Dokkum}
  \& {Labb{\'e}}}{{Williams} et~al.}{2009}]{williams2009}
{Williams} R.~J.,  {Quadri} R.~F.,  {Franx} M.,  {van Dokkum} P.,   {Labb{\'e}}
  I.,  2009, \mn@doi [\apj] {10.1088/0004-637X/691/2/1879}, \href
  {https://ui.adsabs.harvard.edu/abs/2009ApJ...691.1879W} {691, 1879}

\bibitem[\protect\citeauthoryear{{Wolf}, {Meisenheimer}, {Rix}, {Borch}, {Dye}
  \& {Kleinheinrich}}{{Wolf} et~al.}{2003}]{wolf2003}
{Wolf} C.,  {Meisenheimer} K.,  {Rix} H.~W.,  {Borch} A.,  {Dye} S.,
  {Kleinheinrich} M.,  2003, \mn@doi [\aap] {10.1051/0004-6361:20021513}, \href
  {https://ui.adsabs.harvard.edu/abs/2003A&A...401...73W} {401, 73}

\bibitem[\protect\citeauthoryear{{Wright} et~al.,}{{Wright}
  et~al.}{2010}]{wright2010}
{Wright} E.~L.,  et~al., 2010, \mn@doi [\aj] {10.1088/0004-6256/140/6/1868},
  \href {https://ui.adsabs.harvard.edu/#abs/2010AJ....140.1868W} {140, 1868}

\bibitem[\protect\citeauthoryear{{Wright}, {Lagos}, {Davies}, {Power},
  {Trayford}  \& {Wong}}{{Wright} et~al.}{2019}]{wright2019}
{Wright} R.~J.,  {Lagos} C. d.~P.,  {Davies} L. J.~M.,  {Power} C.,  {Trayford}
  J.~W.,   {Wong} O.~I.,  2019, \mn@doi [\mnras] {10.1093/mnras/stz1410}, \href
  {https://ui.adsabs.harvard.edu/abs/2019MNRAS.487.3740W} {487, 3740}

\bibitem[\protect\citeauthoryear{{Wyder} et~al.,}{{Wyder}
  et~al.}{2007}]{wyder2007}
{Wyder} T.~K.,  et~al., 2007, \mn@doi [\apjs] {10.1086/521402}, \href
  {https://ui.adsabs.harvard.edu/abs/2007ApJS..173..293W} {173, 293}

\bibitem[\protect\citeauthoryear{{da Cunha}, {Charlot}  \& {Elbaz}}{{da Cunha}
  et~al.}{2008}]{dacunha2008}
{da Cunha} E.,  {Charlot} S.,   {Elbaz} D.,  2008, \mn@doi [\mnras]
  {10.1111/j.1365-2966.2008.13535.x}, \href
  {https://ui.adsabs.harvard.edu/abs/2008MNRAS.388.1595D} {388, 1595}

\bibitem[\protect\citeauthoryear{pandas~development team}{pandas~development
  team}{2021}]{pandas}
pandas~development team T.,  2021, pandas-dev/pandas: Pandas 1.3.3,
  \mn@doi{10.5281/zenodo.5501881}, \url
  {https://doi.org/10.5281/zenodo.5501881}

\bibitem[\protect\citeauthoryear{{van Dokkum} \& {Franx}}{{van Dokkum} \&
  {Franx}}{1996}]{vandokkum1996}
{van Dokkum} P.~G.,  {Franx} M.,  1996, \mn@doi [\mnras]
  {10.1093/mnras/281.3.985}, \href
  {https://ui.adsabs.harvard.edu/abs/1996MNRAS.281..985V} {281, 985}

\makeatother
\end{thebibliography}

\appendix

\section{Validation of the \prospect\ fit of \shark\ galaxies}\label{app:SHARKfit_qual}

\begin{figure*}
    \centering
    \includegraphics[width=\linewidth]{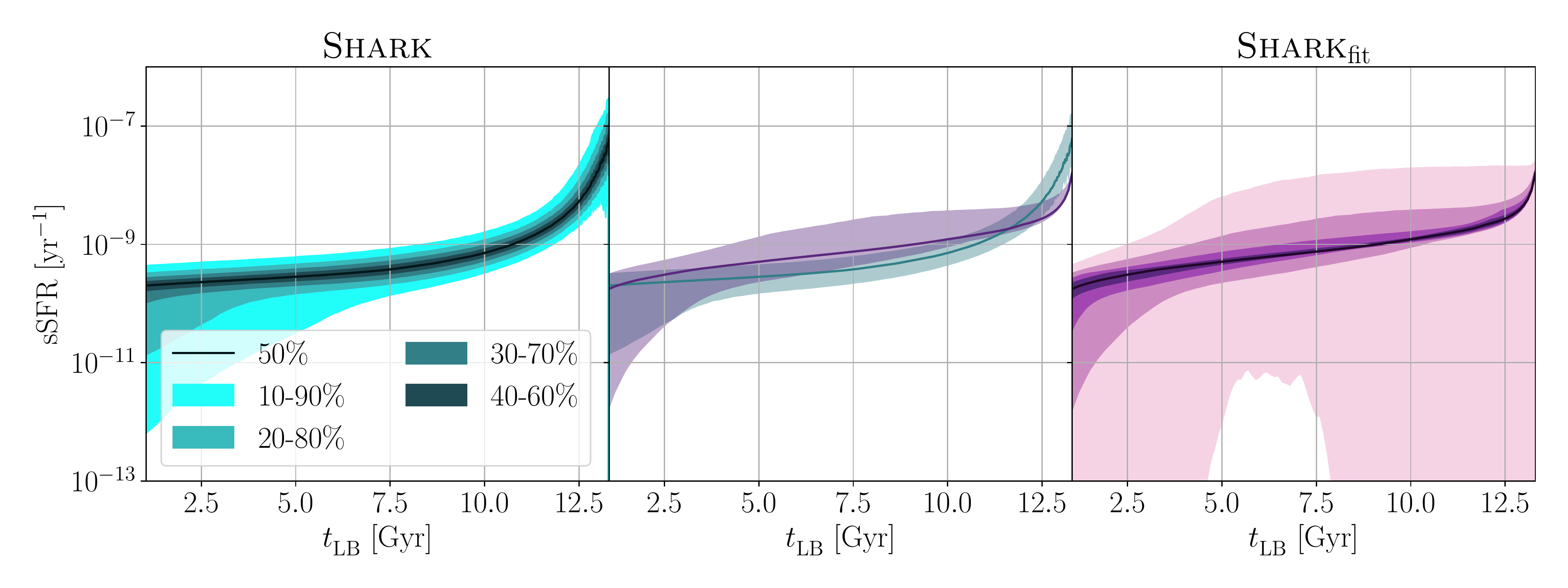}
    \caption{Comparison of the evolution of the sSFRs from \shark\ and \sharkfit.
    Left and right panels: the solid line indicates the running median, while the shaded areas the ranges between different pairs of running percentiles.
    Middle panel: both the running medians and the region between the running $20^\mathrm{th}$-$80^\mathrm{th}$ running percentiles for both \shark\ and \sharkfit.
    Line colours as in Figure \ref{fig:mass_comp}.}
    \label{fig:SHARK_SFH}
\end{figure*}

\begin{figure}
    \centering
    \includegraphics[width=\linewidth]{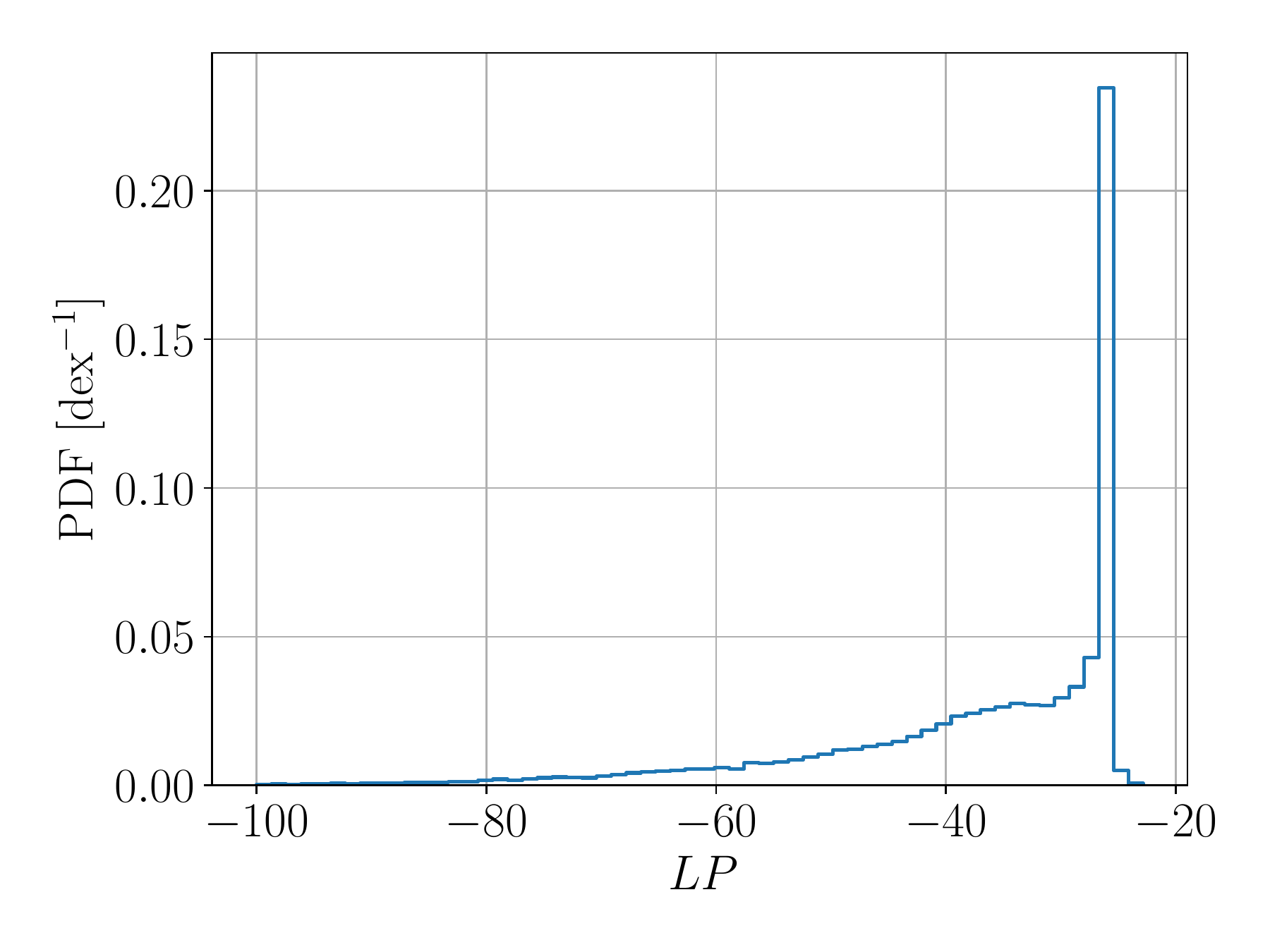}
    \caption{Log-likelihood distribution of the SED fits to our \shark\ galaxy sample.}
    \label{fig:SHARK_lp}
\end{figure}

While we find in general good agreement between the true galaxy properties in \shark\ and the best-fit results from fitting said galaxies with \prospect, Figure \ref{fig:SHARK_SFH} shows that the recovered \shark\ galaxies are too star-forming between $\sim3$ and $\sim10$ Gyr of lookback time.
This poor recovery of the SFH explains what we see in Figure \ref{fig:evol_w}, as the first peak in the stacked SFH in \sharkfit\ sets the early red population that agrees with \shark, but the shift of the stacked SFH towards younger ages delays the downward evolution of the transition mass.
Figure \ref{fig:SHARK_lp} shows no obvious signals of a poor recovery for a subset of the galaxies in the log-likelihood distribution of the fits, e.g., there is no clear evidence for any group of galaxies being stuck in relatively poor fits.

In our extensive exploration of our fitting procedure, we found that we could produce a better recovery of the SFH by fixing the error budget to $10\%$ of the flux (instead of drawing errors from GAMA), but that leads to trade-offs in the recovery of other properties.
The stellar mass recovery proved strongly insensitive to large variations in error budget and number of iterations for the fitting, but colour recovery at all times was in tension with SFH recovery.
This improvement on the SFH recovery also leads to a bimodal log-likelihood distribution, which combined with the previously mentioned tensions lead us to assert that in this work we provide the best SED fitting results for \shark\ galaxies.
The conclusion from these results is that we cannot fully recover the galaxy evolution predicted by \shark\ using \prospect.

The readers familiar with existing literature on SED fitting may recognise that simpler parametric SFH models lead to similar stellar age biases \citep[e.g.,]{iyer2017,leja2019a,lower2020}, which raises the question of whether this is also true for the more flexible skewed-Normal SFHs that we use in \prospect.
While other work using \prospect\ with this SFH model have found good agreement with previous literature on a variety of measurements \citep[e.g.,][]{bellstedt2020b,bellstedt2021,thorne2021}, we acknowledge that we have not directly addressed this topic.
A deep exploration of the differences between models is outside the scope of this work, so here we only offer sufficient evidence to demonstrate that our choice of a parametric SFH is not the cause for our failure to better reconstruct the evolution of \shark\ galaxies.

\subsection{Exploring parametric versus non-parametric SFHs with GAMA galaxies}\label{subapp:PvNP}

To test this, we will use the non-parametric SFH model included in \prospect, which uses 5-segment piece-wise constant SFH\footnote{Called \texttt{massfunc\_b5} in \citet{robotham2020}.}.
This models is formally defined by 11 free parameters, six that define the lookback time limits of each segment plus five for the SFR in each segment, but we leave the former fixed at their default values\footnote{The default values for the transition between segments are: 0, 0.1, 1, 5, 9, and 13 Gyr.} and only fit the latter\footnote{We used a log-Uniform prior range $10^{-7}$--$10^3$ M$_\odot$yr$^{-1}$ for each star formation bin.}. 
A critical challenge for this test is that fitting our \shark\ sample proved to be a factor of $\sim100$ more computationally expensive compared to GAMA galaxies, so we will use the latter for the test.
This decision means that we will not be able to compare to a ground truth, but comparing the skewed-Normal model with the piece-wise constant model should make clear if the former leads to biases in stellar age.

\begin{figure*}
    \centering
    \includegraphics[width=\linewidth]{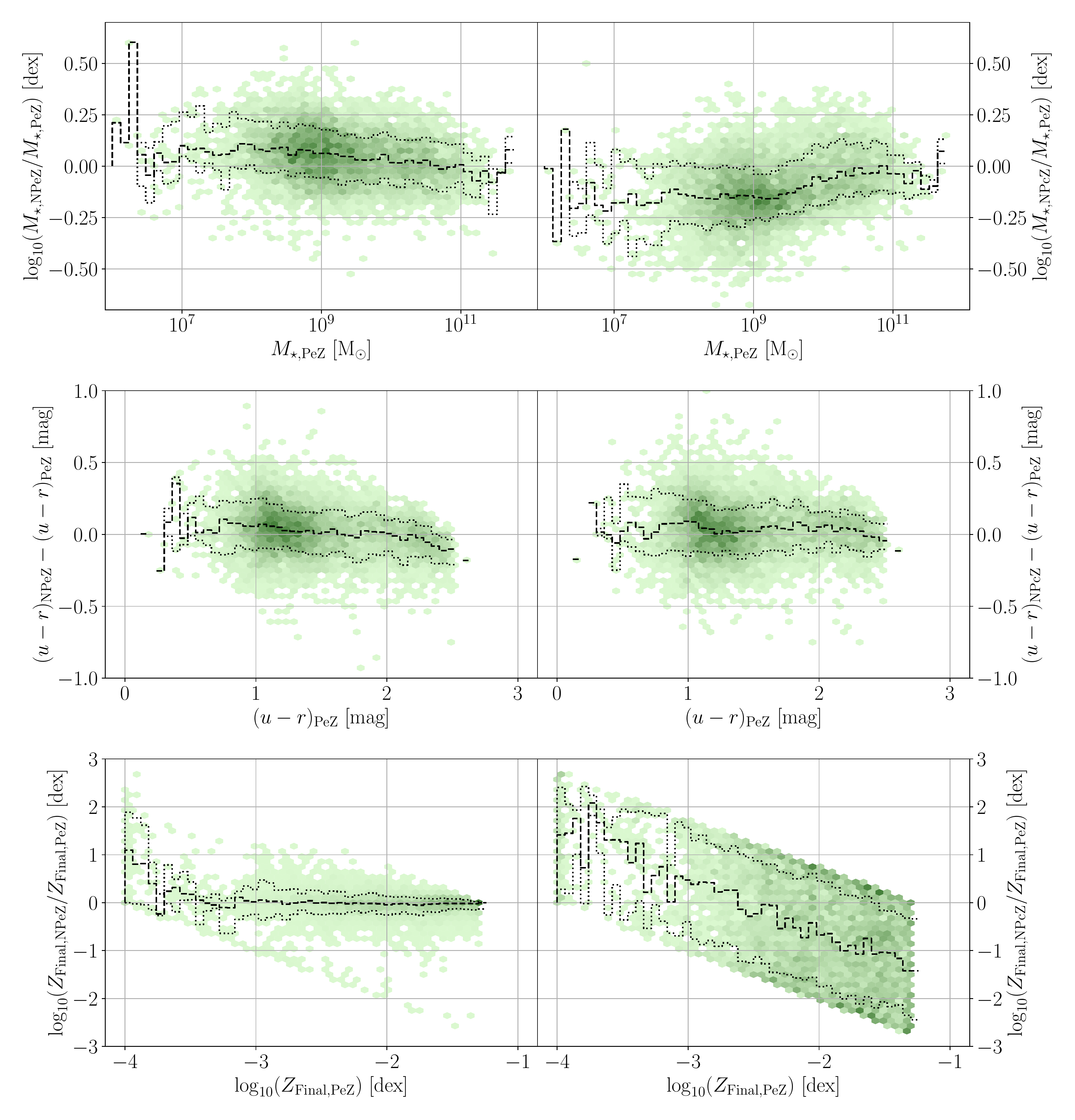}
    \caption{Comparison of the recovered galaxy properties using three different combinations of SFH/$Z$H models for GAMA galaxies.
    The left column shows the comparison between the parametric SFH + evolving $Z$H model (PeZ) and the non-parametric SFH model + evolving $Z$H (NPeZ), the right column a similar comparison between the parametric SFH + evolving $Z$H model and the non-parametric SFH model + constant $Z$H (NPcZ).
    The top row show a comparison for the stellar masses, the middle for the intrinsic (i.e., non-attenuated by dust) $u-r$ colour, and the bottom for the gas-phase metallicity, all at observation time.
    In each panel, bins with a higher number of galaxies are indicated with darker colours, dashed lines indicate the running median, and dotted lines the running 16$^\mathrm{th}$ and 84$^\mathrm{th}$ percentiles.}
    \label{fig:GAMA_SFHprop}
\end{figure*}

Since \prospect\ is unique in the literature in the use of an evolving $Z$H, this is another potential source of differences, so here we compare the results obtained from three SFH/$Z$H combinations:
\begin{itemize}
    \item Parametric + evolving metallicity (PeZ): Skewed-Normal SFH combined with a $Z$H linearly mapped from the mass growth. This is the combination that was the focus of \citet{robotham2020} and used by \citet{bellstedt2020b,bellstedt2021,thorne2021} and this work.
    \item Non-parametric + evolving metallicity (NPeZ): Piece-wise constant SFH combined with a $Z$H linearly mapped from the mass growth. This is a novel combination which enables a direct comparison between the SFH models.
    \item Non-parametric + constant metallicity (NPcZ): Piece-wise constant SFH combined with a constant $Z$H with the metallicity value being a parameter to be fit. This combination offers a direct comparison to prior uses of non-parametric models in the literature.
\end{itemize}

Figure \ref{fig:GAMA_SFHprop} shows a comparison between the recovered stellar masses, colours and gas-phase metallicities for the three combinations of models.
Both PeZ and NPeZ SFH/$Z$H combinations show a remarkable agreement for all properties, with small biases and scatter.
These results show that our choice of SFH does not lead to the disagreements seen in previous literature between parametric and non-parametric SFHs for stellar masses \citep[e.g.,][]{iyer2017,carnall2019,leja2019a,leja2019b,lower2020}.
The comparison between our choice of SFH/$Z$H (PeZ) to that common in literature using non-parametric SFHs (NPcZ) does show tension between stellar masses and gas-phase metallicities, where the stellar masses from the latter combination are weakly biased low compared to the former (by $\sim0.15$ dex), and the metallicities show no correlation between the two models.

\begin{figure}
    \centering
    \includegraphics[width=\linewidth]{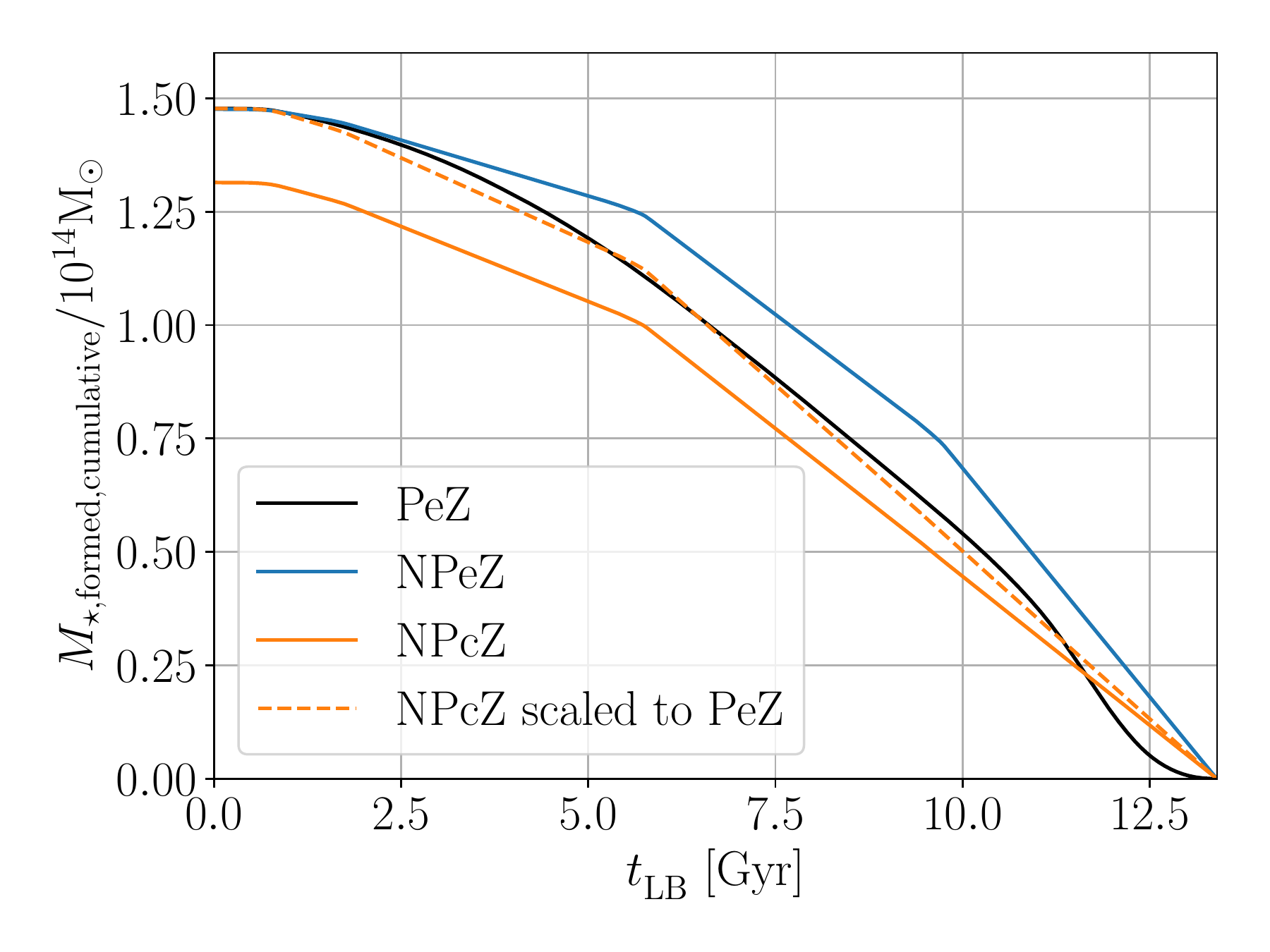}
    \caption{Comparison of the total mass formed between the three combinations of SFH/$Z$H models.
    The parametric SFH + evolving $Z$H is shown in black, the non-parametric SFH + evolving $Z$H in blue, and the non-parametric SFH + constant $Z$H in orange.
    The dashed line in the bottom panel shows the result for the non-parametric SFH + constant $Z$H combination scaled to the same final mass as the parametric SFH + evolving $Z$H combination.}
    \label{fig:GAMA_SFHstack}
\end{figure}

To evaluate for stellar age biases, we show in Figure \ref{fig:GAMA_SFHstack} the total cumulative stellar mass formed for each model combination\footnote{These are the result of integrating the stacked SFH of each model.}.
Once we account for the reduced amount of stellar mass formed by the NPcZ model, both PeZ and NPcZ model are in remarkable agreement.
This indicates that our choice of SFH/$Z$H does not show a bias in stellar ages when compared with literature examples of non-parametric SFHs.
Equally remarkable is that both PeZ and NPcZ are biased toward younger ages when compared with the NPeZ combination.
These results suggests that this combination (non-parametric SFH + evolving $Z$H) deserves to be explored in future work, but for the scope of this work they are enough evidence that the SFH we recover from \shark\ is not a direct consequence of using a parametric SFH.

\subsection{Limitations in the recovery of the \shark\ SFHs}

Our exploration of non-parametric models with GAMA shows that we are not strongly biasing our fits by our choice of SFH model, but that is only true for GAMA and it is possible that rejuvenation plays a larger role in \shark\ galaxies than in GAMA.
To explore this scenario while avoiding the computation cost of re-fitting the SED of our simulated galaxies, we instead use the evolution of the probability of being red ($P^{}_\mathrm{R}$) of \shark\ galaxies.
For this purpose, we choose to define a galaxy as red if $P^{}_\mathrm{R}>0.98$ and blue if $P^{}_\mathrm{R}<0.02$, with galaxies at intermediate values being classified as "uncertain".
The detailed justification for this choice will be given in Bravo et al. (in preparation), but suffice to say here that this is an empirical choice and that small changes to these limits have small effects on our measurements.

For all \shark\ galaxies that have reached the red population ($P^{}_\mathrm{R}>0.98$) at any point in time, we find that $\sim30\%$ of the galaxies that reached the red population have left it at some point, half of which still remained more likely to be red than blue (i.e., $\sim15\%$ crossed $P^{}_\mathrm{R}>0.5$).
Only $\sim10\%$ reached again the blue population ($P^{}_\mathrm{R}<0.02$).
These results show that we are guaranteed to not be able to properly model the evolution of some galaxies in \shark, but this a rare enough occurrence to not affect the overall population evolution\footnote{We would not be justified in our choice of SFH should we be interested in reconstructing the evolution of these galaxies.}, as only $\sim3\%$ of the galaxies become red to then rejuvenate.

Ruling out rejuvenation does not answer why we are not able to recover the evolution of \shark\ galaxies.
Exploring the correlation between true and recovered galaxy properties we found that galaxies above $\sim$\mstar{10.5} are systematically fit with stellar populations that are too red, yet their metallicity are systematically under-predicted.
Their dust opacities are also under-predicted, which to some degree balances out the too-red stellar population.
This is driven by the assumption we make that $\eta^{}_\mathrm{ISM}$ has a fixed value (the same assumption made by \citetalias{bellstedt2020b} and \citetalias{thorne2021}).
This is not true for \shark, as we calculate these values in a per-galaxy basis\footnote{For the birth clouds in \shark\ $\eta^{}_\mathrm{BC}$ is fixed to the same default value in \prospect, $\eta^{}_\mathrm{BC}=-0.7$.}.

Since recovery of dust becomes more challenging for red colours, this leads to the mean of the massive end of the red population being set by our choice of $\eta^{}_\mathrm{ISM}$.
We find that setting a lower value for $\eta^{}_\mathrm{ISM}$ leads to an improved colour recovery.
This improvement is a consequence of massive \shark\ galaxies being bulge-dominated ($B/T\gtrsim0.8$), which are predicted with steeper ISM attenuation slopes than the default value in \prospect\footnote{The disc attenuation slope predicted in \shark\ is in better agreement with the default value.} \citep[see figure 4 of][]{lagos2019}.
We also found that fitting this value instead does not lead to an improvement compared to the defaults assumed for GAMA, suggesting that degeneracies with other parameters may be at play.
We leave for future work to test if whether this is also true for GAMA, and if so what is the best approach to treat these parameters.

\section{Tabulated evolution parameters}\label{app:tables}

Table \ref{t:time_param} shows the best fit parameters for Equations \ref{e:muB1_t}, \ref{e:muB0_t}, \ref{e:muR1_t}, \ref{e:muR0_t}, \ref{e:wMt_t}, \ref{e:wk_t}, \ref{e:sigmaB1_t}, \ref{e:sigmaB0_t}, \ref{e:sigmaR1_t} and \ref{e:sigmaR0_t}.

\begin{table*}
    \centering
    \begin{tabular}{>{$}c<{$}|>{$}r<{\times{}$}@{}>{$}l<{$}|>{$}r<{\times{}$}@{}>{$}l<{$}|>{$}r<{\times{}$}@{}>{$}l<{$}|>{$}r<{\times{}$}@{}>{$}l<{$}}
        \hline
        \multicolumn{1}{c|}{GAMA} & \multicolumn{2}{c|}{$x^{}_{3yz}$} & \multicolumn{2}{c|}{$x^{}_{2yz}$} & \multicolumn{2}{c|}{$x^{}_{1yz}$} & \multicolumn{2}{c}{$x^{}_{0yz}$} \\
        \hline
        \alpha^{}_{\mu\mathrm{B}}    & -2.308 & 10^{-6} &  6.527 & 10^{-3} & -1.064 & 10^{-1} &  5.244 & 10^{-1} \\
        \beta^{}_{\mu\mathrm{B}}     & -7.705 & 10^{-4} &  1.076 & 10^{-2} & -9.390 & 10^{-2} &  1.315 & 10^{ 0} \\
        \alpha^{}_{\mu\mathrm{R}}    &        &         &        &         & -1.766 & 10^{-2} &  3.386 & 10^{-1} \\
        \beta^{}_{\mu\mathrm{R}}     & -8.429 & 10^{-4} &  1.041 & 10^{-2} & -6.740 & 10^{-2} &  2.309 & 10^{ 0} \\
        M^{}_\mathrm{T}              &  6.197 & 10^{-4} & -1.939 & 10^{-2} &  1.543 & 10^{-1} &  1.058 & 10^{ 1} \\
        k                            &  2.477 & 10^{-3} & -2.751 & 10^{-2} &  2.525 & 10^{-1} &  6.593 & 10^{-1} \\
        \alpha^{}_{\sigma\mathrm{B}} & -1.686 & 10^{-4} &  4.211 & 10^{-3} & -2.553 & 10^{-2} &  9.446 & 10^{-2} \\
        \beta^{}_{\sigma\mathrm{B}}  &  3.038 & 10^{-4} & -6.933 & 10^{-3} &  3.596 & 10^{-2} &  1.636 & 10^{-1} \\
        \alpha^{}_{\sigma\mathrm{R}} &        &         & -1.535 & 10^{-3} &  2.504 & 10^{-2} & -1.246 & 10^{-1} \\
        \beta^{}_{\sigma\mathrm{R}}  &  7.031 & 10^{-4} & -1.075 & 10^{-2} &  4.855 & 10^{-2} &  5.125 & 10^{-2} \\
        \hline
        \multicolumn{1}{c|}{\shark}  & \multicolumn{2}{c|}{$x^{}_{3yz}$} & \multicolumn{2}{c|}{$x^{}_{2yz}$} & \multicolumn{2}{c|}{$x^{}_{1yz}$} & \multicolumn{2}{c}{$x^{}_{0yz}$} \\
        \hline
        \alpha^{}_{\mu\mathrm{B}}    &  1.825 & 10^{-3} & -2.527 & 10^{-2} &  5.627 & 10^{-2} &  1.131 & 10^{-1} \\
        \beta^{}_{\mu\mathrm{B}}     &  8.639 & 10^{-7} & -2.171 & 10^{-3} & -9.951 & 10^{-5} &  1.001 & 10^{ 0} \\
        \alpha^{}_{\mu\mathrm{R}}    &        &         &        &         & -3.499 & 10^{-2} &  2.161 & 10^{-1} \\
        \beta^{}_{\mu\mathrm{R}}     & -2.830 & 10^{-5} & -1.293 & 10^{-3} & -4.321 & 10^{-2} &  2.154 & 10^{ 0} \\
        M^{}_\mathrm{T}              &  5.575 & 10^{-3} & -7.713 & 10^{-2} &  3.448 & 10^{-1} &  1.003 & 10^{ 1} \\
        k                            & -1.674 & 10^{-2} &  2.321 & 10^{-1} & -0.702 & 10^{-1} &  2.480 & 10^{ 0} \\
        \alpha^{}_{\sigma\mathrm{B}} &  2.868 & 10^{-4} & -4.080 & 10^{-3} &  1.898 & 10^{-2} &  1.152 & 10^{-1} \\
        \beta^{}_{\sigma\mathrm{B}}  & -2.865 & 10^{-4} &  5.351 & 10^{-3} & -2.889 & 10^{-2} &  1.719 & 10^{-1} \\
        \alpha^{}_{\sigma\mathrm{R}} &        &         &  3.408 & 10^{-3} & -1.670 & 10^{-2} &  4.539 & 10^{-2} \\
        \beta^{}_{\sigma\mathrm{R}}  & -1.301 & 10^{-4} &  4.651 & 10^{-3} & -2.090 & 10^{-2} &  2.443 & 10^{-1} \\
        \hline
    \end{tabular}
    \caption{Values for the time evolution parameterisation of the means (Section \ref{subsubsec:mu}, Equations \ref{e:muB1_t}-\ref{e:muB0_t}), weights (Section \ref{subsubsec:w}, Equations \ref{e:wMt_t}-\ref{e:wk_t}) and standard deviations (Section \ref{subsubsec:sigma}, Equations \ref{e:sigmaB1_t}-\ref{e:sigmaR0_t}).}
    \label{t:time_param}
\end{table*}


\bsp	
\label{lastpage}
\end{document}